%% file: Songetal_SAXS_MS_FR.tex
\documentclass[12pt,amsmath,amssymb,showkeys,prb]{revtex4}
\usepackage{xcolor}
\usepackage{graphicx}
\usepackage{setspace}
\usepackage{bm}
 
\newcommand{\citen}[1]{%
  \begingroup
    \romannumeral-`\x % remove space at the beginning of \setcitestyle
    \setcitestyle{numbers}%
    \cite{#1}%
  \endgroup
}
\newcommand{\ep}{\epsilon_{\rm p}}
\newcommand{\avRgs}{\langle R^2_{\rm g}\rangle}
\newcommand{\avREEs}{\langle R^2_{\rm EE}\rangle}

\begin{document}

$\null$
\hfill {{\bf June 9, 2021}} 
\vskip 0.3in

\begin{center}
{\Large\bf Small-Angle X-Ray Scattering Signatures of}\\ 
\vskip 0.15cm

{\Large\bf Conformational Heterogeneity and}\\
\vskip 0.15cm

{\Large\bf Homogeneity of Disordered Protein Ensembles}\\

\vskip .5in
%
%J. Song and H. S. Chan$^*$\\

{\bf Jianhui S{\footnotesize{\bf{ONG}}}}$^{1,*}$,
{\bf Jichen L{\footnotesize{\bf{I}}}}$^{1}$ and
{\bf Hue Sun C{\footnotesize{\bf{HAN}}}}$^{2,*}$

$\null$
$^1$ School of Polymer Science and Engineering,
Qingdao University of\\ Science and Technology,
53 Zhengzhou Road, Qingdao 266042, China;\\
$^2$ Department of Biochemistry,
University of Toronto,\\
Toronto, Ontario M5S 1A8, Canada

\vskip 1.3cm

%{\tt Submitted to ""}
%{\tt To appear in ""}
%

\end{center}

\noindent
\noindent
$^*$
Corresponding authors.\\
{\phantom{$^*$\ }}Hue Sun Chan.  E-mail: {\tt chan@arrhenius.med.utoronto.ca}\\
{\phantom{$^*$\ }}Jianhui Song.  E-mail: {\tt jhsong@qust.edu.cn}\\

\vskip 3cm

\begin{figure}[ht]
\noindent
--------------------------------------------------------------------------------------------------------------------\\
\vskip 3mm
{\includegraphics[height=45mm,angle=0]{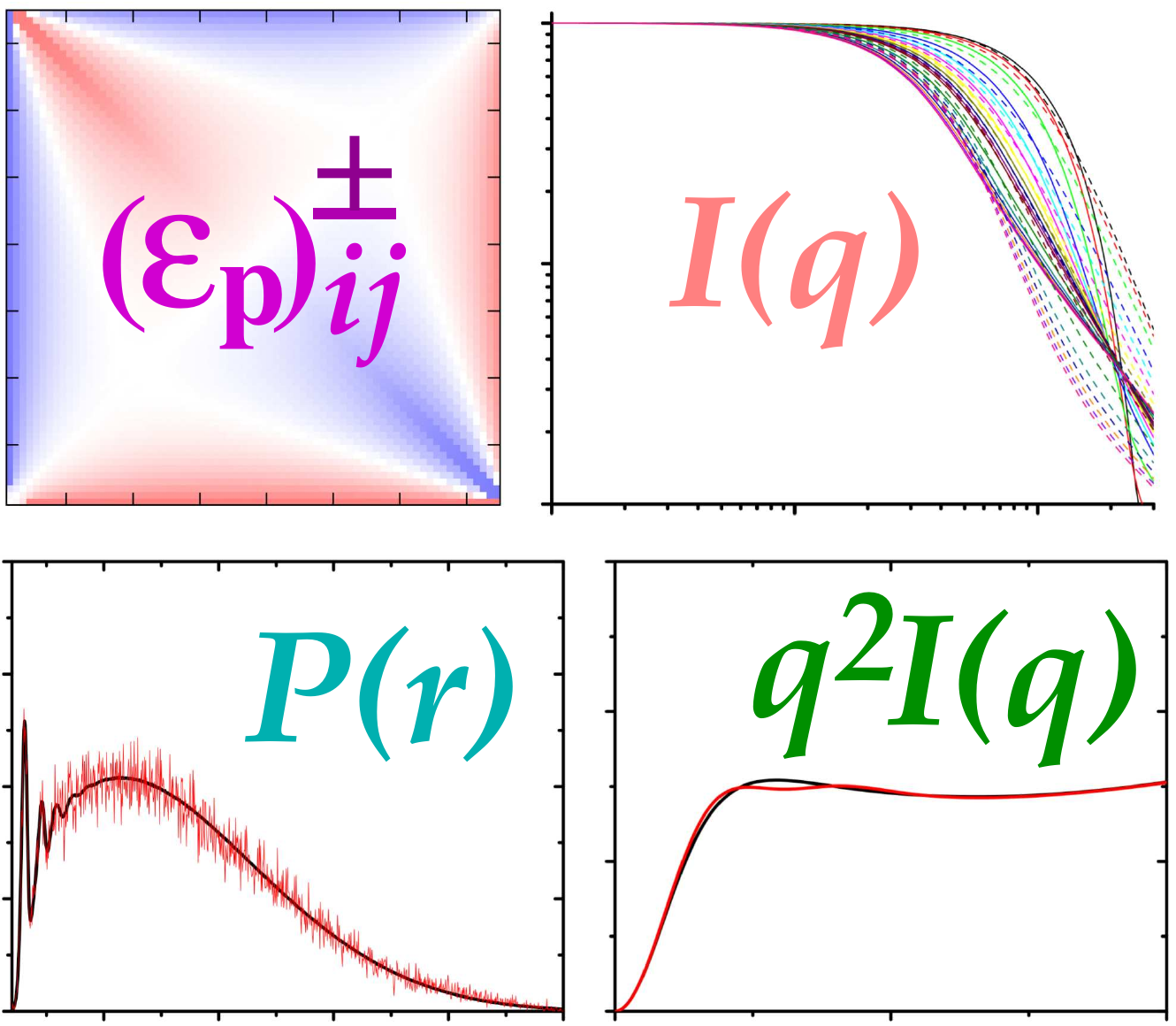}}
\vskip 3mm
\centerline{\large\bf TOC Graphics}
\end{figure}

\vfill\eject
%\endtitlepage
%-----------------------------------------------------------------------------
%\def\thefootnote{\fnsymbol{footnote}}
%\def\thefootnote{$\dagger$}

\noindent
{\large\bf Abstract}\\

An accurate account of disordered protein conformations is of central 
importance to deciphering the physico-chemical basis of biological
functions of intrinsically disordered proteins and the folding-unfolding
energetics of globular proteins. Physically, disordered ensembles of
non-homopolymeric polypeptides are expected to be heterogeneous;
i.e., they should differ from those homogeneous ensembles
of homopolymers that harbor an essentially
unique relationship between average values of
end-to-end distance $R_{\rm EE}$ and radius of gyration $R_{\rm g}$.
It was posited recently, however, that small-angle X-ray scattering (SAXS)
data on conformational dimensions of disordered proteins can be rationalized 
almost exclusively by homopolymer ensembles. Assessing this perspective, 
chain-model simulations are used to evaluate
the discriminatory power of SAXS-determined molecular form factors (MFFs)
with regard to homogeneous versus heterogeneous ensembles. The general
approach adopted here is not bound by any assumption about ensemble 
encodability, in that the postulated heterogeneous ensembles we evaluated 
are not restricted to those entailed by simple interaction schemes. 
Our analysis of MFFs for certain heterogeneous ensembles 
with more narrowly distributed $R_{\rm EE}$ and $R_{\rm g}$ indicates
that while they deviates from MFFs of homogeneous ensembles, the 
differences can be rather small. 
Remarkably, some heterogeneous ensembles with asphericity and 
$R_{\rm EE}$ drastically different from those of homogeneous ensembles 
can nonetheless exhibit practically identical MFFs, demonstrating that SAXS 
MFFs do not afford unique characterizations of basic properties of 
conformational ensembles in general. In other words, the
ensemble to MFF mapping is practically 
many-to-one and likely non-smooth. 
Heteropolymeric variations of the 
$R_{\rm EE}$--$R_{\rm g}$ relationship were further showcased using an
analytical perturbation theory developed here for flexible heteropolymers.
Ramifications of our findings for interpretation of experimental data 
are discussed.
\\

\vfill\eject
%%%%%%%%%%%%%%%%%%%%%%%%%%%%%%%%%%%%%%%%%%%%%%%%%%%%%%%%%%%%%%%%%%%%%%%%%%%%%%%%
\noindent
{\large\bf INTRODUCTION}\\ 

%%%%%%%%%%%%%%%%%%%%%%%%%%%%%%%%%%%%%%
% STARTS HERE - SEP 28, 2020 %%%%%%%%%
%%%%%%%%%%%%%%%%%%%%%%%%%%%%%%%%%%%%%%

A detailed characterization of the conformational properties of disordered 
protein states is essential for many areas of biophysical and biomedical
research. These include, but are not limited to, the thermodynamic balance 
between folded and unfolded states of globular 
proteins\cite{Trewhella1992,julie1995,Plaxco1999,shortle2001,shimizu2002,tobin2004,plaxco2004,rose2004,haran2006,eaton2007,MarshJulie2009,DanRohit2013,raleighBJ2018}
and the relationships between myriad biological functions of the 
increasing repertoire of intrinsically disordered proteins (IDPs) and the 
behaviors of their conformational 
ensembles.\cite{borg07,tanja08,schuler10,tanja2010,fuzzy12,marsh12,baoxu2014,martin2016,zhirong2016a,zhirong2016b,AlexRohitRev2018}
One basic property of disordered conformational ensembles is their
extent in space. This property, often referred to as conformational
dimensions, is of central importance to protein science because it bears
on the very nature of the folding/unfolding cooperativity of globular 
proteins\cite{chanetal2004,fersht2009,chanetal2011,munoz2012,TaoPCCP,Reddy2016,DT_TiBS2-19} 
as well as, for example, the spatial ranges and other 
configurational features of biomolecular interactions involving 
IDPs\cite{PGW2000,zhirong2009,Veronika2017} including those of highly 
disordered ``fuzzy'' dynamic IDP complexes.\cite{sigalov_etal2007,kragelund2008,tanja2014,sigalov2016,wu_etal2017,Amin_etal2020}
Recent advances in the studies of biomolecular 
condensates\cite{cliff2017,rosen2017,lin_biochemrev} suggest further that 
conformational dimensions of individual IDP molecules may serve as an 
indicator of the propensity of the IDP to undergo liquid-liquid phase 
separation.\cite{Amin_etal2020,lin2017,jeetainPNAS,joanJPCL,rohitBJ2020}
Indeed, despite the low-spatial-resolution information they provide directly,
measures of conformational dimensions of disordered protein 
states such as end-to-end distance and radius of gyration (denoted, 
respectively, as $R_{\rm EE}$ and $R_{\rm g}$ hereafter) offer
fundamental insights into the microscopic physical interactions underlying
protein behaviors\cite{DavidShaw2,cosb15} and thus provide critical 
assessments of the extent to which current molecular dynamics force fields 
are adequate for capturing the physics of these 
interactions.\cite{sarah15,sarah17,best2017,shea2017,DEShaw2018,joan2021}

Small-angle X-ray scattering (SAXS) is a commonly utilized technique in 
biophysics\cite{doniachRev2001,SAXS-Complex,AndoChemRev2017} 
to quantify conformational dimensions of disordered proteins by measuring
$R_{\rm g}$ and related ensemble-averaged spatial 
properties.\cite{Trewhella1992,Plaxco1999,tobin2004,plaxco2004,tobinkevinJMB2012,OsmanJMB2014,SvergunFEBSLett,tobinkevinPNAS2015,antonov2016,benPNAS2016,tanja2021} 
Because SAXS takes into account simultaneously 
many positions along the entire chain molecule, SAXS affords information 
complementary to techniques such as F\"orster resonance energy 
transfer (FRET)\cite{haran2006,eaton2007,schuler10,baoxu2014,haran2012,schulerCOSB2013,blanchard2014,rhoades2014,deniz2014,arash2015,rhoades2016,rhoades2016,schulerAnnuRev2016} 
that probe only one or a few relative positions 
at a time. Nonetheless, since scattering intensities are averaged over 
different chain conformations in an ensemble, SAXS data do not provide 
detailed spatial information of individual conformations. Therefore, models 
and assumptions often need to be invoked to relate SAXS data to putative 
conformational ensembles that likely---though not necessarily---underlie 
the experimental data. Recently, the generality of some of these
assumptions, or lack thereof, has been brought into a sharper focus. One of
the reasons is that for several disordered protein states, the $R_{\rm g}$s 
extracted from SAXS using Guinier analysis disagree significantly with 
the $R_{\rm g}$ values inferred from single-molecule FRET (smFRET) data 
by assuming that the underlying conformational distribution 
is Gaussian.\cite{tobinkevinJMB2012,tobinkevinPNAS2015,dt2009,SchulerJCP2013,Songetal2015} 

As we\cite{Songetal2015,Songetal2017} and 
others\cite{lemke2017,raleighPNAS2019} have noted,
besides improving the treatment of excluded volume in the underlying 
baseline homopolymer chain 
model\cite{domb1965,fisher1966,cloizeaux1974,oono_ren81-3} 
used in SAXS and smFRET data analysis\cite{SchulerBestJCP2018,zhengbest2018}
[see, e.g., Eq.~(5.6) of ref.~\citen{oono_ren81-3}],
the apparent mismatches between SAXS- and smFRET-inferred $R_{\rm g}$s 
should be fundamentally reconcilable by recognizing that conformational 
ensembles of proteins are 
heterogeneous\cite{Songetal2017,Vendruscolo2007,Lyle_etal_2013,RuffHolehouse,best2020} 
in that they do not necessarily
resemble those of homopolymers, because proteins are heteropolymeric 
sequences of different amino acid residues. Simply put,
proteins are not homopolymers. Therefore,
the relationship between their average $R_{\rm EE}$ and their 
average $R_{\rm g}$ can differ from that of homopolymers, i.e.,
the relation can be different from that posited by Gaussian chain and 
other homopolymeric models.\cite{Songetal2015,Songetal2017} 
It is possible, then, for heterogeneous 
conformational ensembles to embody both a smFRET-inferred average 
$R_{\rm EE}$ and a SAXS-determined average $R_{\rm g}$ that are not coupled 
in the same way as for 
homopolymers.\cite{Songetal2015,Songetal2017,lemke2017} 
This conceptual framework has 
been applied to rationalize experimental data with apparent 
success.\cite{Songetal2015,lemke2017,raleighPNAS2019,claudiu2016,claudiu2017,GregJACS2020}

In this context, it is notable that recent experimental developments 
emphasize that one should be able to glean more structural information 
about disordered protein ensembles from SAXS data 
than merely extracting the mean square radius of 
gyration, $\langle R^2_{\rm g}\rangle$, by using Guinier analysis, which
relies on scattering intensity, $I(q)$, at small $q$ values, with
$q=|{\bf q}|$ being the magnitude of the scattering vector ${\bf q}$. In 
contrast, some of the recent SAXS studies of disordered protein ensembles
consider Kratky plots, or molecular form factors (MFFs), over a 
substantially wider range of $q$ (refs.~\citen{tobinSci2017,SciComment1,SciResponse2,SciComment3}). 
Logically, the more enriched information provided by MFFs, namely the $I(q)$ 
at larger $q$s, is expected to impose more experimental constraints 
on putative, theoretically constructed conformational ensembles 
beyond merely requiring them to have a given $\langle R^2_{\rm g}\rangle$. 
Accordingly, this recognition raises basic questions as to 
whether theoretical/computational 
heterogeneous conformational ensembles constructed to satisfy a given 
$\langle R^2_{\rm g}\rangle$ and a given $\langle R^2_{\rm EE}\rangle$ 
remain viable when MFFs are taken into account; that is, whether those 
putative heterogeneous ensembles are also consistent with the additional 
information afforded by the larger-$q$ behaviors of 
experimental MFFs. To address this and related 
questions, it should be recognized first that although 
MFFs provide more structural 
information on disordered protein ensembles than Guinier analysis, MFFs still 
involve extensive averaging over individual conformations and therefore
an MFF by itself is far from being able to uniquely define a disordered
conformational ensemble. With these considerations in mind, we seek here 
to clarify the information content of SAXS-determined MFFs by investigating
the compatibility of various putative conformational ensembles with given MFFs. 
As will be apparent below, this delineation is useful toward establishing a
more rigorous perimeter for interpreting SAXS-determined MFFs of disordered 
proteins in terms of heterogeneous conformational ensembles.

To this end, we use explicit-chain simulations of a coarse-grained 
polypeptide model\cite{Songetal2015} 
to construct extensive sets of different heterogeneous ensembles with 
properties selected systematically for the insights they would provide. We
then compare their MFFs with those of full ensembles of homopolymers embodying 
varying degrees of uniform intrachain attractive or repulsive interactions,
paying special attention to identify and evaluate scenarios in which 
the MFFs of heterogeneous and homogeneous ensembles are highly similar.
Building on prior advances,
we commence this effort with a survey of the $(R_{\rm g},R_{\rm EE})$ 
subensembles introduced previously to address the apparent SAXS--smFRET 
mismatches in $R_{\rm g}$ measurement.\cite{Songetal2015,Songetal2017,DICE}
Each of these subensembles is individually a heterogenoues conformational 
ensemble because it is defined to be a small part of a homogeneous ensemble 
and therefore {\it not} a homogeneous ensemble by itself.
Interestingly, while the MFFs of different subensembles sharing
the same $R_{\rm g}$ with the $\langle R^2_{\rm g}\rangle^{1/2}$ 
of the homogeneous ensemble differ among themselves because the MFF depends 
on the subensemble's $R_{\rm EE}$, the MFFs of some subensembles are quite 
similar to the MFF of the full homogeneous ensemble. 

Besides the $(R_{\rm g},R_{\rm EE})$ subensembles, 
other heterogeneous ensembles with more diverse conformations, i.e.,
not limited to a very narrow range of $R_{\rm g},R_{\rm EE}$ values,
are also assessed. Motivated partly by experimental evidence suggesting that 
disordered protein ensembles are heterogeneous in a sequence-sensitive 
manner\cite{DanRohit2013,GregJACS2020}
(sometimes manifested by peculiar forms of 
the inferred $R_{\rm g}$ distribution\cite{tanja2010})
despite their homopolymer-like
overall average $R_{\rm g}$ values,\cite{raleighPNAS2019,raleighBiochem2020}
we study several physically plausible, conformationally diverse 
heterogeneous ensembles with narrower distributions of $R_{\rm g}$ than 
that of a homogeneous ensemble. 
Quite surprisingly, these heterogeneous ensembles nonetheless lead
to MFFs very similar to that of the corresponding homogeneous ensemble. 
Indeed, we even come across other mathematically intriguing cases where
heterogeneous ensembles drastically different from homopolymers 
yet possess MFFs that are essentially identical to MFFs of homopolymers. 
These comparisons indicate that the MFFs of homogeneous ensembles and 
some heterogeneous ensembles can be
practically indistinguishable when experimental uncertainties 
are allowed for, underscoring the desirability of employing additional
experiment techniques complementary to SAXS to better characterize 
disordered protein ensembles.\cite{lemke2017,raleighPNAS2019,GregJACS2020}
Aiming for rudimentary insights into the complex sequence-ensemble 
relationship of heteropolymeric disordered proteins, 
we have also developed an extension of theoretical perturbative 
approaches\cite{fixman1955,fixman1962a,fixman1962b,Edwards1965,tanaka1966,ohta81,Freedbook,ohta82,muthu84,chanJCP89,chanJCP90}
to calculate $\langle R^2_{\rm g}\rangle^{1/2}$,
$\langle R^2_{\rm EE}\rangle^{1/2}$, scattering intensity $I(q)$, 
and MFF$(q\avRgs^{1/2})$ for heteropolymers. Predictions of this 
analytical formulation 
allow for a preliminary understanding of how sequence-specific interactions
may encode heterogeneous ensembles that share the same 
$\langle R^2_{\rm g}\rangle^{1/2}$ but differ in other aspects
of their conformational distributions such as their 
root-mean-square end-to-end distance $\langle R^2_{\rm EE}\rangle^{1/2}$. 
Details of these findings are provided below.
\\

\vfill\eject

\noindent
{\large\bf MODELS AND METHODS}\\ 

The present coarse-grained C$_\alpha$ protein model (one bead per monomer, 
or per residue) and the sampling algorithm are based on the formulation 
used in our previous investigations of smFRET interpretation for disordered
proteins.\cite{Songetal2015,Songetal2017} As before,
a polypeptide chain is modeled as a chain of $n$ beads 
labeled by $i=1,2,\dots,n$ at position ${\bf R}_i$, 
with C$_\alpha$--C$_\alpha$ virtual bond length 
$b=|{\bf r}_{i,i+1}|\equiv |{\bf R}_{i+1}-{\bf R}_i|=3.8$~\AA~between
connected beads.
The bond-angle potential energy $U_{\rm bond} (\{{\bf r}\})
=\sum_{i=2}^{n-1}\epsilon_\theta(\theta_i-\theta_0)^2$,
where $\epsilon_\theta=10.0 k_{\rm B}T$, 
$\theta_i=\cos^{-1}({\bf r}_{i,i-1}\cdot{\bf r}_{i,i+1}/b^2)$ 
is the virtual bond angle at bead $i$, $\theta_0=106.3^\circ$ is the 
reference bond angle corresponding to the most populated virtual bond 
angle in the Protein Data Bank,\cite{levitt1976} $k_{\rm B}$ 
is Boltzmann constant, and $T$ is absolute temperature. 
The potential energy for excluded volume is given by
$U_{\rm SAW}(\{{\bf r}\})
=(1/2)\sum_{i=1}^n\sum_{j=1}^n(U_{\rm SAW})_{ij}(r_{ij})$
where ``SAW'' stands for self-avoiding walk and
$(U_{\rm SAW})_{ij}\equiv\epsilon_{\rm ex}(R_{\rm hc}/r_{ij})^{12}$.
In this expression,
$r_{ij}\equiv |{\bf R}_j-{\bf R}_i|$ is the distance between beads $i$
and $j$, 
and $\epsilon_{\rm ex}=1.0 k_{\rm B}T$ is the 
self-avoiding excluded-volume repulsion strength used in our model.
As in many simulation studies of protein folding\cite{chanetal2011}
and in most of the cases we considered in 
refs.~\citen{Songetal2015}~and~\citen{Songetal2017},
a hard-core repulsion distance $R_{\rm hc}=4.0$~\AA~is adopted.
In addition to the non-bonded excluded-volume repulsive term, here we consider 
also homopolymeric, uniform intrachain non-bonded attractive or repulsive 
interactions given by
$U_{\rm p}(\epsilon_{\rm p},r_0;\{{\bf r}\})
=\sum_{i=1}^n\sum_{j=i+3}^n(U_{\rm p})_{ij}(\epsilon_{\rm p},r_0;r_{ij})$
where $(U_{\rm p})_{ij}(\epsilon_{\rm p},r_0;r_{ij})\equiv
\epsilon_{\rm p}\; \exp[-(r_{ij}/r_0)^2]$
to account for polypeptides under different solvent conditions.
For computational efficiency, all non-bonded potential energy terms are
set to zero for $r_{ij}\ge 10.0$~\AA. 
Illustrative examples of a combination of excluded-volume
and $\ep$-dependent attractive or repulsive interactions 
are provided in Fig.~1. 
In the analysis below, we refer to the $\ep=0$ case as the SAW model, and
the case with $\ep=0$ as well as with $U_{\rm SAW}$ turned off
(effectively setting $\epsilon_{\rm ex}=0$) as the Gaussian chain model.
Theoretical predictions reported in 
this work are for chain length $n=75$, which we have used\cite{Songetal2017} 
to address experimental SAXS and smFRET data on Protein L 
(refs.~\citen{tobinkevinJMB2012,tobinkevinPNAS2015}). By using
this chain length, new findings are amenable to comparison with previous 
results. In the present context, however, $n=75$ is taken
merely as an exemplifying case for proteins of similar lengths
because the focus of this study is on general principles rather than 
any particular protein. 

%%%%%%%%%%%%%%%%%%%%%%%%%%%%%%%%%%%%%%%%%%%%%%%%%%%%%%%%%%%%%%%%%%%%%%%%%%%%%%%%
\begin{figure}[ht]
\vskip -0.5cm
{\includegraphics[height=65mm,angle=0]{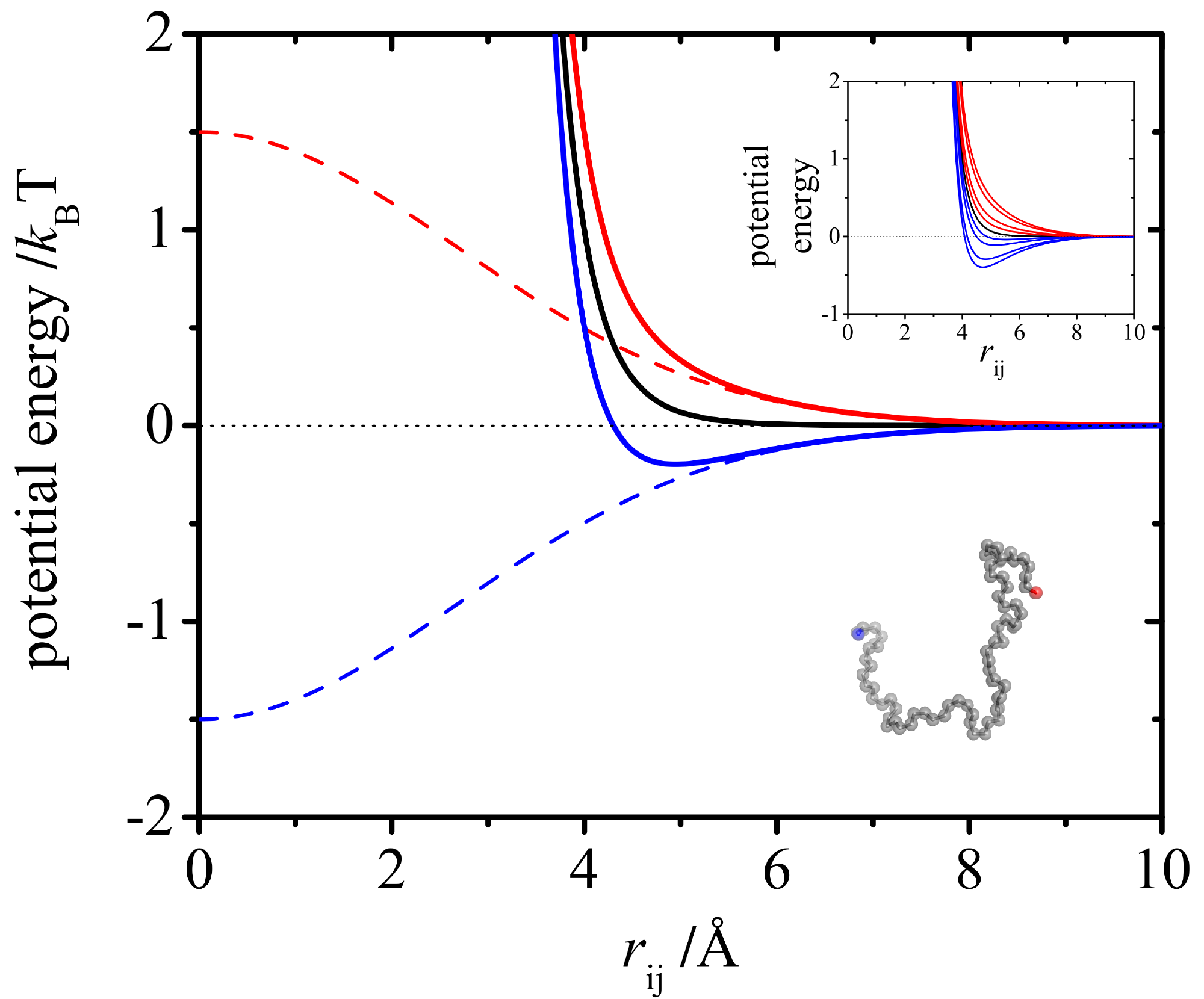}}
\begin{center}
\end{center}
\vspace{-1.5cm}
\caption{Non-bonded interactions in the coarse-grained chain model.
The total pairwise potential energy (solid red or blue curves) between two
monomers $i,j$ that are not directly connected along the polymer chain
($|i-j|>1$) is the sum of the repulsive SAW potential
$(U_{\rm SAW})_{ij}$ (solid black curves) and a pairwise attractive
($\ep<0$, blue) or repulsive ($\ep>0$, red) interaction
$(U_{\rm p})_{ij}(\epsilon_{\rm p},r_0;r_{ij})$ ($r_0=3.8$ \AA, dashed curves).
This interaction reduces to the SAW potential when $\ep=0$
(see text for details).  $(U_{\rm p})_{ij}(\epsilon_{\rm p},r_0;r_{ij})$
is given here for $\ep/k_{\rm B}T=\pm 1.5$ as an example.
The total non-bonded potentials for
$\ep/k_{\rm B}T=\pm 0.5$, $\pm 1.0$, $\pm 2.0$, and $\pm 2.5$ are provided by
the inset as further illustrations. An example conformation is shown for the
$n=75$ model polymer utilized in the present investigation. The red and blue
beads mark the chain termini, positions corresponding to those of FRET dyes
for determining $R_{\rm EE}$ in our previous
study.\cite{Songetal2017}
}
\label{fig1}
\end{figure}
%%%%%%%%%%%%%%%%%%%%%%%%%%%%%%%%%%%%%%%%%%%%%%%%%%%%%%%%%%%%%%%%%%%%%%%%%%%%%%%%
 
Chain conformations are sampled at $T=300$ K using standard 
Metropolis Monte Carlo techniques\cite{MC} 
described before,\cite{song13} wherein equal a priori probabilities
are assigned to pivot and kink jumps,\cite{stockmayer1962,Lal69} 
with acceptance rate of $\approx 30\%$ for the attempted chain 
moves.\cite{Songetal2015,Songetal2017}
For each simulation, the first $10^7$ attempted moves for equilibration
are not used for the calculation of average conformational properties.
Subsequently, $\sim 10^9$ moves are attempted 
to sample $\sim 10^7$ conformations (snapshots taken every 100 attempted
moves) for further analysis.  Among other ensemble properties to
be described in the Results section below, radius of gyration
$R_{\rm g} = [n^{-1}\sum_{i=1}^n |{\bf R}_i-{\bf R}_{\rm cm}|^2]^{1/2}$
(where ${\bf R}_{\rm cm}$ is the center of mass or centroid position, 
${\bf R}_{\rm cm}= n^{-1}\sum_{i=1}^n {\bf R}_i$) and end-to-end 
distance $R_{\rm EE}=|{\bf R}_n-{\bf R}_1|$ 
are computed from the sampled conformations. Examples  
of the distribution of $R_{\rm g},R_{\rm EE}$ populations
for several different values of the intrachain interaction energy parameter
$\ep$ are shown in Fig.~2. As expected, when intrachain interaction is
attractive ($\ep<0$), the distribution shifts to smaller $R_{\rm g}$
and $R_{\rm EE}$ (Fig.~2b,c) relative to the SAW distribution
($\ep=0$, Fig.~2a), whereas the distribution shifts to larger $R_{\rm g}$
and $R_{\rm EE}$ when intrachain interaction is
repulsive ($\ep>0$, Fig.~2d).

%%%%%%%%%%%%%%%%%%%%%%%%%%%%%%%%%%%%%%%%%%%%%%%%%%%%%%%%%%%%%%%%%%%%%%%%%%%%%%%%
\begin{figure}[ht]
{\includegraphics[height=40mm,angle=0]{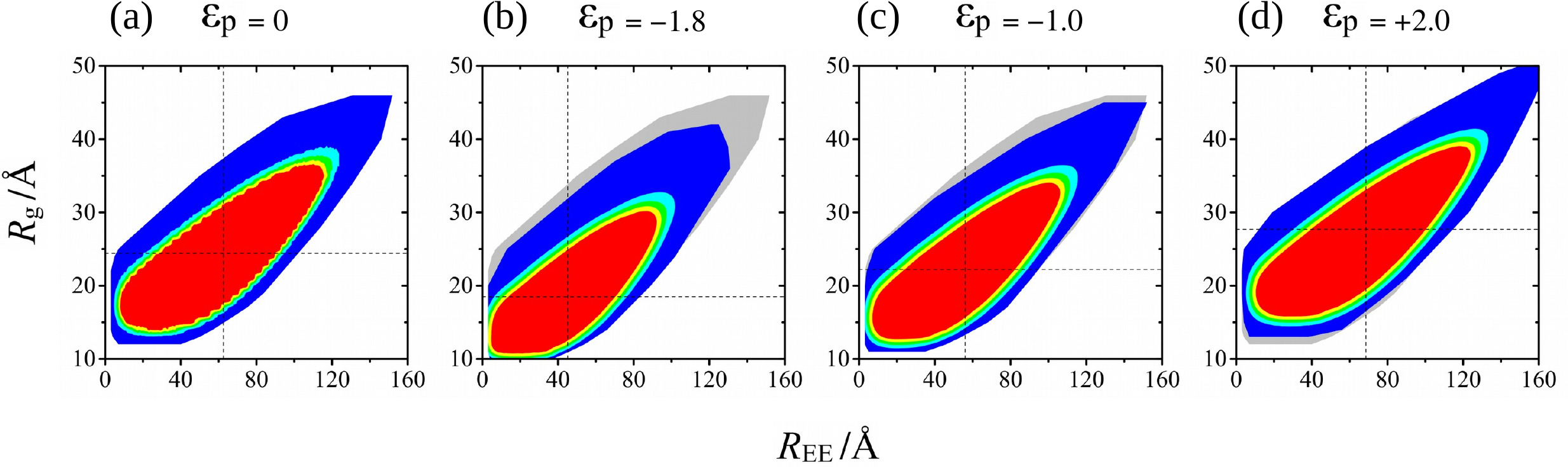}}
\begin{center}
\end{center}
\vspace{-1.3cm}
\caption{Radius of gyration and end-to-end distance distributions of
$\ep$-dependent homopolymers of various compactness.
Joint distributions of $R_{\rm g}$ and $R_{\rm EE}$ (color coded) are
shown for $n=75$ conformational ensembles at representative $\ep$ values
as indicated ($\ep$ given in units of $k_{\rm B}T$ hereafter).
The profile of the SAW [$\ep=0$, (a)] distribution is included
as a gray background in the other three $\ep\neq 0$ distributions [(b)--(d)] 
for comparison. For each ensemble, the root-mean-square
$\langle R_{\rm g}^2\rangle^{1/2}$
and
$\langle R_{\rm EE}^2\rangle^{1/2}$
are marked by the horizontal and vertical dotted lines respectively.
A total of $10^7$ conformations are sampled for each distribution. Numbers
of conformations are recorded for all
$R_{\rm g}$--$R_{\rm EE}$
bins of $1{\rm \AA}\times 1{\rm \AA}$ and are color-coded as follows.
White: 0, blue: [1--200), cyan: [200--400), green: [400--600), yellow:
[600--800), and red: $\geq 800$ conformations.
}
\label{fig2}
\end{figure}
%%%%%%%%%%%%%%%%%%%%%%%%%%%%%%%%%%%%%%%%%%%%%%%%%%%%%%%%%%%%%%%%%%%%%%%%%%%%%%%%

Fig.~3 provides an overview of the properties of the $\ep$-dependent
homogeneous conformational ensembles that are used as baselines in work.
Fig.~3a shows that for these homogeneous ensemble, 
$\avRgs/\avREEs$ is essentially constant at $\approx 0.16$ for SAW ($\ep=0$)
and ensembles with repulsive interactions ($\ep>0$), and the ratio increases
as the ensembles become more compact with attractive interactions ($\ep$
more negative), reaching $\approx 0.36$ for $\ep=-5.0$. This trend is in 
line with the simulated $\langle R_{\rm g}\rangle/\langle R_{\rm EE}\rangle$
values of $\approx 0.41$ for SAWs of comparable lengths and a rough estimate 
of $\langle R_{\rm g}\rangle/\langle R_{\rm EE}\rangle$ of $0.71$ for 
conformations in the shape of a compact sphere\cite{Songetal2015}
because $0.41^2=0.17$ and $0.71^2=0.50$ although
$\avRgs/\avREEs \neq (\langle R_{\rm g}\rangle/\langle R_{\rm EE}\rangle)^2$
($\langle\dots\rangle$ represents averaging over a given ensemble).
The scaling of intrachain distance $r_{ij}$ of these ensembles in the 
form of $\langle r^2_{ij}\rangle^{1/2}\sim |i-j|^\nu$ is shown in Fig.~3b
and the estimated $\ep$-dependent $\nu$ exponents are provided in Fig.~3c.
The tendency for the scaling exponents for smaller $|i-j|$ (red diamonds in
Fig.~3c) to be slightly higher than those for larger $|i-j|$ (black circles 
in Fig.~3c) is in line with that seen in recent simulation results
(e.g., Fig.~3A of ref.~\citen{tobinSci2017}) and is consistent with
excluded-volume effects leading to a lower contact probability when 
the contacting monomers are in the middle of the chain than when the
contact is between the two ends of the chain (with different loop-closure 
exponents).\cite{chanJCP89,chanJCP90,cloizeaux1980}
Because $\avRgs^{1/2}$ is determined by $\ep$ for these homopolymer ensembles,
the essentially one-to-one mapping between $\ep$ and $\nu$ in Fig.~3c 
(aside from the small differences for small and large $|i-j|$s) is translated
into an essentially one-to-one mapping between $\avRgs^{1/2}$ and $\nu$
in Fig.~3d. Disordered protein ensembles inferred from experiments have 
sometimes been characterized by homopolymer scaling exponent $\nu$ as a 
proxy for measured $\avRgs^{1/2}$ in recent 
studies.\cite{SchulerBestJCP2018,zhengbest2018,tobinSci2017,tobinPNAS2019,zheng_etal_JPCL2020}
It should be recognized, however, that for real proteins which are 
heteropolymers, there is no universal correspondence
between $\avRgs^{1/2}$ and $\nu$, as exemplified by recent studies of
the N-terminal domain of the ribosomal protein L9 (NTL9)\cite{raleighPNAS2019}
and the C-terminal domain of the same protein.\cite{raleighBiochem2020}
In principle, when a conformational ensemble is sufficiently heterogeneous, 
$\nu$ may not be well-defined even when the chain dimensions are similar 
to those of SAWs (see below).

%%%%%%%%%%%%%%%%%%%%%%%%%%%%%%%%%%%%%%%%%%%%%%%%%%%%%%%%%%%%%%%%%%%%%%%%%%%%%%%%
\begin{figure}[t]
\vskip -0.8cm
{\includegraphics[height=100mm,angle=0]{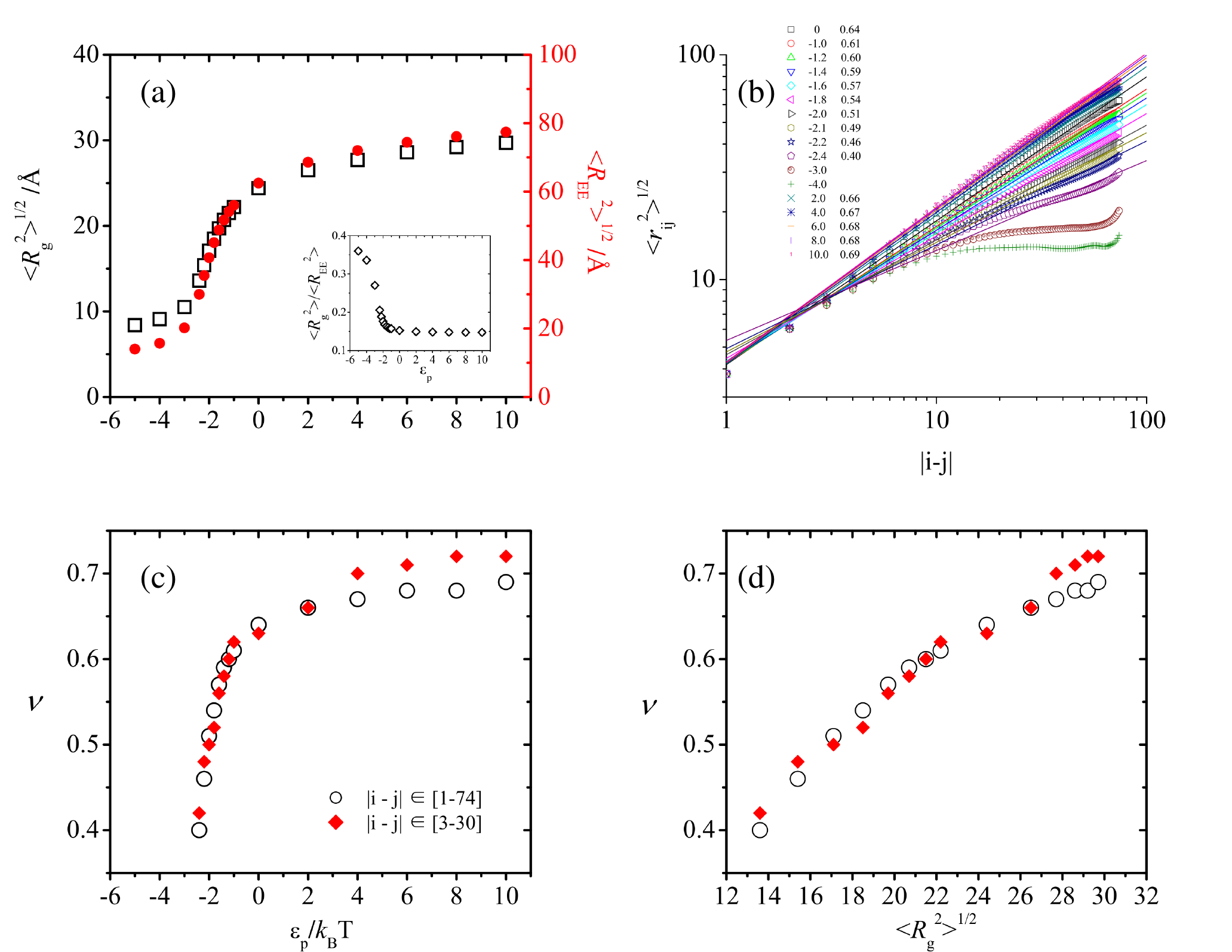}}
\begin{center}
\end{center}
\vspace{-1.5cm}
\caption{Compactness-dependent homopolymer properties.
Dimensional features of conformational ensembles with
$U_{\rm SAW}+U_{\rm p}$ interactions are obtained as functions of $\ep$.
(a) Root-mean-square radius of gyration, $\langle R^2_{\rm g}\rangle^{1/2}$
(black squares, left vertical scale),
and end-to-end distance, $\langle R^2_{\rm EE}\rangle^{1/2}$
(red circles, right vertical scale).
The $\ep$-dependent ratio of mean-square $R_{\rm g}$ to mean-square
$R_{\rm EE}$ is provided in the inset.
(b) Variation of root-mean-square intrachain monomer-monomer distance
$\langle r^2_{ij}\rangle^{1/2}$ with sequence separation $|i-j|$.
Averages are based on $10^5$ sampled conformations for each $\ep$.
The legend provides the symbols (left column) for different $\ep$
values (middle column) together with the scaling exponents, $\nu$ (right
column), obtained by fitting $\langle r^2_{ij}\rangle^{1/2}\sim |i-j|^\nu$
to the plotted data for each $\ep$.
No $\nu$ values are fitted to
the $\langle r^2_{ij}\rangle^{1/2}$ vs $|i-j|$
data for $\ep=-3.0$ and $-4.0$ because of their significant
deviations from linearity.
(c) Scaling exponents $\nu$ from (b) by fitting
$\langle r^2_{ij}\rangle^{1/2}$ data for all sequence
separations ($1\leq |i-j|\leq 74$, black circles) and by fitting only
part of the data for smaller sequence separations
($3\leq |i-j|\leq 30$, red diamonds).
(d) Variation of scaling exponents in (c) with root-mean-square $R_{\rm g}$.
}
\label{fig3}
\end{figure}
%%%%%%%%%%%%%%%%%%%%%%%%%%%%%%%%%%%%%%%%%%%%%%%%%%%%%%%%%%%%%%%%%%%%%%%%%%%%%%%%

The scattering intensity $I(q)$ of the homogeneous and heterogeneous 
conformational ensembles considered in this study is computed using
the Debye formula\cite{AndoChemRev2017}
\begin{equation}
I(q)=4\pi \int_0^\infty dr \; P(r) \frac {\sin (qr)}{qr} \; ,
\label{eq:Iqfirstdef}
\end{equation}
where $q=|{\bf q}|$ is the magnitude of the scattering vector ${\bf q}$,
\begin{equation}
P(r) = \frac {1}{n^2} \sum_{i=1}^n \sum_{j=1}^n \langle \delta(r-r_{ij})\rangle
\label{eq:Prfirstdef}
\end{equation}
is the pair distance distribution function obtained by averaging over
bead-bead distances $r_{ij}$ of sampled
conformations, $\delta$ denotes the Dirac delta function, and $1/n^2$ is
the normalization factor for the total number of $i,j$ pairs.
Because our focus is on general physical principles, we
consider beads in our coarse-grained chain model as simple point-like 
scattering centers, neglecting complexities arising from atomic form 
factors and solvation in computational studies that utilize more atomistic 
representations of the polypeptide chain.\cite{AndoChemRev2017,SaliFoXS}
Inasmuch as a sufficient large number of conformations are used, 
our simulated $I(q)$s are numerically robust, as we have verified by
comparing $I(q)$s computed using $1,000$, $10,000$, or $100,000$ sampled
conformations in selected cases. Practically, the upper limit of $\infty$ 
for the integration in Eq.~\ref{eq:Iqfirstdef} may be replaced by
the longest pairwise distance, $r_{\rm max}$, in the system, which is 
approximately equal to $(n-1)b\sin(\theta_0/2)\approx 225.0$\AA~when 
an $n=75$ chain in our model adopts an all-trans conformation.

The simulated scattering intensities $I(q)$ of the homopolymer ensembles 
in Fig.~3 normalized by $I(0)\equiv I(q=0)=4\pi$ are shown in Fig.~4a, the
$I(q)/I(0)$ for Gaussian chains is also included for comparison. These
curves are similar to those obtained for other models for disordered
proteins.\cite{tobinSci2017}
Their corresponding dimensionless Kratky plots (MFFs) are provided in Fig.~4b.
Here the vertical variable $I(q)/I(0)$ is scaled by $q^2\avRgs$ and 
the horizontal variable $q$ is scaled by $\avRgs^{1/2}$, where 
$\avRgs$ is the mean square radius of gyration determined by the
sampled conformations used for the calculation of $I(q)$ for the
same ensemble.
By definition, all dimensionless Kratky plots
are essentially identical in the small-$q$ Guinier regime 
irrespective of $\avRgs^{1/2}$, as can be seen for the examples
in Fig.~4b, because $I(q)/I(0)\rightarrow \exp(-q^2\avRgs/3)$ for 
$q\rightarrow 0$ (ref.~\citen{AndoChemRev2017}) and therefore
the dimensionless vertical variable always behaves approximately 
as $x^2\exp(-x^2/3)$ for small $x$ where $x=q\avRgs^{1/2}$ is the
dimensionless horizontal variable.

%%%%%%%%%%%%%%%%%%%%%%%%%%%%%%%%%%%%%%%%%%%%%%%%%%%%%%%%%%%%%%%%%%%%%%%%%%%%%%%%
\begin{figure}[t]
\vskip -1cm
{\includegraphics[height=60mm,angle=0]{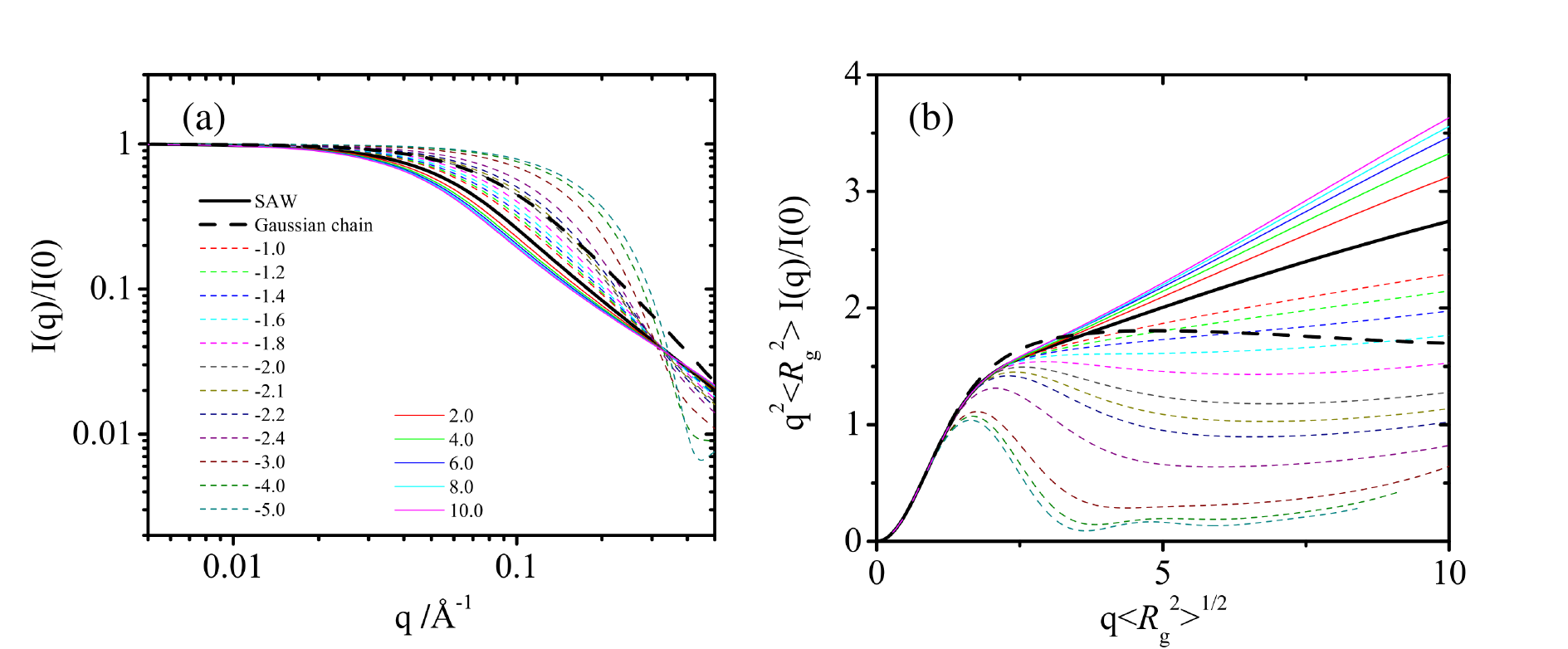}}
\begin{center}
\end{center}
\vspace{-1.5cm}
\caption{Compactness-dependent scattering properties of homopolymer ensembles.
(a) Log-log plot of scattering intensity $I(q)$ (normalized by value
at $q=0$) as a function of the magnitude of the scattering vector, $q$,
and (b) dimensionless Kratky plots (MFFs) are shown for $n=75$
SAW (solid black curve),
Gaussian-chain (dashed black curve), and homopolymer ensembles
defined by various $\ep\neq 0$. Results for the latter ensembles are
depicted as color dashed curves for $\ep<0$ and
color solid curves for $\ep>0$ as indicated by the legend in (a).
}
\label{fig4}
\end{figure}
%%%%%%%%%%%%%%%%%%%%%%%%%%%%%%%%%%%%%%%%%%%%%%%%%%%%%%%%%%%%%%%%%%%%%%%%%%%%%%%%

We note that all the $I(q)/I(0)$ curves for models with excluded volume
in Fig.~4a (all except the Gaussian-chain black dashed curve), irrespective
of their different $\ep$ values, converge in a narrow region around
$q\approx 0.3$\AA$^{-1}$ (though the $I(q)/I(0)$ values do not converge at
exactly the same $q$). This behavior of the model may be understood by
recognizing that these models, even for very different $\ep$, should
share essentially identical probabilities for two shortest distances dictated
by the local bond structure that are independent of or minimally
affected by the global $\ep$-dependent conformational compactness. These 
distances are the virtual bond length between two sequential beads 
($r_{i,i+1}=b=3.8$\AA) and the distance between two beads separated by a 
single bead along the chain sequence 
($r_{i-1,i+1}\approx 2b\sin(\theta_0/2)=6.08$\AA). The virtual bond length 
is a constant in the model, whereas small variations 
in $r_{i-1,i+1}$ are possible 
because the virtual bond angle $\theta_i$ fluctuates around $\theta_0$ in 
accordance with a harmonic potential.
The essential identical probabilities of these short distances among
the models translate into a near-coincidence of their $I(q)/I(0)$ values
around $q\approx\pi/2r_{i,i+1}= 0.41$\AA$^{-1}$ and
$q\approx \pi/2r_{i,i+2}\approx 0.26$\AA$^{-1}$ in reciprocal space, averaging
to $q\approx 0.34$\AA$^{-1}$ which is consistent with the
approximate convergence observed in Fig.~4a. An approximate convergence of
theoretical $I(q)/I(0)$ curves has also been seen in other studies, e.g.,
in Fig.~3B of ref.~\citen{tobinSci2017}. In the latter case,
the convergence is at $\approx 0.2$\AA$^{-1}$, indicating that there
are similarly probable distances 
$\approx \pi/(2\times 0.2$\AA$^{-1}$)$=7.9$\AA~between scattering centers 
among the chain models in ref.~\citen{tobinSci2017}
for different conformational compactness.

Examples of the homopolymer pair distance distribution functions $P(r)$ 
underlying the $I(q)$ functions in Fig.~4 are shown in Fig.~5. 
A notable shared feature among the different $P(r)$ plots in Fig.~5 as
well as subsequent $P(r)$ plots in this article
is the local $P(r)$ peaks at small $r$ values corresponding to the
$r_{i,i+1}=b=3.8$\AA~and the 
$r_{i-1,i+1}\approx 2b\sin(\theta_0/2)=6.08$\AA~distances 
discussed above. As expected, aside from these common local peaks at small
$r$s, the overall peak of the $P(r)$ distribution is 
shifted to smaller $r$ for increasingly
negative $\ep$ with a concomitant narrowing of the distribution. However,
the shift to larger $r$ relative to the distribution for our SAW model is quite
small for $\ep>0$ because short-spatial-range contact-like repulsive 
potentials like those in our $\ep>0$ models are essentially enhanced 
excluded volume interactions which, unsurprisingly, do not expand chain 
dimensions much beyond those of a SAW that already possesses a sizable 
excluded volume repulsion. 
\\

%\vfill\eject

%%%%%%%%%%%%%%%%%%%%%%%%%%%%%%%%%%%%%%%%%%%%%%%%%%%%%%%%%%%%%%%%%%%%%%%%%%%%%%%
\begin{figure}[t]
\vskip -1cm
{\includegraphics[height=65mm,angle=0]{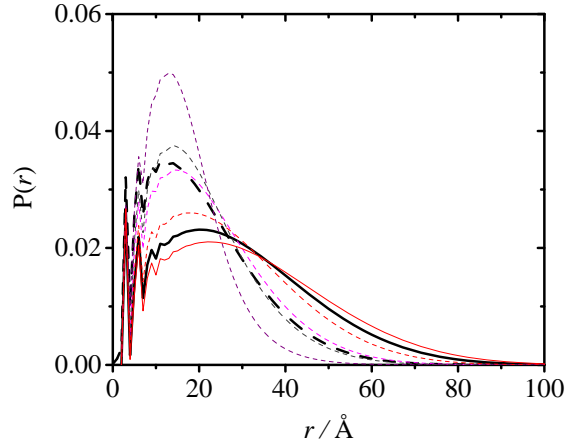}}
\begin{center}
\end{center}
\vspace{-1.5cm}
\caption{$\ep$-dependent (with excluded volume) 
and Gaussian-chain distributions of intrachain
monomer-monomer distances for $n=75$ homopolymers. As in Fig.~\ref{fig4},
solid and dashed black curves are for, respectively, the SAW ($\ep=0$,
$\epsilon_{\rm ex}=1.0k_{\rm B}T$) and
Gaussian-chain ($\ep=\epsilon_{\rm ex}=0$)
conformational ensembles. The correspondence between color curve styles
and $\ep\neq 0$ ensembles is identical to that in Fig.~\ref{fig4} as well;
but, for clarity, $P(r)$ is plotted only for $\ep=-2.4$, $-2.0$, $-1.8$,
$-1.0$, and $+2.0$ in the present figure.
}
\label{fig5}
\end{figure}
%%%%%%%%%%%%%%%%%%%%%%%%%%%%%%%%%%%%%%%%%%%%%%%%%%%%%%%%%%%%%%%%%%%%%%%%%%%%%%%

\noindent
{\large\bf RESULTS}\\

{\bf MFFs of heteropolymeric and homopolymeric SAWs are sometimes
clearly distinct; but MFFs of select heterogeneous 
ensembles with very narrow ranges of ${\bm{R}}_{\bf g}$ and/or 
${\bm{R_{\rm EE}}}$ can be very similar to MFFs of homopolymeric SAWs.}
We begin our analysis with three different hypothetical subensembles
put forth previously as possible heterogeneous model conformational 
ensembles for Protein L with very narrow ranges of $R_{\rm EE}$ 
values\cite{Songetal2017} consistent with the experimental FRET 
efficiencies of $E\approx 0.75$ at [GuHCl] $=$ 1 M and $E\approx 0.45$ 
at [GuHCl] $=$ 7 M (Fig.~6).
Two of the MFFs in Fig.~6 are for two subensembles 
sharing the same narrow range of $R_{\rm g}\approx 23.5$\AA~($\avRgs^{1/2}$
$\approx$ $R_{\rm g}$ because of the narrow range) but with
significantly different $R_{\rm EE}$s, namely
$R_{\rm EE}\approx 46.5$\AA~for [GuHCl] $=$ 1 M (solid magenta curve)
and $R_{\rm EE}\approx 56.9$\AA~for [GuHCl] $=$ 7 M (solid blue curve).
These subensembles are of interest as examples of $R_{\rm g}$--$R_{\rm EE}$
decoupling,\cite{Songetal2017,lemke2017} in that the ensembles have the 
same $R_{\rm g}$ despite having very different $R_{\rm EE}$. Fig.~6 shows that
their MFFs are distinct but similar in some notable respects. The MFF for the 
subensemble with smaller $R_{\rm EE}$ (magenta) is more oscillatory than that 
for the subensemble with larger $R_{\rm EE}$ (blue) 
for $2\lesssim q\avRgs^{1/2}\lesssim 7$ while the two MFFs converge 
at larger $q\avRgs^{1/2}$ values. Moreover, the MFFs of both of these
subensembles---which are heterogeneous ensembles by construction---are
quite similar to the MFF of a homopolymer ensemble
with the same $\avRgs^{1/2}$ and an $\avREEs^{1/2}$ corresponding
approximately to the [GuHCl] $=$ 7 M case (black dashed curve). The
differences among the MFFs are small except for 
$2\lesssim q\avRgs^{1/2}\lesssim 4$ which one may refer to as the
``shoulder region'' of the dimensionless Kratky curves. 
These comparisons indicate that, for some heterogeneous ensembles,
different heterogenous and homogeneous ensembles entail different MFFs.
In principle, therefore, these ensembles are distinguishable by 
SAXS measurements alone even though the ensembles share the same 
average $R_{\rm g}$. However, as seen in Fig.~6, the differences among 
some of the theoretical MFFs can be subtle.  Thus, detectability of such 
differences may still be limited practically by uncertainties in 
experimental measurements.  In contrast, the MFF for a subensemble with a 
smaller $\avRgs^{1/2}\approx 21.6$\AA~(dashed red curve) is clearly 
distinguishable from the other three MFFs because of its substantially lower 
$q^2\avRgs I(q)/I(0)$ for $q\avRgs^{1/2}\gtrsim 2$.

%%%%%%%%%%%%%%%%%%%%%%%%%%%%%%%%%%%%%%%%%%%%%%%%%%%%%%%%%%%%%%%%%%%%%%%%%%%%%%
\begin{figure}[t]
{\includegraphics[height=60mm,angle=0]{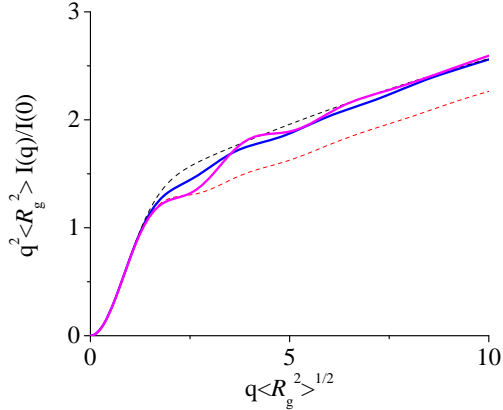}}
\begin{center}
\end{center}
\vspace{-1.5cm}
\caption{Theoretical MFFs pertinent to the case of Protein L
at [GuHCl] = 1 and 7 M analyzed previously.\cite{Songetal2017}
Dimensionless Kratky plots for $n=75$ heterogeneous conformational ensembles
with (i) a sharp, $\delta$-function-like distribution with a narrow range
of $R_{\rm g}$ centered at $23.5$ \AA~and a similarly sharp distribution of
$R_{\rm EE}$ at $46.5$ \AA~(consistent with FRET efficiency $E=0.745$ when
F\"orster radius $R_0=55$ \AA, solid magenta curve),
(ii) a sharp distribution of
$R_{\rm EE}$ at $46.5$~\AA~(corresponding to $E=0.745$) but no restriction on
$R_{\rm g}$ otherwise
($\langle R_{\rm g}^2\rangle^{1/2}=21.6$~\AA, dashed red curve),
and (iii) a sharp distribution of $R_{\rm EE}$ at
$56.9$~\AA~(consistent with $E=0.447$) but no restriction on
$R_{\rm g}$ otherwise
($\langle R_{\rm g}^2\rangle^{1/2}=23.4$~\AA, solid blue curve) are
compared with that for an $\ep=-0.5$ homogeneous conformational ensemble
with $\langle R_{\rm g}^2\rangle^{1/2}=23.5$~\AA~and
$\langle R_{\rm EE}^2\rangle^{1/2}\approx 59.9$~\AA~(dashed black curve).
}
\label{fig6}
\end{figure}
%%%%%%%%%%%%%%%%%%%%%%%%%%%%%%%%%%%%%%%%%%%%%%%%%%%%%%%%%%%%%%%%%%%%%%%%%%%%%%

Fig.~7 provides a systematic comparison of SAXS signatures of SAW homopolymers 
with a broad distribution of $R_{\rm g}$ and $R_{\rm EE}$
on one hand against SAW subensembles each with a narrow range of 
$R_{\rm g}$ and/or a narrow range of $R_{\rm EE}$ on the other. 
Here we focus on subensembles with 
an $\avRgs^{1/2}\approx 24.5$\AA~(Fig.~7a) equals to the
$\avRgs^{1/2}$ of the full homopolymeric SAW ensemble. 
As emphasized above, these subensembles are, by construction, 
heterogeneous conformational ensembles. It is noteworthy that 
despite the differences among 
the subensembles themselves and their drastically different 
$R_{\rm g}$--$R_{\rm EE}$ distributions vis-\`a-vis that of the full 
homopolymer ensemble (Fig.~7a), their $I(q)/I(0)$ versus $q$ plots 
appear to be quite similar aside from seemingly minor differences around
$q\sim 0.1$\AA$^{-1}$ (Fig.~7b). When presented as dimensionless
Kratky plots (Fig.~7d), the differences in the shoulder region of 
the plots ($2\lesssim q\avRgs^{1/2}\lesssim 4$) for the different 
heterogeneous subensembles are more discernible.
The subensembles' different $I(q)$s are a reflection of their different 
pair distance distribution functions (Fig.~7c, Eq.~\ref{eq:Iqfirstdef}).
Among the subensembles with the same narrow range of 
$R_{\rm g}$ but different narrow ranges of $R_{\rm EE}$ in Fig.~7c, 
the peaks of the $P(r)$ of small-$R_{\rm EE}$ subensembles (dashed color
curves) shift to larger $r$ values relative to the peak of the $P(r)$ 
for the homopolymer ensemble (solid black curve). Interestingly, the
$P(r)$ of the subensemble with the largest $R_{\rm EE}\approx 88.5$\AA~among
the subensembles considered is almost identical to the $P(r)$ 
of the homopolymer ensemble. 
Accordingly, their dimensionless Kratky plots essentially overlap (Fig.~7d,
solid black and dashed dark-purple curves).
In other words, quite remarkably, the SAXS signatures of these two very 
different disordered conformational ensembles---a highly heterogeneous ensemble 
with a narrow range of $R_{\rm g}$ as well as a narrow range of $R_{\rm EE}$ 
on one hand, and a homogeneous ensemble with a broad distribution of both 
$R_{\rm g}$ and $R_{\rm EE}$ on the other---are practically indistinguishable.
Because $R_{\rm EE}\approx 88.5$\AA~is substantially larger
than the $\avREEs^{1/2}\approx 62.5$\AA~for the full homopolymer ensemble,
it appears that inasmuch as effects on $P(r)$ are concerned, narrowing
the broad distribution of $R_{\rm g}$ of the full homopolymer ensemble
to a sharply peaked distribution around its $\avRgs^{1/2}$ value can
be compensated by replacing the broad distribution of $R_{\rm EE}$ of the
full homopolymer ensemble with a sharply peaked distribution around
an $R_{\rm EE}$ value that is significantly larger than the $\avREEs^{1/2}$ 
of the homopolymer ensemble.

%%%%%%%%%%%%%%%%%%%%%%%%%%%%%%%%%%%%%%%%%%%%%%%%%%%%%%%%%%%%%%%%%%%%%%%%%%%%%%%%
\begin{figure}[ht]
\vskip -0.8cm
{\includegraphics[height=100mm,angle=0]{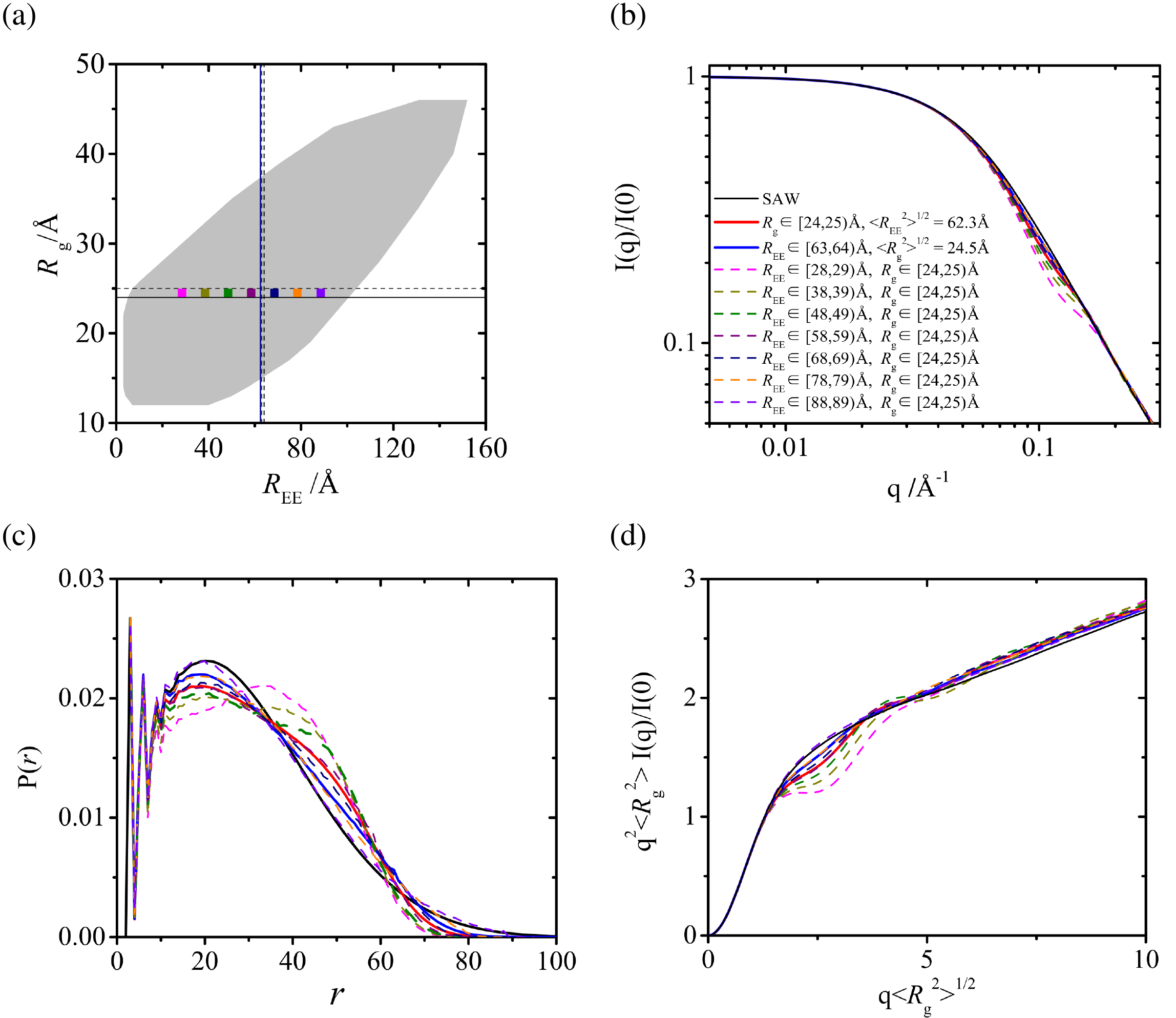}}
\begin{center}
\end{center}
\vspace{-1.3cm}
\caption{SAXS properties of subensembles of SAW chains.
(a) The complete SAW homopolymer ensemble is represented by the
gray area corresponding to the overall profile for $\ep=0$ in Fig.~\ref{fig2};
its $\langle R_{\rm EE}^2\rangle^{1/2}$ is indicated by
the vertical blue solid line.
The subensemble with a narrow $R_{\rm g}$ range,
$24\leq R_{\rm g}/$\AA~$<25$, but is otherwise unrestricted, is
marked by the horizontal lines. Similarly, the subensemble with
a narrow $R_{\rm EE}$ range, $63\leq R_{\rm EE}/$\AA~$<64$,
but is otherwise unrestricted, is marked by the vertical dashed lines.
Subensembles defined by narrow ranges for both $R_{\rm g}$
and $R_{\rm EE}$ are indicated by the small color squares,
which are slightly wider than the actual
ranges of $R_{\rm EE}$ to enhance legibility.
(b) Log-log plot of scattering intensity for the SAW homopolymer ensemble and
various subensembles as specified by the legend.
The color code of the dashed lines for the subensembles with narrow ranges
for both $R_{\rm g}$ and $R_{\rm EE}$ (dashed lines) is the same as that
for the small squares in (a).
(c) Distributions of intrachain monomer-monomer distances, and
(d) dimensionless Kratky plots (MFFs) for the SAW homopolymer ensemble and
the subensembles in (b), plotted using the same line styles
as those in (b).
}
\label{fig7}
\end{figure}
%%%%%%%%%%%%%%%%%%%%%%%%%%%%%%%%%%%%%%%%%%%%%%%%%%%%%%%%%%%%%%%%%%%%%%%%%%%%%%%%

Aiming to generalize the above analysis to conformations that are more compact 
or even more open, we have also compared the SAXS signatures of $\ep=0$ SAW 
subensembles with narrow ranges of $\avRgs^{1/2}\approx 22.5$, $18.5$, 
and $26.5$\AA, which are equal, respectively, to the $\avRgs^{1/2}$ of 
the homopolymer ensembles with $\ep=-1.0$, $-1.8$, $+2.0$ (the 
$R_{\rm g}$--$R_{\rm EE}$ distributions of which are illustrated in Fig.~2). 
The results of the analysis 
are documented in Figs.~S1--S3 of the Supporting Information. They indicate
that while the trend observed in Fig.~7c of a shift of the $P(r)$ peak 
to higher $r$ values relative to that of the homopolymer ensemble
for subensembles with smaller $R_{\rm EE}$ persists in Figs.~S1c, S2c, and
S3c, none of the subensembles considered---including those with large 
$R_{\rm EE}$s---has a $P(r)$ that matches closely
with the $P(r)$ of the corresponding homopolymer ensembles. Consequently,
all of these subensembles entail dimensionless Kratky plots that are
quite clearly distinguishable from that of their homopolymer counterparts
(Figs.~S1d, S2d, and S3d), thus offering examples for which heterogeneous 
and homogeneous conformational ensembles with the same $\avRgs^{1/2}$ can 
be distinguished by SAXS-determined MFFs alone. At the same time, the 
observation from Figs.~S1--S3 suggests that heterogeneous ensembles
constructed as subensembles of homopolymers with different overall compactness
may be different even if the subensembles themselves feature the same narrow
range of $R_{\rm g}$ and narrow range of $R_{\rm EE}$. 
This issue will be further explored below.

Root-mean-square of intrachain distance $r_{ij}$
as a function of contour length separation $|i-j|$ are shown for 
representative subensembles in Fig.~S4 of Supporting Information.
For subensembles with compact conformational dimensions such as 
those in Fig.~S4d, the $\langle r_{ij}^2\rangle^{1/2}$ versus $|i-j|$ 
relationship is nonlinear, similiar to the corresponding relationships
exhibited by the homopolymer ensembles with $\ep\lesssim -2.4$ in Fig.~3b.
Indeed, the $\langle r_{ij}^2\rangle^{1/2}$ versus $|i-j|$ relationship 
can be highly nonlinear for heteropolymers,\cite{pappu13,kings2015,huihui2020}
in which cases no $\nu$ can be reasonably defined. More recent examples
of simulated heteropolymers lacking an approximate 
$\langle r_{ij}^2\rangle^{1/2}\sim |i-j|$ scaling include results shown in 
Figs.~2 and 4 of ref.~\citen{huihui2020} and Fig.~S3 of 
ref.~\citen{zheng_etal_JPCL2020} even though usage of $\nu$ in lieu of
$\langle R^2_{\rm g}\rangle$ is advocated in ref.~\citen{zheng_etal_JPCL2020}.
Here, for the subensembles with the same $R_{\rm g}$ as the relatively 
open $\ep=0$ SAW homopolymers studied in Fig.~7, the 
$\langle r_{ij}^2\rangle^{1/2}$ versus $|i-j|$ plots in Fig.~S4a are largely 
linear except for $|i-j|\approx n=75$, but they do exhibit other, more minor 
variations despite the subensembles sharing the same $R_{\rm g}$. 
The divergent behaviors for large $|i-j|$s, which correspond to distances 
between two ends of the chain, are stemming from the narrow ranges of 
$R_{\rm EE}$ imposed by the definition of the subensembles. Aside from
that, variations are also noticeable for $|i-j|$ $\sim 7$---$40$. 
A zoomed-in version of the $\langle r_{ij}^2\rangle^{1/2}\sim |i-j|^\nu$ 
plots for these subensembles in Fig.~8 indicates that the apparent scaling 
exponent $\nu$ of the subensembles ranges from $\approx 0.62$ to $0.71$.
While these mathematically constructed subensembles are hypothetical
as to their physical realizability, they do serve to underscore that for
heterogeneous disordered conformational ensembles, the apparent scaling 
exponent $\nu$ is not necessarily a proxy for $\avRgs^{1/2}$, as has been
demonstrated recently in combined theoretical/experimental studies of
real disordered proteins.\cite{raleighPNAS2019,raleighBiochem2020}
As it stands, $\nu$ is largely a model parameter that is currently not 
amenable to direct experimental determination. Using such a parameter to 
replace the experimentally measured $\langle R^2_{\rm g}\rangle$ as a 
descriptor of ensemble properties does not appear to be well advised.
\\

%%%%%%%%%%%%%%%%%%%%%%%%%%%%%%%%%%%%%%%%%%%%%%%%%%%%%%%%%%%%%%%%%%%%%%%%%%%%%%%%
\begin{figure}[t]
\vskip -0.8cm
{\includegraphics[height=50mm,angle=0]{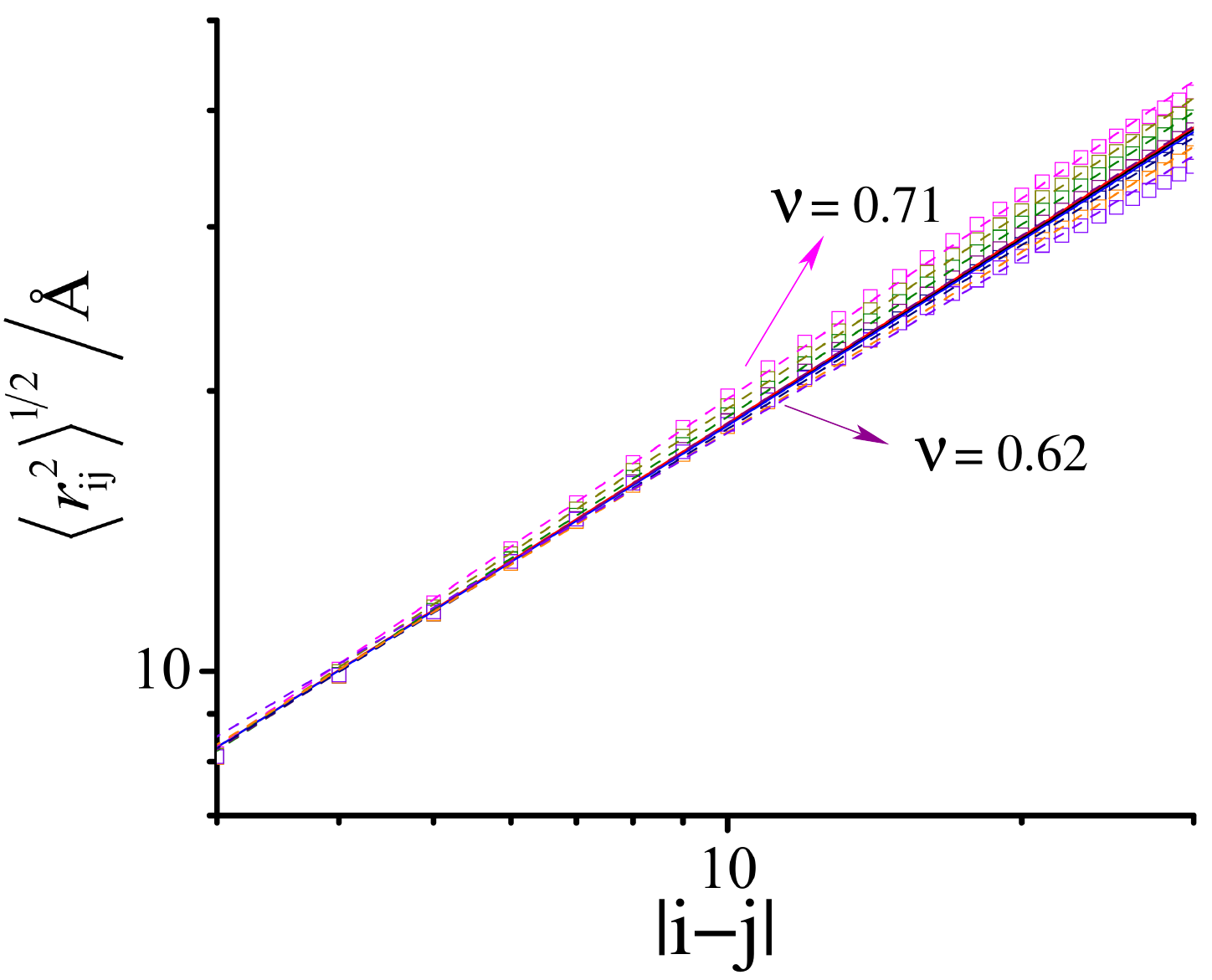}}
\begin{center}
\end{center}
\vspace{-1.3cm}
\caption{Variation of root-mean-square intrachain monomer-monomer distance
$\langle r^2_{ij}\rangle^{1/2}$ with sequence separation $|i-j|$
for the SAW homopolymer ensemble and various
subensembles in Fig.~\ref{fig7}, plotted using
the same color code. The maximum and minimum scaling exponents ($\nu$)
obtained from the fitted straight lines are indicated.
}
\label{fig8}
\end{figure}
%%%%%%%%%%%%%%%%%%%%%%%%%%%%%%%%%%%%%%%%%%%%%%%%%%%%%%%%%%%%%%%%%%%%%%%%%%%%%%%%

{\bf MFFs of select heterogeneous ensembles with very narrow ranges of 
${\bm{R}}_{\bf g}$ and ${\bm{R_{\rm EE}}}$ can be very similar to 
MFFs of compact homopolymers.} Building on the initial results in
Figs.~S1--S3 (Supporting Information) discussed above, we explore
whether and, if so, what heterogeneous ensembles may possess SAXS 
signatures practically indistinguishable from those of homopolymer 
ensembles that are more compact or more open than the $\ep=0$ SAW 
homopolymers. Now, instead of constructing heterogeneous ensembles by 
selecting from the $\ep=0$ homopolymer conformations as in Figs.~S1--S3, 
we construct heterogeneous ensembles by selecting from 
the conformations of an $\ep\neq 0$ homopolymer ensemble.
Interestingly, among the heterogeneous subensembles constructed in this 
manner to cover a narrow range of $R_{\rm EE}$ and a narrow range 
of $R_{\rm g}$ values around the $\avRgs^{1/2}$ of the $\ep\neq 0$ 
homopolymer, some of the heterogeneous subensembles with 
relatively large $R_{\rm EE}$s can have $P(r)$s very similar to 
the $P(r)$ of the corresponding homopolymer ensemble with the same $\ep\neq 0$.
Consequently, their MFFs are also extremely similar.
Examples of such heterogeneous and $\ep\neq 0$ homogeneous ensembles
with closely matching dimensionaless Kratky plots are provided in
Figs.~9a,b, and c between subensembles with
$R_{\rm EE}\approx 70.5$, $65.5$, and $84.5$\AA, respectively, and 
their corresponding homopolymer ensembles with different compactness
as specified by $\ep=-1.0$, $-1.8$, and $+2.0$.
In line with the $\ep=0$ situation in Fig.~7, the $R_{\rm EE}$ of these 
subensembles in Figs.~9a--c with homopolymeric SAXS signatures are considerably 
higher than the $\avREEs^{1/2}\approx 50.5$, $18.5$, and $74.5$\AA,
respectively, of their 
corresponding $\ep=-1.0$, $-1.8$, and $+2.0$ homopolymer ensembles.
It is instructive to contrast these subensembles of $\ep\neq 0$
homopolymers in Figs.~9a--c exhibiting homopolymeric SAXS signatures 
with those subensembles with the same 
narrow $R_{\rm g},R_{\rm EE}$ ranges but constructed from
$\ep=0$ chains in Figs.~S1--S3 (Supporting Information) that do
not exhibit similar SAXS signatures. This observation indicates that 
MFFs of disordered chains can be sensitive to $\ep$-dependent
conformational preferences even when variations in $R_{\rm g}$ and 
$R_{\rm EE}$ in the ensembles are highly restricted.
In any event, the examples of SAXS signature matching in Figs.~9a,b
affirm that MFFs of at least some highly heterogeneous 
ensembles can be practically identical to MFFs of compact homopolymers
and therefore these disordered conformational ensembles cannot be 
distinguished solely by their SAXS spectra.

To facilitate systematic computation and evaluation of MFFs of a large 
number of heterogeneous ensembles (see below), 
$\Delta_{\rm Kratky}(1,2)\equiv \int_0^{x_{\rm max}} dx |y_1(x)-y_2(x)|$,
where $y=q^2\avRgs I(q)/I(0)$ and $x=q\avRgs^{1/2}$, is hereby defined 
(Fig.~9d) to quantify the difference in SAXS signature between two 
conformational ensembles (labeled 1 and 2). As illustrated in Fig.~9d,
$\Delta_{\rm Kratky}$ may be viewed as a simple measure of mismatch between two
dimensionless Kratky plots. We use $x_{\rm max}=10$ for the present study.
Besides the present application, we note that $\Delta_{\rm Kratky}$
may also be useful in future investigations to optimize 
the match between experimental MFFs and those computed from
inferred conformational ensembles.\cite{GregJACS2020,choy2001}
\\

%%%%%%%%%%%%%%%%%%%%%%%%%%%%%%%%%%%%%%%%%%%%%%%%%%%%%%%%%%%%%%%%%%%%%%%%%%%%%%%
\begin{figure}[ht]
\vskip -0.8cm
{\includegraphics[height=90mm,angle=0]{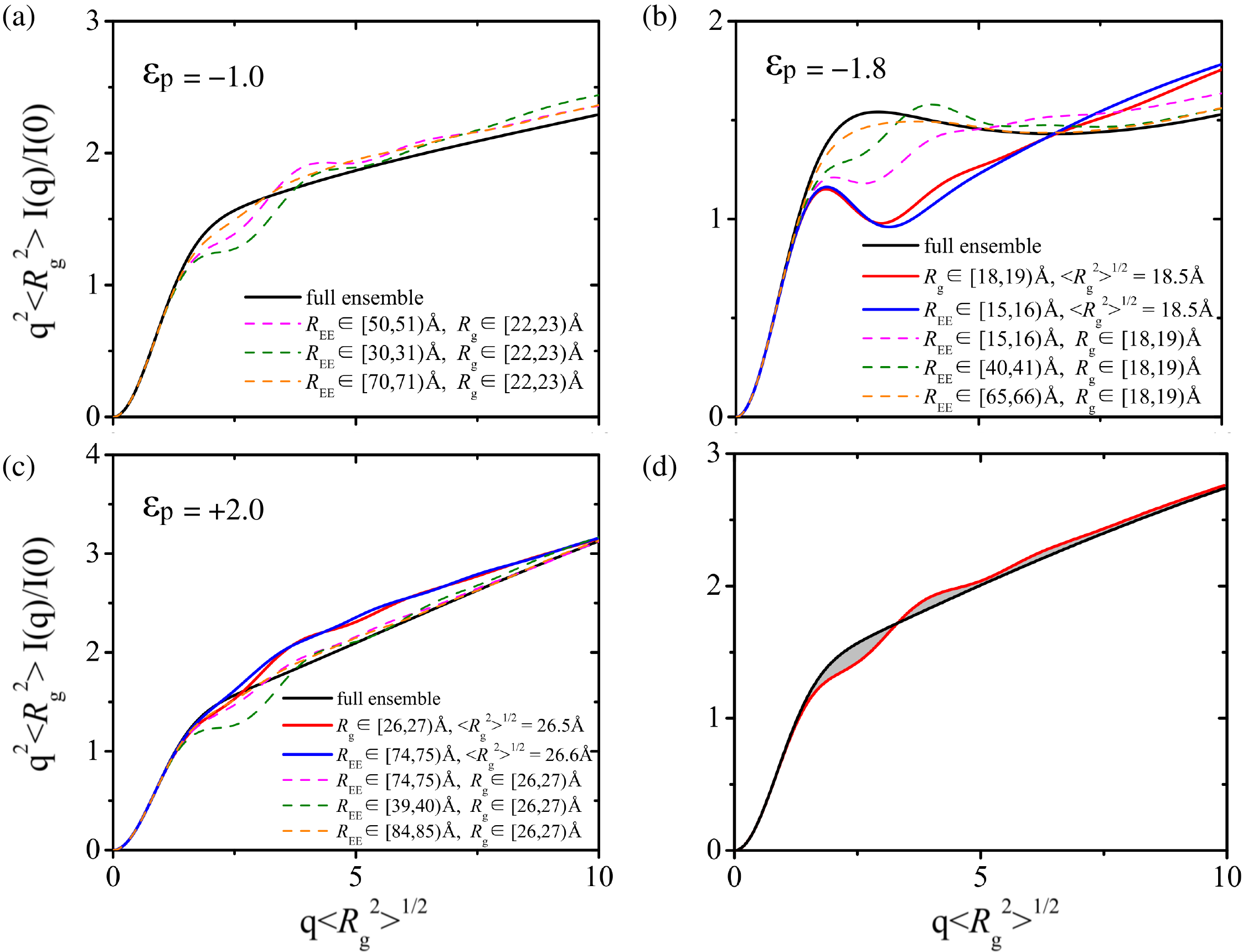}}
\begin{center}
\end{center}
\vspace{-1.3cm}
\caption{Comparing MFFs of homogeneous and heterogeneous ensembles
sharing essentially the same $\langle R^2_{\rm g}\rangle^{1/2}$.
(a)--(c) Dimensionless Kratky plots (MFFs) of $\ep\neq 0$ homogeneous ensembles
are compared with those of subensembles that are
subsets of the full ensemble for the given $\ep$ with
$R_{\rm g}$ and/or $R_{\rm EE}$ ranges specified by the inset legends.
(d) Schematic of a pairwise difference measure for {\it any} two dimensionless
Kratky plots referred to as $\Delta_{\rm Kratky}$ (see text)
and defined as the total
area (shaded) between the two plots within a given range of
$q\langle R^2_{\rm g}\rangle^{1/2}$ (e.g., from
$q\langle R^2_{\rm g}\rangle^{1/2}=0$ to $10$ as in this depiction).
The two Kratky plots shown in (d) are hypothetical and for illustration
only. They do not correspond to the simulated Kratky plots in (a)--(c).
}
\label{fig9}
\end{figure}
%%%%%%%%%%%%%%%%%%%%%%%%%%%%%%%%%%%%%%%%%%%%%%%%%%%%%%%%%%%%%%%%%%%%%%%%%%%%%%%

{\bf MFFs of heterogeneous ensembles with significantly less variation
in ${\bm{R}}_{\bf g}$ can still be very similar to MFFs of homopolymeric SAWs.}
As shown by the above analysis, subensembles defined by narrow ranges 
of $R_{\rm g}$ and $R_{\rm EE}$ are instrumental---as individual 
heterogeneous ensembles---in identifying scenarios in which the SAXS 
signatures of heterogeneous and homogeneous ensembles are clearly 
distinguishable or remarkably similar, or somewhere in between. 
In aggregate, these subensembles may be seen as components of 
a basis set for a ``conformational ensemble space'' by which other 
heterogeneous ensembles can 
be constructed as weighted combinations of component ensembles. In this regard,
the homopolymer ensemble is a special case of such an ensemble in which
the components are weighted by their fractional populations in the
homopolymer ensemble. Adopting this conceptual framework,
we now extend our consideration to heterogeneous ensembles, constructed
as weighted combinations of $(R_{\rm g},R_{\rm EE})$ subensembles,
that encompass a broader variety of conformations. Instead of restricting 
to narrow ranges $R_{\rm g}$ and $R_{\rm EE}$, 
nonzero weights are assigned to all conformations in a homopolymer ensemble
in the construction of these ensembles.
In particular, we are interested in ensembles
with a tighter distribution of $R^2_{\rm g}$ than the SAW homopolymer 
ensemble. Such heterogeneous ensembles are apparently less 
artifical than the individual $(R_{\rm g},R_{\rm EE})$ subensembles.
Intuitively, they are physically plausible, especially when the distribution
$P(R^2_{\rm g})$ of the ensemble is at most moderately tighter than
that of a homopolymer ensemble. Conceivably, sequence-dependent effects 
of IDPs may encode a tighter $P(R^2_{\rm g})$ for biological function,
as envisioned, e.g., in the proposed polyelectrostatic binding between 
Sic1 and Cdc4 (ref.~\citen{borg07}).
It would be useful, therefore, to ascertain theoretically whether such 
heterogeneous ensembles are distinguishable from homopolymer ensembles by 
SAXS-measured MFFs alone or measurements by complementary techniques 
are needed.

Toward this aim, reweighted ensembles with the same $\avRgs$ as 
the homopolymers but with a narrower distribution of $R^2_{\rm g}$
are constructed. Here we let
$\delta R^2_{\rm g} \equiv R^2_{\rm g} - \langle R^2_{\rm g} \rangle$
and introduce $\alpha$, $\alpha^\prime\ge 0$ as scaling factors for
controlling the broadness of the reweighted $R^2_{\rm g}$ distribution.
For a given homopolymeric $P(R^2_{\rm g})$, the reweighted
$R^2_{\rm g}$ distribution is obtained by the modification
$P(R^2_{\rm g})\rightarrow$
${\cal N}_0^{-1}\exp[-\alpha (\delta R^2_{\rm g})]P(R^2_{\rm g})$
for $\delta R^2_{\rm g}<0$
and
$P(R^2_{\rm g})\rightarrow$
${\cal N}_0^{-1}\exp[-\alpha^\prime (\delta R^2_{\rm g})]P(R^2_{\rm g})$
for $\delta R^2_{\rm g}>0$
where ${\cal N}_0=$ 
$2\int_{-\infty}^0 d{\delta}R^2_{\rm g} \; 
\exp[-\alpha (\delta R^2_{\rm g})]P(R^2_{\rm g})$ 
is the overall normalization factor for the reweighted (modified) 
$P(R^2_{\rm g})$, and $\alpha^\prime$ is related to $\alpha$ by
the equation
$\int_{-\infty}^0 d{\delta}R^2_{\rm g} \; 
\exp[-\alpha (\delta R^2_{\rm g})]P(R^2_{\rm g})$ 
$=$ ${\cal N}_0/2=$
$\int_0^\infty d{\delta}R^2_{\rm g} \; 
\exp[-\alpha^\prime (\delta R^2_{\rm g})]P(R^2_{\rm g})$, which determines
$\alpha^\prime$ numerically for any given $\alpha$.

The dimensionless Kratky plots of the SAW ($\ep=0$) homopolymer ensemble
and three reweighted ensembles---which are heterogeneous ensembles---are
shown in Fig.~10a. Notably, despite the heterogeneous ensembles'
very different $P(R^2_{\rm g})$ distributions, ranging from being 
only slightly narrower than that of the homopolymer ($\alpha=0.001$) 
to being highly peaked ($\alpha=0.1$, Fig.~10b), their dimensionless Kratky
plots are extremely similar (Fig.~10a), exhibiting little variation even in
the shoulder regions where considerable variation was observed in
Fig.~7d among the dimensionless Kratky plots of the $(R_{\rm g},R_{\rm EE})$ 
subensembles. Hence,
the results in Fig.~10 suggest strongly that certain physically plausible 
heterogeneous ensembles 
of disordered proteins with narrower distributions of $R_{\rm g}^2$
can hardly be distinguishable by their SAXS signatures from a homopolymer 
ensemble with the same $\avRgs$. 
We will extend the study of similarly reweighted
ensembles to $\ep=-1.0$, $-1.8$, and $+2.0$ below.
\\

%%%%%%%%%%%%%%%%%%%%%%%%%%%%%%%%%%%%%%%%%%%%%%%%%%%%%%%%%%%%%%%%%%%%%%%%%%%%%%%
\begin{figure}[t]
{\includegraphics[height=45mm,angle=0]{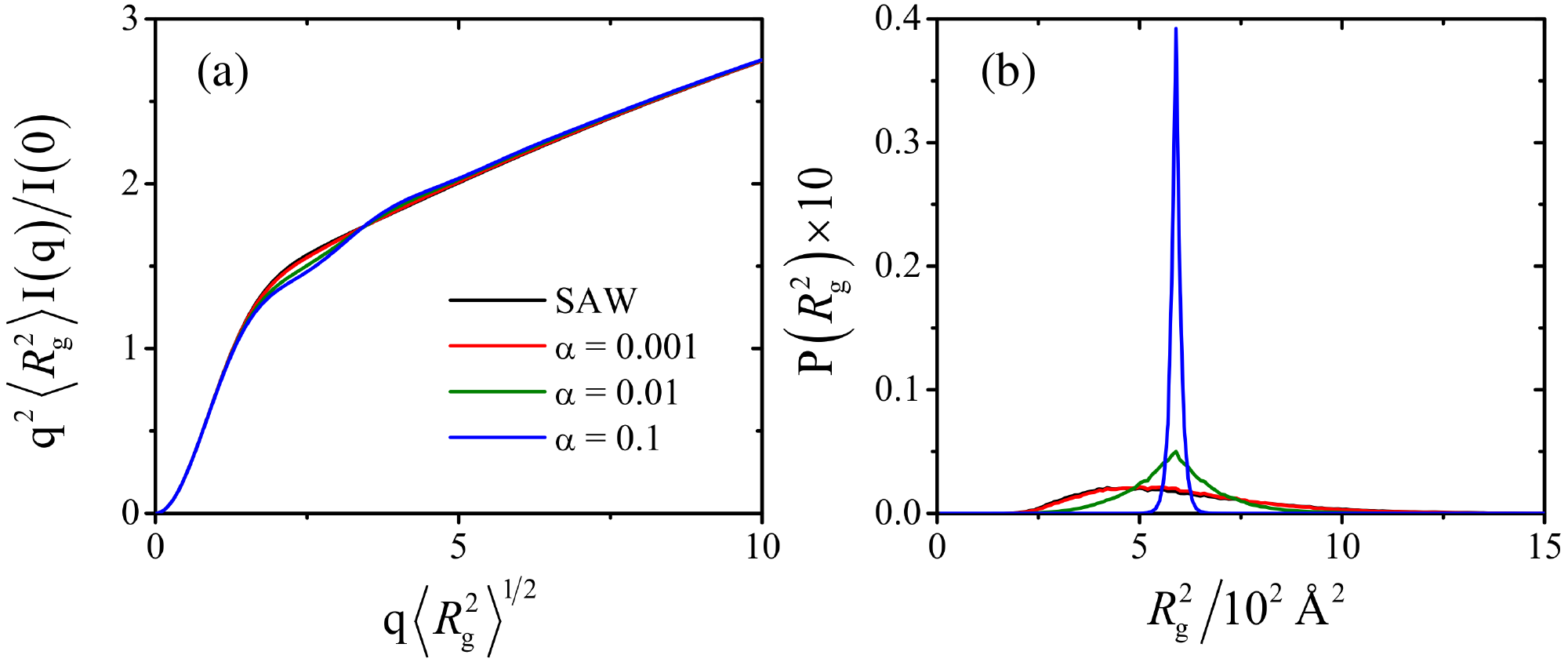}}
\begin{center}
\end{center}
\vspace{-1.3cm}
\caption{SAXS spectra of conformational ensembles with the same
$\langle R^2_{\rm g}\rangle$ but different $R^2_{\rm g}$ distributions.
(a) MFFs (dimensionless Kratky plots) of the homogeneous SAW ($\ep=0$
homopolymer) ensemble
and heteropolymeric reweighted ensembles with $\alpha=0.1$, $0.01$, and
$0.001$ (the corresponding $\alpha^\prime$ values are, respectively, $0.09351$,
$0.007153$, and $0.0006099$). (b) $R^2_{\rm g}$ distributions of the
ensembles in (a) plotted using the same color code.
The standard deviations of $R^2_{\rm g}$ for the SAW ($\alpha=0$),
$\alpha=0.001,$ $0.01$, and $0.1$ distributions are, respectively,
218.7, 205.8, 117.8, and 14.8 \AA$^2$, their corresponding square roots are
14.8, 14.3, 10.9, and 3.85 \AA.
}
\label{fig10}
\end{figure}
%%%%%%%%%%%%%%%%%%%%%%%%%%%%%%%%%%%%%%%%%%%%%%%%%%%%%%%%%%%%%%%%%%%%%%%%%%%%%%%

{\bf MFFs of heterogeneous and homogeneous ensembles can be practically 
identical despite significant differences in
${\bm{R_{\rm EE}}}$, asphericity, and ${\bm{R}}_{\bf g}$ distributions.}
We have now demonstrated that homopolymer ensembles and certain subensembles 
defined by very narrow $R_{\rm g},R_{\rm EE}$ ranges and relatively 
large $R_{\rm EE}$s can lead to essentially indistinguishable dimensionless 
Kratky plots (Figs.~7d and 9a--c). Separately, high degrees of similarity 
can also be seen between the dimensionless Kratky plots of homopolymers and 
those of conformationally more diverse heterogeneous ensembles with slightly 
to significantly narrower $R^2_{\rm g}$ distributions (Fig.~10). 
In view of these findings, we next consider, in a systematic manner, 
an extensive set of heterogeneous ensembles with 
narrower-than-homopolymer distributions of $R_{\rm g}$ similar to those 
studied in Fig.~10 and also narrower-than-homopolymer distributions 
of $R_{\rm EE}$ peaking at different $R_{\rm EE}$ values to catalog these 
heterogeneous ensembles' SAXS signatures. As such,
these heterogeneous ensembles may be viewed as ``smeared'' versions of the 
$(R_{\rm g},R_{\rm EE})$ subensembles with very narrow $R_{\rm g},R_{\rm EE}$ 
ranges. In this regard, these heterogeneous ensembles are intuitively 
more plausible to be physically encodable by heterpolymeric amino acid 
sequences and therefore of more immediate relevance to potential experimental 
situations.

We construct these ensembles by additional reweighting of the above-described
reweighted ensembles with narrower $R^2_{\rm g}$ distributions, now
applied also to chains with $\ep\neq 0$
[$P(R^2_{\rm g})$ parameterized by $\ep$ and $\alpha$], 
to further bias the final ensembles toward a select 
$R_{\rm EE}$ value. Let $P(R^2_{\rm g},R_{\rm EE})$ be the 
fractional population density with square radius of gyration $R^2_{\rm g}$
and end-to-end distance $R_{\rm EE}$. By definition, the above 
$\alpha$-dependent reweighted $P(R^2_{\rm g})$ may be written as an integral
over $P(R^2_{\rm g},R_{\rm EE})$, viz.,
\begin{equation}
P(R^2_{\rm g}) = \int_0^\infty d R_{\rm EE} \; P(R^2_{\rm g},R_{\rm EE})
\; .
\end{equation}
We now reweight each $P(R^2_{\rm g},R_{\rm EE})$ as follows:
\begin{equation}
P(R^2_{\rm g},R_{\rm EE}) \rightarrow
{\cal N}(R_{\rm g},\gamma, R^0_{\rm EE})^{-1}
P(R^2_{\rm g},R_{\rm EE})
\exp[-\gamma(R_{\rm EE}-R^0_{\rm EE})^2] \; , 
\label{eq:rw1}
\end{equation}
where the normalization factor
${\cal N}(R_{\rm g},\gamma, R^0_{\rm EE})$, defined by
\begin{equation}
\int_0^\infty dR_{\rm EE} 
P(R^2_{\rm g},R_{\rm EE})
\exp[-\gamma(R_{\rm EE}-R^0_{\rm EE})^2] 
= {\cal N}(R_{\rm g},\gamma, R^0_{\rm EE}) 
\int_0^\infty dR_{\rm EE} P(R^2_{\rm g},R_{\rm EE})
\; ,
\label{eq:rw2}
\end{equation}
preserves the weight of each individual $R^2_{\rm g}$ value, and therefore
the overall $\langle R^2_{\rm g}\rangle$ of the reweighted ensemble
remains the same as that of the original ensemble.
In Eqs.~\ref{eq:rw1} and \ref{eq:rw2},
$R^0_{\rm EE}$ is an input parameter, a reference value of $R_{\rm EE}$
toward which the ensemble is biased to a degree parameterized by $\gamma$.
It should be noted that the actual $\langle R_{\rm EE}\rangle$ or
$\langle R^2_{\rm EE}\rangle^{1/2}$ of the final reweighted ensemble
depends on $\alpha$, $\gamma$ as well as the value of $R^2_{\rm g}$ and 
is therefore not expected to be exactly equal to $R^0_{\rm EE}$.

Using the MFF difference measure $\Delta_{\rm Kratky}$ defined above (Fig.~9d),
we have conducted an extensive exploration of the $\alpha,\gamma,R^0_{\rm EE}$ 
parameter space to identify heterogeneous ensembles with MFFs 
closely matching those of homopolymers. Because it is combinatorically
impractical to examine all three parameters exhaustively, we focus 
on several representative $R^0_{\rm EE}$ values. We do so by first examining
$\Delta_{\rm Kratky}$ between homopolymer ensembles and an extensive set 
of heterogeneous ensembles parametrized by combinations of $\alpha,\gamma$ 
values for $R^0_{\rm EE}=$ $30$, $60$, and $90$\AA. As shown in
Fig.~S5 of Supporting Information, the results of this calculation
suggest that ensembles with $R^0_{\rm EE}=90$\AA~are more likely to lead
to small $\Delta_{\rm Kratky}$, especially when
putative optimal values for $\gamma\approx 0.05$, $0.06$, $0.03$,
and $0.1$ that minimize $\Delta_{\rm Kratky}$ are chosen, 
respectively, for chains with intrachain interaction
$\ep=0$, $-1.0$, $-1.8$, and $+2.0$. 

We then proceed to explore the $\alpha,R^0_{\rm EE}$ parameter space while
using these putative optimal $\gamma$ values as given (Fig.~11). 
The variations of
the resulting $\Delta_{\rm Kratky}$ between homopolymer ensembles and 
the heterogeneous ensembles as functions of $\alpha$ and $R^0_{\rm EE}$
are depicted by the contour/heat plots in Figs.~11a--d. Interestingly, while
minimal $\Delta_{\rm Kratky}$ is found in a region of
relatively large $R^0_{\rm EE}$ $\sim 70$--$100$\AA~in
every case studied here (most lightly shaded region in Figs.~11a--d)---as
one might expected because selection of the $\gamma$ parameters used 
in Fig.~11 is based on ensembles with $R^0_{\rm EE}=90$\AA~in Fig.~S5,
a second minimal-$\Delta_{\rm Kratky}$ region is also observed
in Fig.~11c around $R^0_{\rm EE}\approx 30$\AA~and $\alpha\approx 0.01$.

%%%%%%%%%%%%%%%%%%%%%%%%%%%%%%%%%%%%%%%%%%%%%%%%%%%%%%%%%%%%%%%%%%%%%%%%%%%%%%%%
\begin{figure}[t]
\vskip -1cm
{\includegraphics[height=120mm,angle=0]{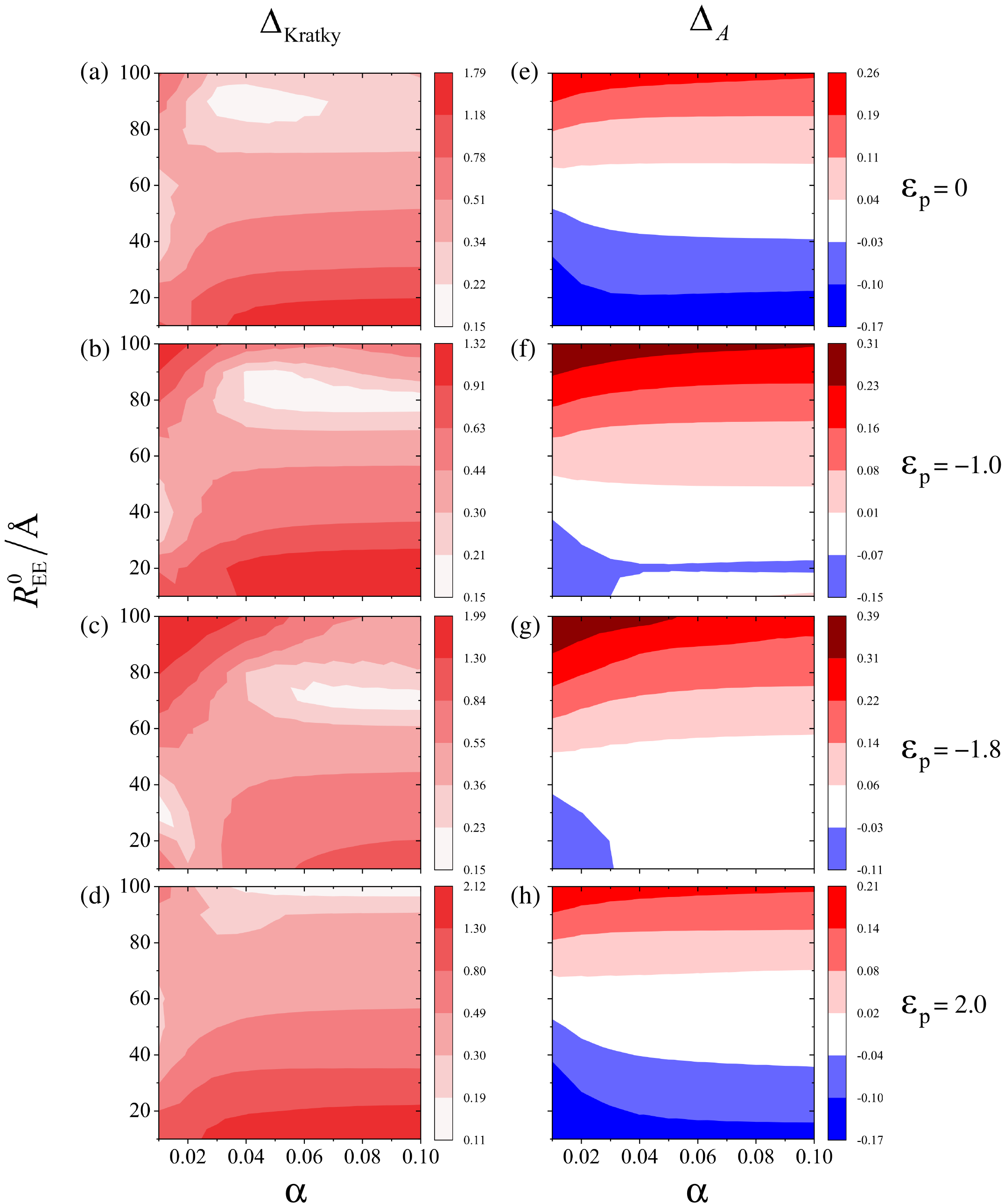}}
\begin{center}
\end{center}
\vspace{-1.3cm}
\caption{Variation of SAXS behaviors and conformational asphericity among
heteropolymeric ensembles with the same $\langle R^2_{\rm g}\rangle$.
(a)--(d) $\Delta_{\rm Kratky}$ and (e)--(h) difference in average asphericity,
$\Delta_A$, between heteropolymeric $(R^0_{\rm EE},\alpha,\gamma)$-defined
reweighted ensembles
($\gamma=0.05$, $0.06$, $0.03$, and $0.1$, respectively, for $\ep=0$,
$-1.0$, $-1.8$, and $+2.0$)
and the homogeneous ensemble for the given $\ep$
values are computed for a $10\times 10$ grid
of ($R^0_{\rm EE},\alpha$) values to produce each of the contour plots.
$\Delta_A$ is the average asphericity of the heteropolymeric ensemble
minus that of the homogeneous ensemble.
}
\label{fig11}
\end{figure}
%%%%%%%%%%%%%%%%%%%%%%%%%%%%%%%%%%%%%%%%%%%%%%%%%%%%%%%%%%%%%%%%%%%%%%%%%%%%%%%%

To quantify the structural differences between the reweighted 
heterogeneous ensembles constructed for chains with different intrachain 
interaction energy $\ep$ with their corresponding homopolymer ensemble, we
compute the asphericity\cite{kuhn1934} for each chain conformation,
defined as\cite{Rudnick1986}
\begin{equation}
A\equiv 1 -
\frac {3(\lambda_1\lambda_2+\lambda_1\lambda_3+\lambda_2\lambda_3)}{(\lambda_1
+\lambda_2+\lambda_3)^2}
\; ,
\end{equation}
where $\lambda_1,\lambda_2,\lambda_3$ (all $\ge 0$) are the eigenvalues
of the gyration tensor $S_{\mu\nu}\equiv n^{-1}\sum_{i=1}^n
({\bf R}_i-{\bf R}_{\rm cm})_\mu ({\bf R}_i-{\bf R}_{\rm cm})_\nu$, 
where $\mu,\nu=1,2,3$ label the Cartesian axes (the $\nu$ index here should
not be confused with the exponent $\nu$ for pair distance scaling), 
and $R_{\rm g}^2 = \lambda_1+\lambda_2+\lambda_3$.
The asphericity quantity $A$ has been used to analyze folded and disordered
states of proteins,\cite{DT2004,Rohit2006}
including recent theoretical applications to better understand smFRET and SAXS
signatures of disordered proteins.\cite{Songetal2015,DT2019}
Here we determine the average asphericity, $\langle A \rangle$, for each
ensemble of interest by averaging $A$ over the (weighted) conformations
in the ensemble and, as measure for one aspect of structural differences 
between a heterogeneous ensemble and a homopolymer ensemble, we
define $\Delta_A$ as the difference between $\langle A\rangle$ of a 
heterogeneous ensemble and that of a homopolymer ensemble. 
The variations of $\Delta_A$ among the reweighted heterogeneous ensembles
as functions of $\alpha$ and $R^0_{\rm EE}$ are depicted by the
contour/heat plots in Figs.~11e--h.

%%%%%%%%%%%%%%%%%%%%%%%%%%%%%%%%%%%%%%%%%%%%%%%%%%%%%%%%%%%%%%%%%%%%%%%%%%%%%%%
\begin{figure}[t]
\vskip -0.5cm
{\includegraphics[height=70mm,angle=0]{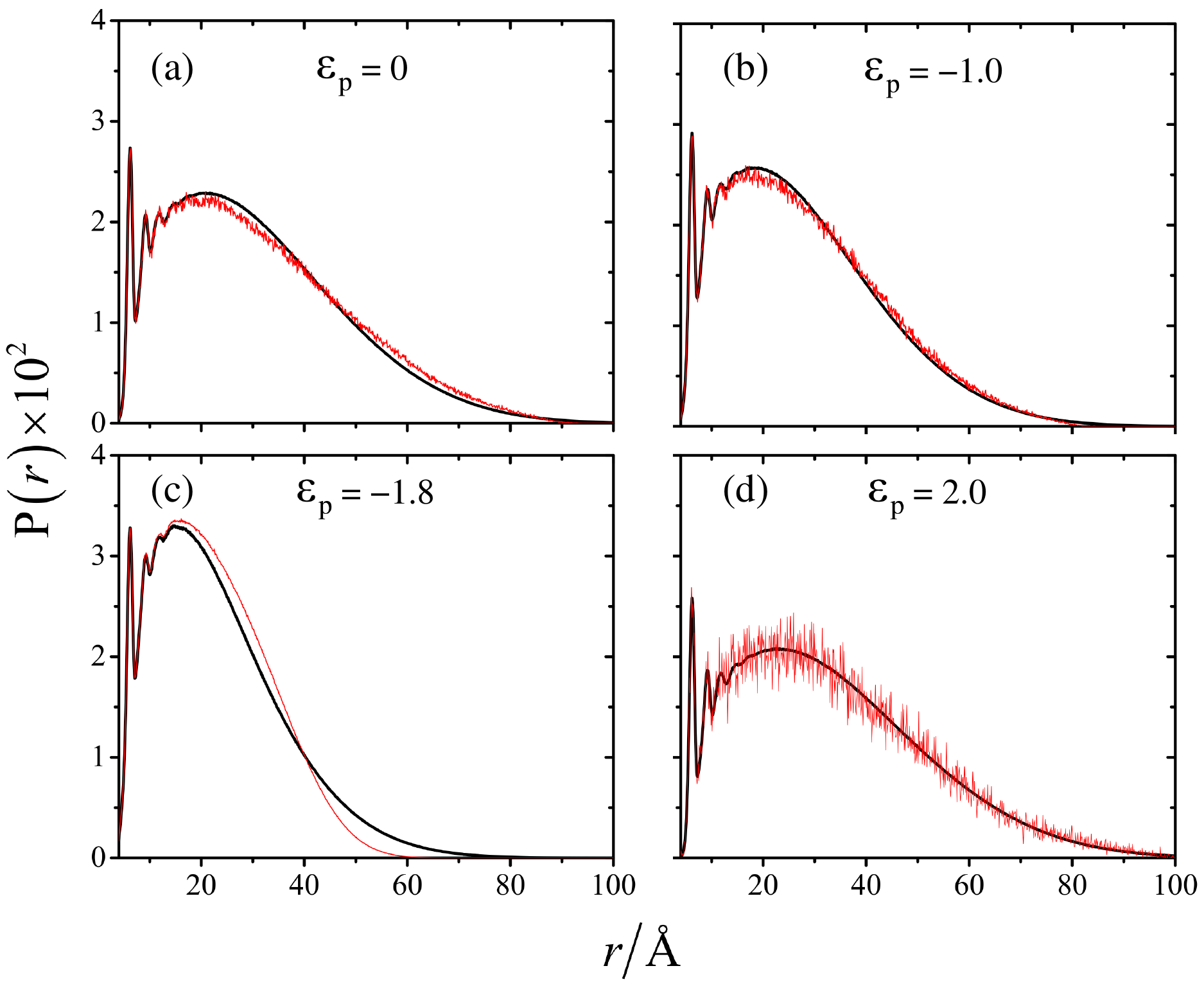}}
\begin{center}
\end{center}
\vspace{-1.3cm}
\caption{Homogeneous and heterogeneous ensembles can have essentially identical
distributions of intrachain monomer-monomer distances.
$P(r)$s are shown for the homogeneous ensembles (black curves) and the
heteropolymeric reweighted ensembles with minimized $\Delta_{\rm Kratky}$
deduced from Fig.~\ref{fig11} (red curves).
The reweighting parameters $(R^0_{\rm EE}/{\rm \AA},\alpha,\gamma)$
for minimum $\Delta_{\rm Kratky}$ are:
(a) (90.0, 0.04, 0.05),
(b) (80.0, 0.08, 0.06),
(c) (30.0, 0.01, 0.03), and
(d) (110.0, 0.1, 0.1).
}
\label{fig12}
\end{figure}
%%%%%%%%%%%%%%%%%%%%%%%%%%%%%%%%%%%%%%%%%%%%%%%%%%%%%%%%%%%%%%%%%%%%%%%%%%%%%%%

%%%%%%%%%%%%%%%%%%%%%%%%%%%%%%%%%%%%%%%%%%%%%%%%%%%%%%%%%%%%%%%%%%%%%%%%%%%%%%%
\begin{figure}[ht!]
\vskip -1cm
{\includegraphics[height=100mm,angle=0]{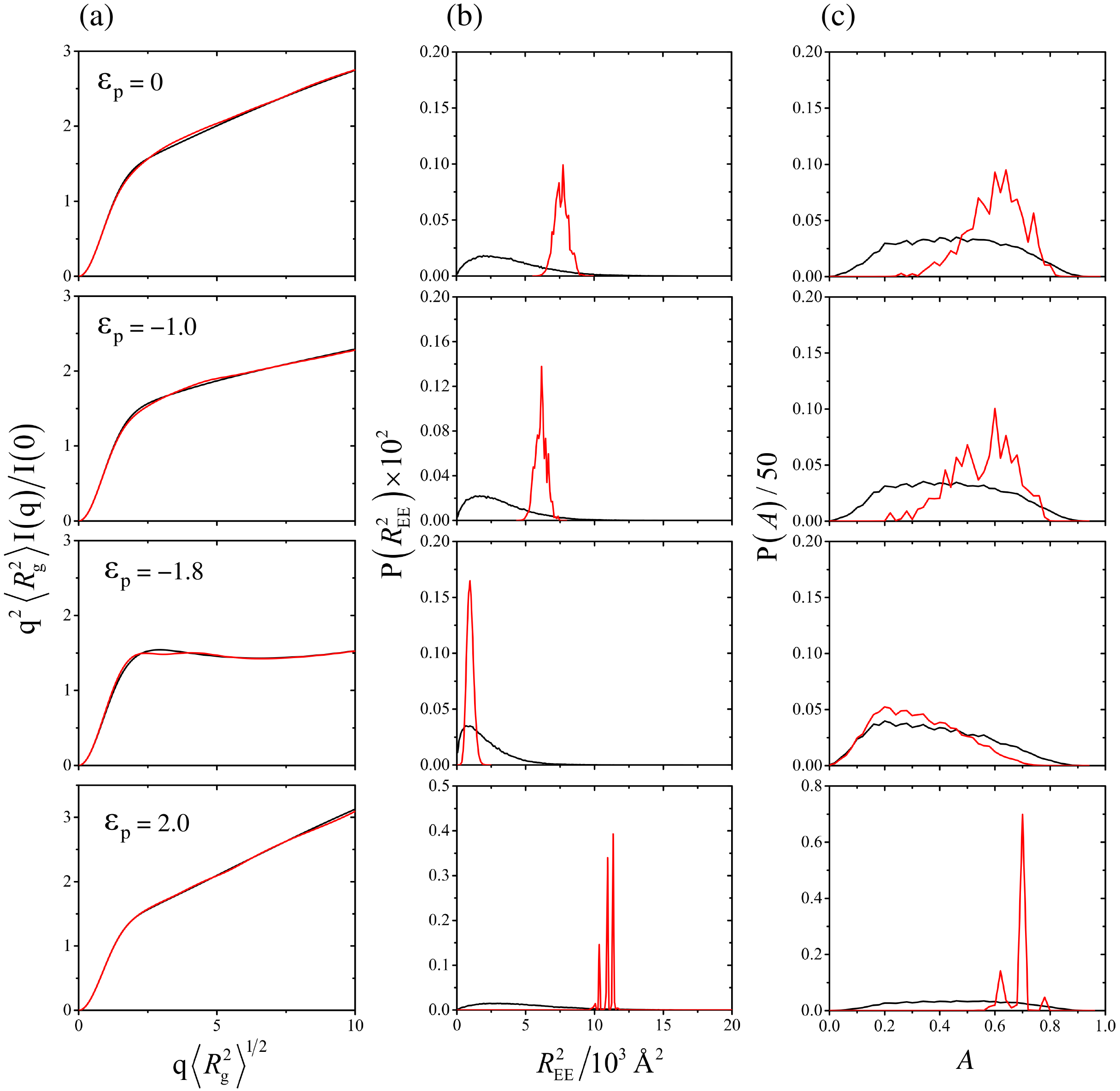}}
\begin{center}
\end{center}
\vspace{-1.3cm}
\caption{Drastically different disordered conformational ensembles can
lead to essentially identical MFFs.
(a) Dimensionless Kratky plots (MFFs), (b) distributions of square end-to-end
distances $R^2_{\rm EE}$, and (c) distributions of conformational
asphericity $A$ are shown for the homogeneous ensembles (black curves)
and heteropolymeric minimum-$\Delta_{\rm Kratky}$ reweighted ensembles
(red curves) in Fig.~\ref{fig12}.
The standard deviation of the $R^2_{\rm g}$ distribution (in units of
\AA$^2$), $\langle R_{\rm EE}^2\rangle^{1/2}/{\rm \AA}$, and $\langle A\rangle$
for the homogeneous $\ep=0$ ensemble are 218.7, 62.5, 0.461, respectively;
whereas the
corresponding values for the $\ep=0$ minimum-$\Delta_{\rm Kratky}$ ensemble
are 42.8, 87.3, 0.610.
For $\ep=-1.0$, the corresponding numbers are
190.1, 56.0, 0.442; 19.0, 78.1, 0.566.
For $\ep=-1.8$:
149.8, 45.1, 0.386; 70.1, 31.1, 0.331,
and for $\ep=+2.0$:
248.8, 68.6, 0.473; 31.9, 105.0, 0.696.
The $10^2$ (b) and $1/50$ (c) in the vertical variables are normalization
factors.
}
\label{fig13}
\end{figure}
%%%%%%%%%%%%%%%%%%%%%%%%%%%%%%%%%%%%%%%%%%%%%%%%%%%%%%%%%%%%%%%%%%%%%%%%%%%%%%%

Figs.~11a--d show that there are low-$\Delta_{\rm Kratky}$ regions of
considerable extent in $(\alpha,R^0_{\rm EE})$ parameter space, but these
regions do not coincide with the low-$\Delta_A$ white regions in
Figs.~11e--h. This observation indicates that there are a wide variety
of heterogeneous ensembles with conformational properties significantly
different from those of homopolymers but nonetheless possess SAXS signatures
essentially indistinguishable from those of homopolymers.
Examples of how the pair distance distribution function $P(r)$ of some of
these heterogeneous ensembles with minimal $\Delta_{\rm Kratky}$ match
closely with the $P(r)$ of homopolymer ensembles are provided in Fig.~12.
As far as MFFs are concerned,
it follows that the dimensionless Kratky plots of the four example
heterogeneous ensembles in Figs.~12a--d, constructed from chains with
intrachain interaction energy $\ep=0$, $-1.0$, $-1.8$, and $+2.0$, and
therefore are of different conformational compactness characterized by
$\avRgs^{1/2}=24.4$, $22.2$, $18.5$, and $26.5$\AA, respectively,
are hardly distinguishable from the dimensionless Kratky plots of
their homopolymeric counterparts (Fig.~13a).
Despite the near-coincidence of their SAXS signatures of these pairs
of heterogeneous and homogeneous ensembles, their conformational properties
are drastically different, as can be seen by their very different
distributions of mean square end-to-end distance (Fig.~13b) and
average asphericity (Fig.~13c).
Interestingly, among the four examples highlighted,
the average asphericities of the heterogeneous and homogeneous ensembles
are least dissimilar for the $\ep=-1.8$ case ($\avRgs^{1/2}=18.5$,
Fig.~13, third row from top). This
feature may be a result of contraints imposed by the
overall conformational compactness, or it may be related to our choice
of this particular heterogeneous ensemble with an
$R_{\rm EE}^2$ distribution peak that coincides approximately with that of
the homopolymer ensemble, though the $R_{\rm EE}^2$ distribution itself
is much narrower for the heterogeneous ensemble than for the homopolymer
ensemble. In any event, the above extensive cataloging of SAXS signatures
of heterogeneous ensembles (Fig.~11) and the explicit examples in Figs.~12 and
13 demonstrate that, across conformational ensembles of different overall
compactness, certain heterogeneous ensembles with conformational properties
dramatically different from those of homopolymers can nonetheless exhibit
essentially identical SAXS signatures as homopolymers.
\\

{\bf Approximate analytical theory for sequence-dependent MFFs of 
disordered heteropolymers.}
The heterogeneous ensembles considered above are tools for logically
delineating the information content of SAXS signatures. To serve 
as examples and counterexamples, it suffices to
define these ensembles mathematically, as long as the ensembles are
in principle physically realizable, without demonstrating 
how the presumed protein ensembles might be encoded exactly by amino 
acid sequences. Nonetheless, it is intuitively plausible
that disordered protein ensembles similar to some---though not all---of these 
constructs, such as the reweighted ensembles encompassing diverse 
conformations but with distributions of $R^2_{\rm g}$ somewhat narrower 
than those of homopolymers (Fig.~10), may be encodable by specific
amino acid sequences. Given the current inadequate
understanding of the physical interactions governing conformational 
properties of disordered proteins,\cite{cosb15}
few insights, if any, exist as to the encodability of 
heterogeneous disordered protein ensembles, i.e., there is very little 
knowledge about what ensembles are physically realizable and what 
ensembles are not. In this light, while a recent analysis 
of computed MFFs of heteropolymers with hydrophobicity-like interactions 
led the authors to conclude that ``$R_{\rm g}$ and $R_{\rm EE}$ remain 
coupled even for heteropolymers'',\cite{tobinPNAS2019} it should be noted
that their study covered only a limited regime of heteropolymeric 
interactions, leaving the likely vast possibilities of disordered 
conformational heterogeneity allowable by polypeptide sequences unsurveyed.
Here we take a rudimentary step to further explore these possibilities by
developing an extension of the perturbative techniques in polymer theory for 
treating excluded volume effects\cite{fixman1955,fixman1962a,fixman1962b,Edwards1965,tanaka1966,ohta81,Freedbook,ohta82,muthu84,chanJCP89,chanJCP90}
to incorporate heterogeneous pairwise intrachain interactions.
Although our analytical formulation is restricted 
to contact-like interactions with a short spatial range and thus 
subtle effects of polypeptide interactions cannot be addressed,
results below from this computationally efficient approach are instructive
in offering a glimpse of how various properties of heterogeneous disordered 
ensembles---including decoupling of $R_{\rm g}$ and $R_{\rm EE}$
in some cases---might be encoded.

Our analytical formulation is based on the path-integral representation
of the polymer partition function
\begin{equation}
{\cal Q}(N,\{{\tilde{v}}\})= \int [ {\cal D} {\bf R} ] \* 
{\rm e}^{-{\cal H}(N,\{{\tilde{v}}\},\{ {\bf R}\})} \;
\; ,
\label{eq:Q1}
\end{equation}
where $N$ is the total contour length of the polymer and ${\bf R}(\tau)$
is the spatial position of the point labeled by the contour length
variable $\tau$ along the polymer. 
If a bond connecting two monomer beads is identified with a chain 
segment of length $l$, the total number of beads in each chain 
modeled by Eq.~\ref{eq:Q1} is equal to $n=N/l+1$.
For notational convenience, $l$ is set to unity ($l=1$) unless specified 
otherwise. The Hamiltonian ${\cal H}$ is given by
\begin{equation}
{\cal H}(N,\{{\tilde{v}}\},\{ {\bf R}\}) = \frac {1}{2}
\int_0^N d\tau \; \Bigl\vert\frac {d{\bf R}(\tau)}{{d\tau}}\Bigr\vert^2 +
\int_a^N d\tau\int_0^{\tau-a} d\tau^\prime \;
{\tilde{v}}(\tau,\tau^\prime)\delta[{\bf R}(\tau)-{\bf R}(\tau^\prime)] 
\; ,
\label{eq:Q2}
\end{equation}
where
${\tilde{v}}(\tau,\tau^\prime)$ is the pairwise interaction energy between
the points labeled by $\tau$ and $\tau^\prime$ along the polymer chain,
and $a \sim 1$ is a cutoff in contour length to remove unphysical self 
interaction of any chain segment with itself.
In general, the energy function ${\tilde{v}}(\tau,\tau^\prime)$ depends on 
$\tau,\tau^\prime$ and thus the formulation describes a heteropolymer.
In the special case when ${\tilde{v}}(\tau,\tau^\prime)$ 
$={\tilde{v}}_0>0$, i.e., when $\tilde{v}$ is independent of $\tau$ and 
$\tau^\prime$, Eqs.~\ref{eq:Q1} and \ref{eq:Q2} reduce to those
for a homopolymer with uniform excluded volume interactions
[see, e.g., Eqs.~(5.1) and (5.2) of ref.~\citen{chanJCP89} and
Eqs.~(4.1) and (4.2) of ref.~\citen{chanJCP90}].
It should also be noted that although we allow individual pairwise interactions 
in Eq.~\ref{eq:Q2} to be neutral, attractive or repulsive, for simplicity,
we do not employ a three-body repulsion term (as in some other analytical
formulations, see, e.g., ref.~\citen{huihui2020}) to account 
for excluded volume when ${\tilde{v}}$ is attractive in the present 
perturbative treatment. 

Diagrammatic perturbation expansions in the present heteropolymer formulation
proceed largely along the description in ref.~\citen{chanJCP89} 
except $v_0$ in this reference is now replaced by
${\tilde{v}}(\tau,\tau^\prime)$ which is then placed inside the 
$\int d\tau\int d\tau^\prime$ integrals as a factor of the integrand. 
As well, the ${\bf c}(\tau)=\sqrt{d}{\bf R}(\tau)$ rescaling of the 
spatial coordinates in ref.~\citen{chanJCP89} is not applied here because 
it offers no advantage for our present applications which are all 
for $d=3$ spatial dimensions. As a result, the
propagator given by Eq.~(5.10) of ref.~\citen{chanJCP89} is now
replaced by $(3/2\pi)^{3/2} (1/|\tau-\tau^\prime|)^{3/2}
\exp[-3|{\bf R} - {\bf R}^\prime|^2/(2|\tau-\tau^\prime|)]$.
To simplify the expressions to be presented below, we define
$v(\tau,\tau^\prime)$ $\equiv$ $(3/2\pi)^{3/2}{\tilde{v}}(\tau,\tau^\prime)$
and $v_0\equiv (3/2\pi)^{3/2}{\tilde{v}}_0$. After all these notational
modifications are taken into account, 
the ``$v_0$'' in ref.~\citen{chanJCP90} is seen to be
equivalent to $(2\pi)^{3/2}v_0$ in the present formulation.

%%%%%%%%%%%%%%%%%%%%%%%%%%%%%%%%%%%%%%%%%%%%%%%%%%%%%%%%%%%%%%%%%%%%%%%%%%%%%%%%
\begin{figure}[t]
{\includegraphics[height=45mm,angle=0]{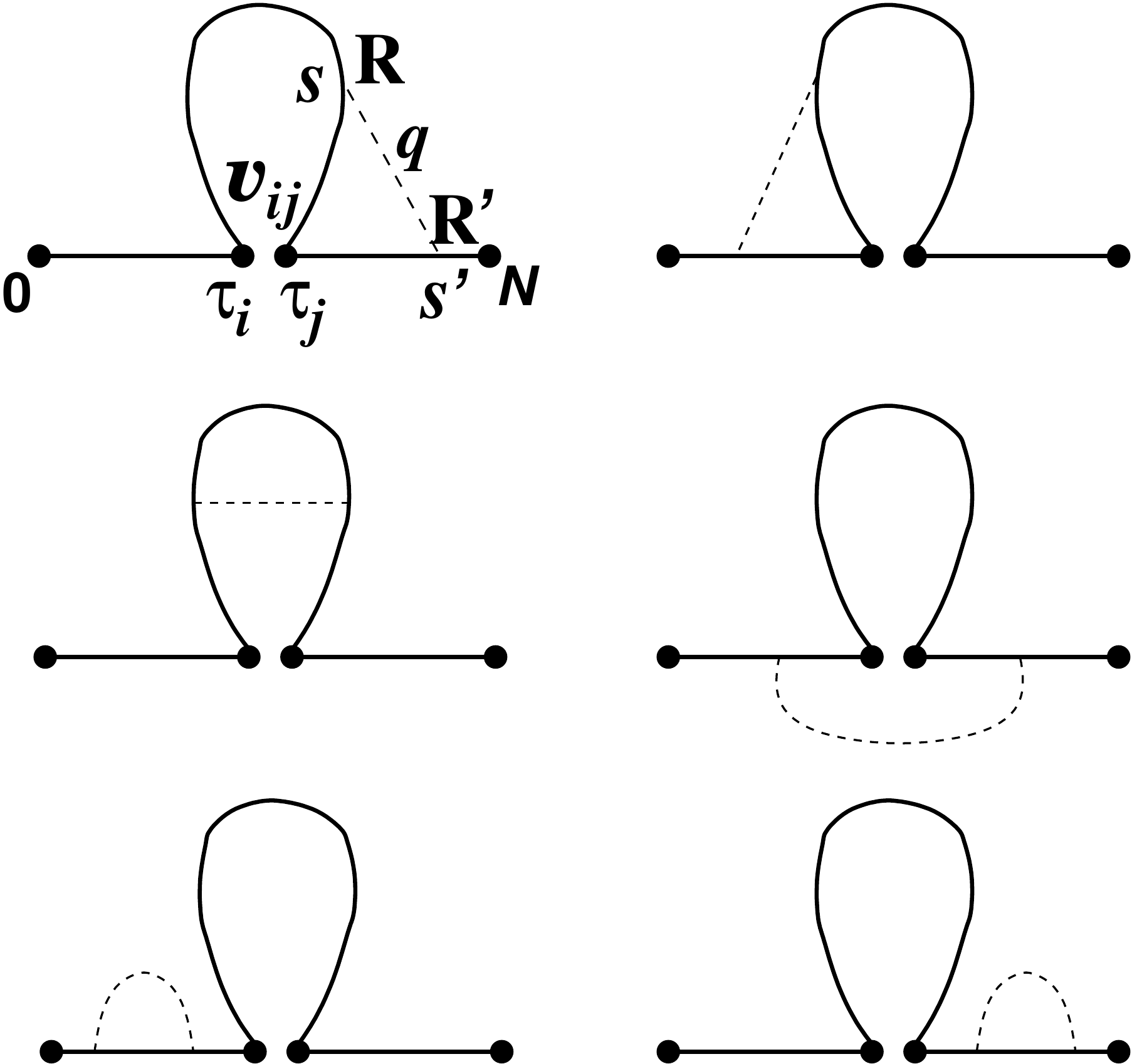}}
\begin{center}
\end{center}
\vspace{-1.0cm}
\caption{Diagrams for perturbative calculations of scattering intensities
for heteropolymers. $v_{ij}$ is the interaction energy between contour
positions $\tau_i$ and $\tau_j$. Contour positions $s$ and $s^\prime$
at spatial positions ${\bf R}$ and ${\bf R}^\prime$
(connected by a dashed line)
are associated with the scattering vector ${\bf q}$.
}
\label{fig14}
\end{figure}
%%%%%%%%%%%%%%%%%%%%%%%%%%%%%%%%%%%%%%%%%%%%%%%%%%%%%%%%%%%%%%%%%%%%%%%%%%%%%%%%

In the present notation, the standard perturbative formula for
mean square end-to-end distance\cite{fixman1955,muthu84} 
is now generalized to 
\begin{equation}
\langle R_{\rm EE}^2\rangle = 
N + \int_a^N d\tau_j \int_0^{\tau_j-a} d\tau_i 
\frac {{v}_{ij}}{\sqrt{\Delta\tau_{ij}}} + O(v_{ij}^2) \;
\label{REE-vij}
\end{equation}
for heteropolymer, where
$\Delta\tau_{ij}\equiv \tau_j-\tau_i$ 
is the contact order\cite{chanJCP89,chanJCP90} 
of the $\tau_i,\tau_j$ contact, and
$v_{ij}=v(\tau_i,\tau_j)$.
For $v_{ij}=v_0$, i.e., for the special case in which the chain is a 
homopolymer, the above expression reduces to
\begin{equation}
\langle R_{\rm EE}^2\rangle = 
N \biggl\{1 + v_0 \biggl[ \frac {4}{3} \sqrt{N} 
- \biggl( 2 \sqrt{a} - \frac {2a^{3/2}}{3N} \biggr)
\biggr]\biggr\} + O(v_0^2) 
\; ,
\label{REE-v0}
\end{equation}
as in refs.~\citen{fixman1955} and \citen{muthu84}.
For heteropolymers with discrete sequences, we replace the 
integral in Eq.~\ref{REE-vij} by summing over
a discrete interaction matrix $v_{ij}$---which may be viewed as
containing the net energetic effects of ``hard-core'' 
excluded volume repulsions and short-spatial-ranged sequence-dependent 
attractive or repulsive interactions, viz.,
\begin{equation}
\langle R_{\rm EE}^2\rangle = 
N + \sum_{j=2}^n \sum_{i=1}^{j-1}
\frac {v_{ij}}{\sqrt{\Delta_{ij}}} + O(v_{ij}^2) 
\;
\label{eq:REE_discrete1}
\end{equation}
where $\Delta_{ij}\equiv j - i$, and $a=1$ is used for the
double summations here in Eq.~\ref{eq:REE_discrete1}.

Application of the above-described perturbative formalism to the Feynman-type 
diagrams in Fig.~14 for heteropolymer scattering intensities
(which correspond to the six diagrams in Fig.~1 
of Ohta etal.\cite{ohta82} for homopolymer scattering intensities)
leads to the following perturbative expression for the scattering intensity:
\begin{eqnarray}
I(q) & = & \frac {2}{\alpha} \biggl [ N + \frac {1}{\alpha}
\Bigl (e^{-\alpha N} - 1 \Bigr ) \biggr ]
\nonumber \\
&&
- 2 \int_a^N d\tau_j \int_0^{\tau_j-a} d\tau_i 
\frac {v_{ij}}{\Delta\tau_{ij}^{3/2}} 
\Biggl \{ \frac {1}{\alpha} \biggl [ N - \Delta\tau_{ij} +
\frac{1}{\alpha} \Bigl (e^{-\alpha (N-\Delta\tau_{ij})} - 1 \Bigr ) \biggr ]
\nonumber \\
&& \quad + \Delta\tau_{ij}
\biggl [ \frac {1}{\alpha}
\biggl ( 2-e^{-\alpha \tau_i} -e^{-\alpha(N-\tau_j)}\biggr ) + 
\frac {\Delta\tau_{ij}}{2}\biggr ] 
{\cal F}\Bigl(\frac {\sqrt{\alpha\Delta\tau_{ij}}}{2}\Bigr) \Biggr \}
+ O(v_{ij}^2) \;
\nonumber \\
\label{Iq-eq0}
\end{eqnarray}
where $\alpha\equiv q^2/6$, and
\begin{equation}
{\cal F}(z) \equiv \frac {e^{-z^2}}{z}\int_0^z dt\; e^{t^2} 
\; 
\label{Fz-eq}
\end{equation}
is plotted in Fig.~S6a of the Supporting Information.
We have verified by a rather involved algebraic comparison of the 
expressions in Eqs.~(\ref{Iq-eq0}) and (\ref{Fz-eq}) for the 
$v_{ij}=v_0$ special case of homopolymers against the
results provided by Eqs.~(3.3)--(3.8) in Ohta et al.\cite{ohta82} 
that our first-order perturbative results for $I(q)$ 
are consistent with theirs except for several likely typographical 
errors in ref.~\citen{ohta82}, as described in the Supporting 
Information of the present article. In this regard, we should also
note that the subject matter and goals of the two efforts are different: 
whereas ref.~\citen{ohta82} studies universal homopolymeric behaviors 
in the limit of infinite chain length through 
renormalization group analysis\cite{oono_ren81-1} 
(see the ``RN'' Kratky plot in Fig.~S2A of ref.~\citen{tobinSci2017} and 
the $I(x)$ curve in Fig.~2 of ref.~\citen{ohta82}), the main focus of 
the present work is on sequence-specific properties of finite-length 
heteropolymers.

Proceeding now from Eq.~\ref{Iq-eq0} above, 
as $q\rightarrow 0$, i.e., in the $\alpha\rightarrow 0$ limit,
the expression in Eq.~\ref{Iq-eq0} becomes
\begin{equation}
I(0) = N^2 \biggl [ 1 
-  \int_a^N d\tau_j \int_0^{\tau_j-a} d\tau_i 
\frac {v_{ij}}{\Delta\tau_{ij}^{3/2}} \biggr] 
+ O(v_{ij}^2) \;
\end{equation}
because 
${\cal F}(z)= 1 - 2z^2/3 +4z^4/15 -8z^6/105 + O(z^8)$,
thus
${\cal F}(\sqrt{\alpha\Delta\tau_{ij}}/2) =
1 - \alpha\Delta\tau_{ij}/6 + \alpha^2(\Delta\tau_{ij})^2/60 + O (\alpha^3)$
and hence
$\lim_{\alpha\to 0} {\cal F}(\sqrt{\alpha\Delta\tau_{ij}}/2) = 1$. 
It follows that
\begin{eqnarray}
\frac {I(q)}{I(0)} & = & \frac {2}{\alpha N} \biggl [ 1 + \frac {1}{\alpha N}
\Bigl (e^{-\alpha N} - 1 \Bigr ) \biggr ]
\nonumber \\
&&
- 2 \int_a^N d\tau_j \int_0^{\tau_j-a} d\tau_i 
\frac {v_{ij}}{\Delta\tau_{ij}^{3/2}} 
\Biggl \{ \frac {1}{\alpha N} 
\Bigl [ -\frac {\Delta\tau_{ij}}{N} + \frac {1}{\alpha N}
\Bigl(e^{-\alpha (N-\Delta\tau_{ij})} - e^{-\alpha N}\Bigr ) \Bigr]
\nonumber \\
&& \quad + \frac {\Delta\tau_{ij}}{N}
\biggl [ \frac {1}{\alpha N}
\biggl ( 2-e^{-\alpha \tau_i} -e^{-\alpha(N-\tau_j)}\biggr ) + 
\frac {\Delta\tau_{ij}}{2N}\biggr ] 
{\cal F}\Bigl(\frac {\sqrt{\alpha\Delta\tau_{ij}}}{2}\Bigr) \Biggr \}
+ O(v_{ij}^2) 
%\; 
.
\label{Iq-eq1}
\end{eqnarray}
Using the standard formula for Guinier's approximation,\cite{AndoChemRev2017}
\begin{equation}
\langle R_{\rm g}^2 \rangle = -3 \frac {d}{dq^2} [I(q)/I(0)] \vert_{q=0}
= -\frac{1}{2} \frac{d}{d\alpha} [I(q)/I(0)] \vert_{\alpha=0}
\end{equation}
as well as the expansion
$d{\cal F}(z)/dz= -4z/3 + 16z^3/15 - 16z^5/35 + O(z^7)$ and thus
$d{\cal F} (\sqrt{\alpha\Delta\tau_{ij}}/2)/d\alpha
= -\Delta\tau_{ij}/6 + \alpha(\Delta\tau_{ij})^2/30 + O(\alpha^2)$
and therefore 
$\lim_{\alpha \to 0} d{\cal F}(\sqrt{\alpha\Delta\tau_{ij}}/2)/d\alpha= 
-\Delta\tau_{ij}/6$, we obtain, in (implicit) units of $l^2$:
\begin{eqnarray} 
\langle R^2_{\rm g} \rangle
& = & \frac {N}{6} 
-  \int_a^N d\tau_j \int_0^{\tau_j-a} d\tau_i 
\frac {v_{ij}}{\Delta\tau_{ij}^{3/2}} 
\biggl \{
 \frac {\Delta\tau_{ij}}{2} \biggl 
[ \frac {\tau_i^2 + (N-\tau_j)^2}{N^2} -1 \biggr]
+ \frac {2\Delta\tau_{ij}^2}{3N}
- \frac {\Delta\tau_{ij}^3}{4N^2}
\biggr \}
\nonumber \\
&& + O(v_{ij}^2) \; .
\label{Rg-hetero-eq}
\end{eqnarray} 
For the special case of homopolymer, $v_{ij}=v_0$, 
and this expression reduces to
\begin{equation}
\langle R_{\rm g}^2\rangle
= \frac {N}{6} \biggl \{ 1 + v_0 \biggl [ \frac {134}{105}\sqrt{N}
- \biggl ( 2\sqrt{a} - \frac {2a^{3/2}}{3N} 
+ \frac {a^{5/2}}{5N^2} 
+ \frac {a^{7/2}}{7N^3} 
\biggr )
\biggr ] \biggr \} + O (v_0^2) \; ,
\label{Rg-v0}
\end{equation}
which is consistent with the result of Fixman.\cite{fixman1955}
For heteropolymers, combining Eq.~\ref{REE-vij} with
Eq.~\ref{Rg-hetero-eq} yields
\begin{eqnarray} 
\frac {\langle R^2_{\rm g} \rangle}{\langle R^2_{\rm EE} \rangle}
& = & \frac {1}{6} 
-  \int_a^N d\tau_j \int_0^{\tau_j-a} d\tau_i 
\frac {v_{ij}}{\Delta\tau_{ij}^{3/2}} 
\biggl \{
 \frac {\Delta\tau_{ij}}{2N} \biggl 
[ \frac {\tau_i^2 + (N-\tau_j)^2}{N^2} +1 \biggr]
+ \frac {2\Delta\tau_{ij}^2}{3N^2}
- \frac {\Delta\tau_{ij}^3}{4N^3}
\biggr \}
\nonumber \\
&& + O(v_{ij}^2) 
\; .  
\label{Rg-REE-hetero-eq}
\end{eqnarray} 
The corresponding expression for homopolymers is obtained by
combining Eq.~\ref{REE-v0} and Eq.~\ref{Rg-v0}:
\begin{equation}
\frac {\langle R_{\rm g}^2\rangle}{\langle R_{\rm EE}^2\rangle}
= \frac {1}{6} \biggl \{ 1 - v_0 \biggl [ \frac {2}{35}\sqrt{N}
+ \biggl ( \frac {a^{5/2}}{5N^2} + \frac {a^{7/2}}{7N^3} \biggr )
\biggr ] \biggr \} + O (v_0^2) \: .
\label{eq:RgREE-0}
\end{equation}
For heteropolymers with discrete sequences, the integrals
in Eq.~\ref{Iq-eq1} for $I(q)/I(0)$, Eq.~\ref{Rg-hetero-eq} for
$\langle R^2_{\rm g}\rangle$, and Eq.~\ref{Rg-REE-hetero-eq}
for $\langle R^2_{\rm g}\rangle/\langle R^2_{\rm EE}\rangle$
are replaced by summmations with a discrete pairwise interaction matrix
$v_{ij}$ that replaces the $v_{ij}=v(\tau_i,\tau_j)$ in the
continuum, viz.,
\begin{equation}
\int_a^N d\tau_j \int_0^{\tau_j-a} d\tau_i 
\longleftrightarrow
\sum_{j=a+1}^n \sum_{i=1}^{j-a}
\; ,
\label{eq:discretizeALL}
\end{equation}
where, in most cases, we take $a=1$ as in Eq.~\ref{eq:REE_discrete1}
for $\langle R_{\rm EE}^2\rangle$.
Practically, the same discretization is also used for numerical calculations
of $I(q)/I(0)$ for homopolymers when $v_{ij}=v_0$. In general,
it follows from Eq.~\ref{Iq-eq1} that
\begin{equation}
\frac {I(q)}{I(0)} = \biggl[\frac {I(q)}{I(0)}\biggr]_0  
+{\widetilde{G}}_2(N,\{v_{ij}\},q;a)
+O(v_{ij}^2)
\; ,
\label{eq:Iq-expansion-hetero}
\end{equation}
where\cite{ohta81}
\begin{equation}
\biggl[\frac {I(q)}{I(0)}\biggr]_0  
=  \frac {2}{\alpha N} \biggl [ 1 + \frac {1}{\alpha N}
\Bigl (e^{-\alpha N} - 1 \Bigr ) \biggr ]
\;
\label{eq:IqI0_0}
\end{equation}
follows from the 
$P(r)=(8\pi r^2/N^2)\int_0^N dx (N-x) (3/2\pi x)^{3/2}\exp(-3r^2/2x)$
pair distance distribution function for Gaussian chains, and
\begin{eqnarray}
{\widetilde{G}}_{2}(N,\{v_{ij}\},q;a)
& \equiv & 
- \frac {2}{N^2} 
\sum_{j=a+1}^n \sum_{i=1}^{j-a}
\frac {v_{ij}}{\Delta_{ij}^{3/2}} 
\Biggl \{ \frac {1}{\alpha} \biggl [ - \Delta_{ij} + \frac {1}{\alpha}  
\Bigl (e^{-\alpha (N-\Delta\tau_{ij})} - e^{-\alpha N}
 \Bigr ) \biggr ]
\nonumber \\
&& \quad + \Delta_{ij}
\biggl [ \frac {1}{\alpha} \biggl ( 
2-e^{-\alpha i} -e^{-\alpha(N-j)} \biggr) + 
\frac {\Delta_{ij}}{2}\biggr ] 
{\cal F}\Bigl(\frac {\sqrt{\alpha\Delta_{ij}}}{2}\Bigr) \Biggr \}
\; 
\label{eq:G2_a-general}
\end{eqnarray}
in the discretized form. The continuum form
for ${\widetilde{G}}_{2}(N,\{v_{ij}\},q;a)$ is readily obtainable by replacing 
the double summations by the double integrals in Eq.~\ref{eq:G2_a-general} 
in accordance with the correspondence
specified in Eq.~\ref{eq:discretizeALL}.
To show $I(q)/I(0)$ in a logarithmic scale, which is a common practice
as in Figs.~4a and 7b, we use the standard expansion of $\ln(1+x)=x+O(x^2)$ 
to recast Eq.~\ref{eq:Iq-expansion-hetero} as
\begin{equation}
\ln\biggl[\frac {I(q)}{I(0)}\biggr] =  
\biggl[\frac {I(q)}{I(0)}\biggr]_0  + {\widetilde{F}}_2(N,\{v_{ij}\},q;a)
+O(v_{ij}^2)
\;
\label{eq:logI-0-general}
\end{equation}
where
\\
\begin{equation}
{\widetilde{F}}_2(N,\{v_{ij}\},q;a) \equiv 
\frac {{\widetilde{G}}_2(N,\{v_{ij}\},q;a)}{[I(q)/I(0)]_0}
\; .
\label{eq:F2-general}
\end{equation}
For the special case of homopolymers, $v_{ij}=v_0$, we have
\begin{equation}
\frac {I(q)}{I(0)} = \biggl[\frac {I(q)}{I(0)}\biggr]_0  
+v_0 G_2(N,q;a)
+O(v_0^2)
\; ,
\label{eq:Iq-expansion}
\end{equation}
where
\begin{equation}
G_2(N,q;a) \equiv 
{\widetilde{G}}_{2}(N,\{v_{ij}\},q;a)\vert_{v_{ij}=1}
\; 
\label{eq:G2-def}
\end{equation}
is the expression given by Eq.~\ref{eq:G2_a-general} with 
all $v_{ij}$ set to unity, and
\begin{equation}
\ln\biggl[\frac {I(q)}{I(0)}\biggr] =  
\biggl[\frac {I(q)}{I(0)}\biggr]_0  +v_0 F_2(N,q;a)
+O(v_0^2)
\;
\label{eq:logI-0}
\end{equation}
where
\begin{equation}
F_2(N,q;a) \equiv 
\frac {G_2(N,q;a)}{[I(q)/I(0)]_0}
\; .
\label{eq:F2}
\end{equation}

To compare these theoretical predictions with our simulated explicit-chain
model results, it is necessary to determine the effective Kuhn length,
referred to as ${\tilde b}$ below, of the explicit-chain model
because the polypeptide-mimicking bond-angle potential of the model 
(see Models and Methods) entails ${\tilde b}\neq b$ where $b=3.8$~\AA~is
the C$_\alpha$--C$_\alpha$ virtual bond length. Here, we determine
${\tilde b}$ and the effective chain length $\tilde N$ of the polypeptide
model by equating the 
the mean-square radius of gyration, $\langle R^2_{\rm g}\rangle_0$, of
the Gaussian ($\epsilon_{\rm ex}=0$) version of our model
with $\tilde N \tilde b^2 /6$ while keeping the total contour length
${\tilde N} {\tilde b} = Nb$ unchanged.
For our Gaussian chain model, $N=n-1=74$, 
$\langle R^2_{\rm g}\rangle_0^{1/2}=17.4$~\AA~(corresponding
$\langle R^2_{\rm EE}\rangle_0^{1/2}=42.6$~\AA,
$\langle R^2_{\rm g}\rangle_0/\langle R^2_{\rm EE}\rangle_0=6.0$).
This calculation yields $\tilde b=1.7b=6.46$~\AA~and
${\tilde N}=43.5$. Based on this determination, we use
$N\rightarrow {\tilde N}\rightarrow \lfloor \tilde N \rfloor =43$ 
and $n\rightarrow {\tilde N}+1=44$ in the applications of our discretized 
formulation below.

%%%%%%%%%%%%%%%%%%%%%%%%%%%%%%%%%%%%%%%%%%%%%%%%%%%%%%%%%%%%%%%%%%%%%%%%%%%%%%%%
\begin{figure}[t]
\vskip -0.6cm
{\includegraphics[height=48mm,angle=0]{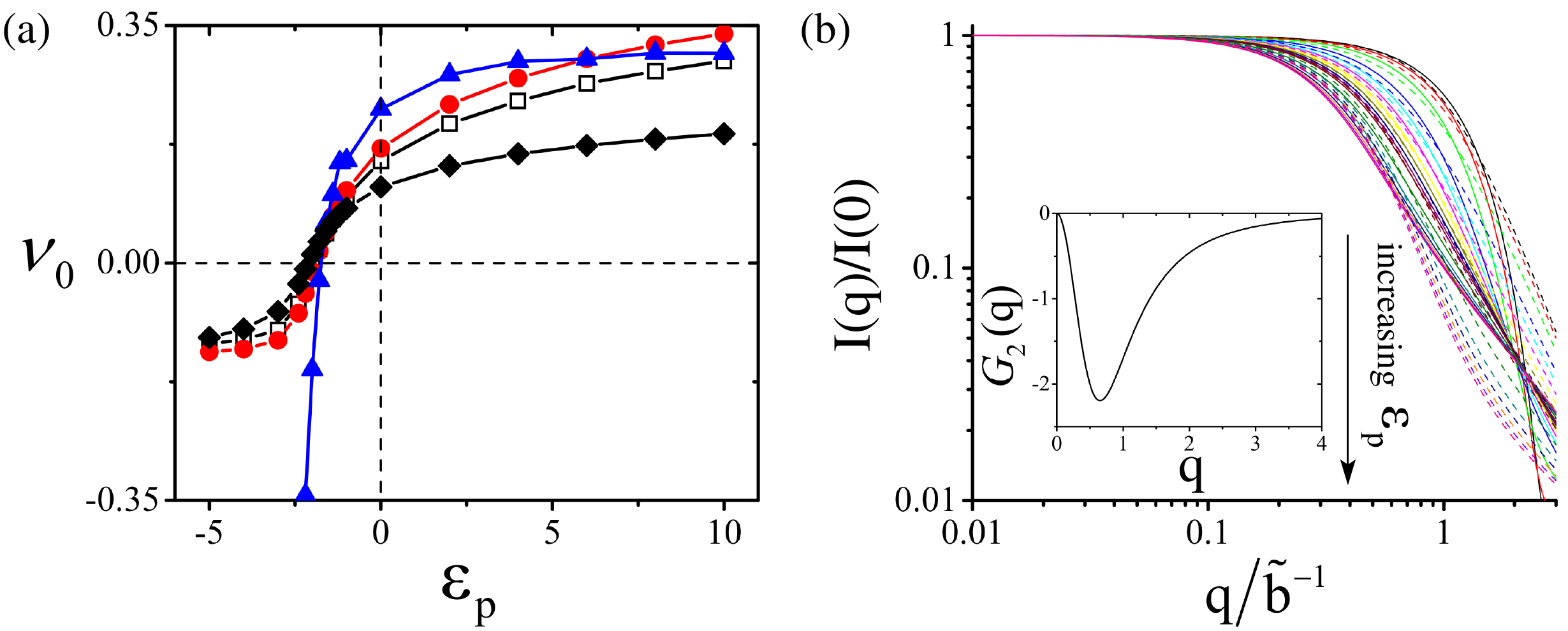}}
\begin{center}
\end{center}
\vspace{-1.3cm}
\caption{Perturbation theory-predicted and explicit-chain model simulated
conformational and SAXS properties of homogeneous ensembles.
(a) Plotted $v_0$ values are obtained by fitting perturbation theory
results calculated using effective chain length
${\tilde{N}}=43$, effective Kuhn length ${\tilde{b}}=l=6.46$~\AA,~and $a=1$
to explicit-chain simulated $\langle R^2_{\rm g}\rangle$
(black squares), $\langle R^2_{\rm EE}\rangle$ (red circles),
$\langle R^2_{\rm g}\rangle/\langle R^2_{\rm EE}\rangle$ (blue triangles),
and $I(q)/I(0)$ (black diamonds) for homopolymers with excluded volume
($\epsilon_{\rm ex}=1.0 k_{\rm B}T$, Figs.~3 and 4) and
$\ep=-5.0$, $-4.0$, $-3.0$, $-2.4$, $-2.2$, $-2.0$, $-1.8$, $-1.6$, $-1.4$,
$-1.2$, $-1.0$, $0.0$, $+2.0$, $+4.0$, $+6.0$, $+8.0$, and $+10.0$.
Lines joining data points are merely guides for the eye.
(b) The explicit-chain simulated $I(q)/I(0)$ (solid curves, from Fig.~4a)
are compared with their fitted perturbation theory-predicted $I(q)/I(0)$
calculated using Eq.~\ref{eq:Iq-expansion} (dashed curves, same color code).
As indicated by the vertical arrow,
$\ep$ increases monotonically (becoming less attractive or more repulsive)
from the highest to the lowest $I(q)/I(0)$ curves plotted. The inset shows
the function $G_2(q)=G_2({\tilde N}=43,q;a=1)$ in
Eq.~\ref{eq:G2-def}.
}
\label{fig15}
\end{figure}
%%%%%%%%%%%%%%%%%%%%%%%%%%%%%%%%%%%%%%%%%%%%%%%%%%%%%%%%%%%%%%%%%%%%%%%%%%%%%%%%

We first compare predictions of our analytical formulation for
homopolymers with the corresponding explicit-chain simulation results.
Setting the chain length $N$ to ${\tilde N}=43$ and 
the length unit $l$ from unity to $l={\tilde b}=6.46$~\AA~in 
Eq.~\ref{REE-v0} for $\langle R^2_{\rm EE}\rangle$,
Eq.~\ref{Rg-v0} for $\langle R^2_{\rm g}\rangle$,
Eq.~\ref{eq:RgREE-0} for 
$\langle R^2_{\rm g}\rangle/\langle R^2_{\rm EE}\rangle$,
and Eqs.~\ref{eq:G2_a-general}, \ref{eq:Iq-expansion}, and \ref{eq:G2-def}
for $I(q)/I(0)$, we obtain, separately for each of these $v_0$-dependent
quantities, the $v_0$ values in the analytical formulation that optimize
fitting to the corresponding explicit-chain simulated results for 
these quantities in Figs.~3 and 4 (Fig~15a).
The relationship between the $\ep$ in the explicit-chain model 
(horizontal variable in Fig.~15a) and the optimized
$v_0$ in the analytical formulation (vertical variable in Fig.~15a) 
is monotonic, as one would expect, but is clearly nonlinear, exhibiting
a sigmoidal-like increase around $\ep\approx -2.0$ and $v_0\approx 0$.
The trends for the four conformational properties tested are largely 
consistent, supporting, at least to a degree, the effectiveness of the theory.
But the optimally fitted $v_0$ values for $I(q)/I(0)$ are appreciably
lower than those for $\langle R^2_{\rm EE}\rangle$, 
$\langle R^2_{\rm g}\rangle$, and 
$\langle R^2_{\rm g}\rangle/\langle R^2_{\rm EE}\rangle$, indicating
that more subtle structural and energetic features of the explicit-chain 
models are not captured by the analytical formulation.
Nonetheless, with the optimized $v_0$ values for $I(q)/I(0)$, Fig.~15b
shows that the analytical predictions fit the explicit-chain results
quite well overall. 
The $G_2(q)=G_2(N={\tilde N}=43,q;a=1)$ function (Eq.~\ref{eq:G2-def}) used
for computing the theoretical scattering curves is shown in the inset.
The fit in Fig.~15b is excellent for $I(q)/I(0)\gtrsim 0.2$ 
($q\lesssim 0.9$--$2.0$).
Mismatches appear for $I(q)/I(0)\lesssim 0.2$ in that the 
explicit-chain-simulated curves converge but the analytical curves do not,
for the simple reason that the analytical formulation---unlike the 
explicit-chain model (see discussion of the results in Fig.~4a above)---does 
not entail a near-universal minimum nonzero intrachain pairwise distance 
for all the chain conformations in any given ensemble.
For a more extensive survey of the theoretical formulation
developed here, the $G_2(N,q;a)$ and $F_2(N,q;a)$ (Eq.~\ref{eq:F2})
for several other $N$ and $a$ values are provided in
Figs.~S6b,c of the Supporting Information.

In Fig.~15a, we notice that the optimally fitted $v_0$ excluded-volume 
parameters for the explicit-chain simulated 
$\langle R^2_{\rm EE}\rangle$, $\langle R^2_{\rm g}\rangle$,
$\langle R^2_{\rm g}\rangle/\langle R^2_{\rm EE}\rangle$, and $I(q)/I(0)$
of $\ep=0$ SAW homopolymers, all in the neighborhood of $v_0\sim 0.2$,
are considerably larger than the corresponding optimally fitted theoretical
excluded-volume parameter for intrachain contact probabilities
obtained previously [see, e.g., Eq.~(4.4) of ref.~\citen{chanJCP90}].
As an example, the simulated
$\avRgs/\avREEs\approx (24.4$\AA$/62.5$\AA$)^2=0.1524$
for $\ep=0$ SAW homopolymers of length $N\rightarrow {\tilde N}=43$,
which according to Eq.~\ref{eq:RgREE-0} yields a fitted $v_0=0.228$ for $a=1$.
For intrachain probabilities, a value of $v_0\simeq 0.41$ was 
estimated,\cite{chanJCP90} which translates, as explained above, into 
a much smaller $v_0\rightarrow v_0/(2\pi)^{3/2}\approx 0.026$
in the present unit for $v_0$.
A possible cause of this difference---which deserves to be studied
further---is that the perturbative terms for the quantities in Fig.~15a 
is of order $v_0 N^{1/2}$ whereas the perturbative terms for the contact 
reduction factors in ref.~\citen{chanJCP90} is of order $v_0N^0=v_0$.
Interestingly, while the simulated $\avRgs/\avREEs$ ratio of $\approx 0.1524$
is less than the Gaussian-chain value of $1/6=0.1667$ as one would expect from
Eq.~\ref{eq:RgREE-0}, our simulated ratio for an effective SAW chain length
of ${\tilde N}=43$ is also less than the renormalization-group-predicted
universal ratio of $\avRgs/\avREEs\approx (1/6)(95/96)=0.1649$ for SAWs.
The latter ratio follows from the expansion\cite{ohta82,kosmas1981} 
$\avRgs/\avREEs=(1/6)(1-\epsilon/96)+O(\epsilon^2)$ 
[Eq.~(4.6) of ref.~\citen{ohta82}],
where $\epsilon=4-d$, and the number of spatial dimensions, $d$, is equal to 3
for our model systems and thus $\epsilon=1$ (the dimensional 
parameter $\epsilon$ here is not to be confused with an energy parameter). 
\\

{\bf Explicit-chain simulations of heteropolymers with theory-inspired 
interactions exemplify
${\bm{\langle R_{\bf g}^2\rangle}}$--${\bm{\langle R_{\bf EE}^2\rangle}}$
decoupling in heterogeneous ensembles.}
We are now in a position to apply the analytical formulation to explore
heteropolymer sequences that would likely lead to significant decoupling
of $\langle R_{\bf g}^2\rangle$ and $\langle R_{\bf EE}^2\rangle$,
beginning with a class of heteropolymeric interactions that leads to
pairs of heteropolymers predicted by our perturbation theory
to have the same $\langle R_{\bf EE}^2\rangle$ but 
different $\langle R_{\bf g}^2\rangle$s.

Consider the $O(v_{ij})$ term for $\langle R^2_{\rm EE}\rangle$ 
in Eq.~\ref{REE-vij}. By changing the contour variables $\tau_i,\tau_j$ 
to $\tau_i,\Delta\tau_{ij}$ and thus rewriting $v_{ij}=v(\tau_i,\tau_j)=
v(\tau_i,\Delta\tau_{ij})$, the $O(v_{ij})$ term for 
$\langle R^2_{\rm EE}\rangle$ may be expressed in the equivalent form 
\begin{equation}
\int_a^{N} d\Delta\tau_{ij}
\frac {1}{\sqrt{\Delta\tau_{ij}}}
\int_0^{N-\Delta\tau_{ij}} d\tau_i \; 
v(\tau_i,\Delta\tau_{ij}) \; .
\label{REE-rewrite-eq}
\end{equation}
It follows that for a given $\Delta\tau_{ij}$, all variations of
$v(\tau_i,\Delta\tau_{ij})$ over $\tau_i$ that
leave the $\int_0^{N-\Delta\tau_{ij}} d\tau_i$ integral over
$v(\tau_i,\Delta\tau_{ij})$ in Eq.~\ref{REE-rewrite-eq} unchanged would 
result in the same predicted $\langle R^2_{\rm EE}\rangle$.
However, according to Eq.~\ref{Rg-hetero-eq}, such variations can 
change $\langle R_{\rm g}^2\rangle$. Therefore, different heteropolymers
represented by different $v(\tau_i,\Delta\tau_{ij})$ functions 
that nevertheless yield the same 
$\int_0^{N-\Delta\tau_{ij}} d\tau_i \; v(\tau_i,\Delta\tau_{ij})$
for all $\Delta\tau_{ij}$
are predicted to have the same $\langle R^2_{\rm EE}\rangle$ but they
can have different $\langle R_{\rm g}^2\rangle$ values. In other words, 
by Eq.~\ref{Rg-REE-hetero-eq}, heteropolymers can share the
same $\langle R^2_{\rm EE}\rangle$ but have different 
$\langle R^2_{\rm g}\rangle/\langle R^2_{\rm EE}\rangle$.
As an illustration, consider two heteropolymers with model interaction schemes
$v_{ij}=v^+_1$ and $v^-_1$ defined by 
\begin{equation}
v^\pm_1(\tau_i,\Delta\tau_{ij})=v_0+u^\pm(\tau_i,\Delta\tau_{ij}) \; ,
\label{eq:v0}
\end{equation}
where
\begin{equation}
u^\pm(\tau_i,\Delta\tau_{ij})=
\begin{cases}
\pm v[4\tau_i/(N-\Delta\tau_{ij}) - 1] \; , & 
\mbox{for } 0\le \tau_i \le (N-\Delta\tau_{ij})/2 ; \\
\mp v[4\tau_i/(N-\Delta\tau_{ij}) - 3] \; , & 
\mbox{for } (N-\Delta\tau_{ij})/2 \le \tau_i \le N-\Delta\tau_{ij} \; .
\end{cases}
\label{v1-eq0}
\end{equation}
Here $v^+_1$ and $v^-_1$ are given, respectively, by the above expressions 
carrying the upper and lower signs, and $v$ is a constant.
Because $\int_0^{N-\Delta\tau_{ij}} d\tau_i \; u^\pm(\tau_i,\Delta\tau_{ij})=0$
and therefore
$\int_0^{N-\Delta\tau_{ij}} d\tau_i \; v^\pm_1(\tau_i,\Delta\tau_{ij})$
$=(N-\Delta\tau_{ij})v_0$, the $\langle R^2_{\rm EE}\rangle$ values
predicted by Eq.~\ref{REE-vij} for these two heteropolymers,
$\langle R^2_{\rm EE}\rangle_{v^+_1}$ and 
$\langle R^2_{\rm EE}\rangle_{v^-_1}$,
are identical, i.e., 
$\langle R^2_{\rm EE}\rangle_{v^+_1}=\langle R^2_{\rm EE}\rangle_{v^-_1}$
for any value of $v$. 
However, the $\langle R^2_{\rm g}\rangle$ values 
predicted by Eq.~\ref{Rg-hetero-eq} for these two heteropolymers,
$\langle R^2_{\rm g}\rangle_{v^+_1}$ and $\langle R^2_{\rm g}\rangle_{v^-_1}$,
are not identical, as it can readily be shown that
\begin{equation}
\langle R^2_{\rm g}\rangle_{v^+_1}-\langle R^2_{\rm g}\rangle_{v^-_1}
= v N^{3/2} \biggl [ \frac {8}{105} + O(\sqrt{a/N})
\biggr] \; .
\label{eq:vpm-comp}
\end{equation}
A discrete version of the model heteropolymer interaction schemes 
in Eq.~\ref{v1-eq0} for an explicit chain model with a background
excluded-volume interaction (corresponding to $v_0>0$) may be
implemented by assigning the additional pairwise interaction energies
between monomers $i,j$ as follows:
\begin{equation}
[(\epsilon_{\rm p})_1^\pm]_{ij} =
\begin{cases}
\pm v\{2(i-1)/[(n-\Delta_{ij})/2 -1]-1\} \; , & \\
 & \hskip -7.8cm 
\mbox{for }(n-\Delta_{ij}) \mbox{ even \& } 1\le i\le (n - \Delta_{ij})/2\; ;\\
\mp v\{2[i-(n-\Delta_{ij})/2 -1]/[(n-\Delta_{ij})/2 -1]-1\} \; , & \\ 
 & \hskip -7.8cm 
\mbox{for }(n - \Delta_{ij}) \mbox{ even \& } 
\{[(n - \Delta_{ij})/2]+1\}\le i\le (n - \Delta_{ij}) \; ;\\
\pm v\{2(i-1)/[(n-\Delta_{ij}-1)/2]-1\} \; , & \\
 & \hskip -7.8cm 
\mbox{for }(n-\Delta_{ij}) \mbox{ odd \& } 1\le i\le (n - \Delta_{ij}+1)/2 \; ;\\
\mp v\{2[i-(n-\Delta_{ij}+1)/2]/[(n-\Delta_{ij}-1)/2]-1\} \; , & \\ 
 & \hskip -7.8cm 
\mbox{for }(n - \Delta_{ij}) \mbox{ odd \& } 
[(n - \Delta_{ij}+1)/2)]\le i\le (n - \Delta_{ij}) \;  \\
\end{cases}
\label{v1-eq-discrete}
\end{equation}
for $\Delta_{ij}=|j-i|\ge 3$;
$(\epsilon_{\rm p})^\pm_{ij} =0$ for $\Delta_{ij}< 3$,
and $[(\epsilon_{\rm p})_1^\pm]_{ij} =
[(\epsilon_{\rm p})_1^\pm]_{ji}$ by definition.

%%%%%%%%%%%%%%%%%%%%%%%%%%%%%%%%%%%%%%%%%%%%%%%%%%%%%%%%%%%%%%%%%%%%%%%%%%%%%%%%
\begin{figure}[t]
\vskip -0.3cm
{\includegraphics[height=40mm,angle=0]{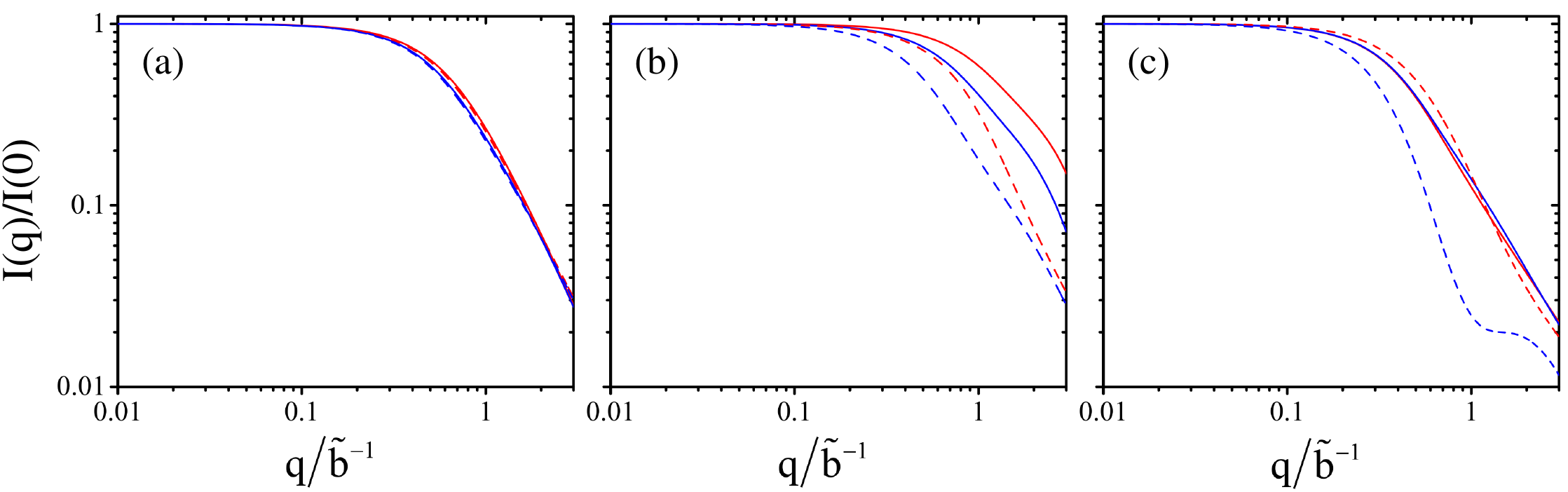}}
\begin{center}
\end{center}
\vspace{-1.3cm}
\caption{Comparing heteropolymer perturbation theory-predicted and
explicit-chain simulated scattering intensities in the
$v_1^\pm$ and $(\epsilon_{\rm p})_1^\pm$ interaction schemes. Here
$I(q)/I(0)$ is computed using $v_1^+$ and $(\epsilon_{\rm p})_1^+$
(blue curves) as well as $v_1^-$ and $(\epsilon_{\rm p})_1^-$ (red curves),
with $v>0$.
Perturbation theory-predicted results (dashed curves) are calculated
according to Eq.~\ref{eq:logI-0-general} using
effective chain length ${\tilde{N}}=43$, effective Kuhn length
${\tilde{b}}=6.46$ \AA, $a=1$, and
Eqs.~\ref{eq:v0} and \ref{v1-eq0}
with (a) $v_0=0$, $v=0.1$ for $v_1^\pm$, (b) $v_0=0$, $v=0.5$ for $v_1^\pm$,
and (c) $v_0=0.1121$---which is the background repulsion
corresponding to the $\ep=0$ SAW according to the
$I(q)/I(0)$-fitting in Fig.~\ref{fig15}a---together with
$v=2.5$ for $v_1^+$ and $v=0.5$ for $v_1^-$.
Explicit-chain simulation results (solid curves) are obtained
using Eq.~\ref{v1-eq-discrete}, with
(a) $\epsilon_{\rm ex}=0$, $v=0.1$ for $(\epsilon_{\rm p})_1^\pm$, (b)
$\epsilon_{\rm ex}=0$, $v=0.5$ for $(\epsilon_{\rm p})_1^\pm$,
and (c) $\epsilon_{\rm ex}=1.0 k_{\rm B}T$ (full excluded volume) together
with $v=2.5$ for $(\epsilon_{\rm p})_1^+$ and
$v=0.5$ for $(\epsilon_{\rm p})_1^-$.
}
\label{fig16}
\end{figure}
%%%%%%%%%%%%%%%%%%%%%%%%%%%%%%%%%%%%%%%%%%%%%%%%%%%%%%%%%%%%%%%%%%%%%%%%%%%%%%%%

Comparisons of theory-predicted and explicit-chain-simulated
scattering intensities are provided in Fig.~16 for heteropolymers
embodying examples of these $v_1^\pm$ or $(\ep)_1^\pm$ interactions.
Because the baseline (zeroth order term) of our analytical perturbative 
formula for $I(q)/I(0)$ is that of a Gaussian chain
(Eqs.~\ref{eq:IqI0_0}--\ref{eq:F2-general}), we first compare
theoretical predictions with simulation results of heteropolymers
with no hardcore excluded volume ($\epsilon_{\rm ex}=0$, Figs.~16a,b).
In these cases, we find good agreement between theory and simulation
when the heteropolymeric interactions are relatively weak ($v=0.1$, Fig.~16a),
indicating that the theoretical formulation is effective at a 
rudimentary level. However, an offset between theoretical and simulated 
results is seen for stronger heteropolymeric interactions
($v=0.5$, Fig.~16b) although the rank orderings of the $I(q)/I(0)$
entailed by the $v_1^\pm$ and $(\ep)_1^\pm$ interaction schemes
are nonetheless consistent (red curves are higher than blue curves 
for both the solid and dashed curves in Fig.~16b). This mismatch is
probably related to the nonlinear relationship between the $v_0$-like 
and $\ep$-like energy parameters in the theoretical formulation and 
the explicit-chain model, respectively, as has been observed for 
homopolymers in Fig.~15b. In this regard, a sizable theory-simulation
mismatch is also observed in Fig.~16c, where the simulated $I(q)/I(0)$s 
are seen to be practically identical for two different SAW heteropolymers 
($\epsilon_{\rm ex}=10k_{\rm B}T$) with essentially the 
same $\avRgs^{1/2}\approx 24.1$~\AA~(solid red and blue curves in Fig.~16c),
but the theory-predicted $I(q)/I(0)$s 
by assuming a linear relationship between the $v_0$-like and $\ep$-like 
energy parameters differ significantly (dashed curves in Fig.~16c).
Because the theory-simulation mismatch in Fig.~16c is still quite small for
the smaller $v=0.5$ (red curves) and becomes significant only for the 
larger $v=2.5$ (blue curves),
the mismatch here is also likely attributable to a nonlinear relationship
between the $v_0$-like and $\ep$-like energy parameters as suggested above
for the results in Fig.~16b. A resolution of this issue will 
significantly broaden the utility of the present analytical formulation
for heteropolymers and thus deserves to be further investigated in 
future efforts.

%%%%%%%%%%%%%%%%%%%%%%%%%%%%%%%%%%%%%%%%%%%%%%%%%%%%%%%%%%%%%%%%%%%%%%%%%%%%%%%%
\begin{figure}[t]
\vskip -0.5cm
{\includegraphics[height=60mm,angle=0]{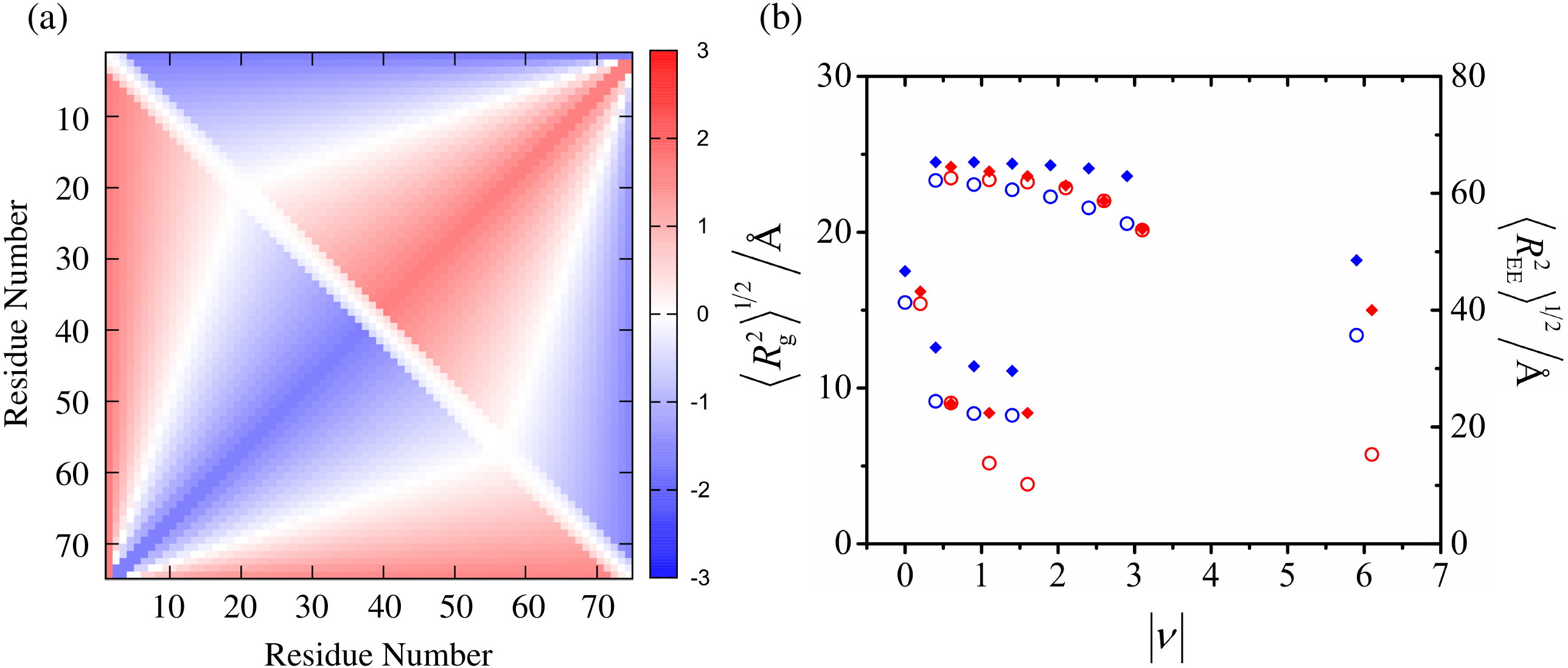}}
\begin{center}
\end{center}
\vspace{-1.3cm}
\caption{Variations of explicit-chain-simulated root-mean-square radius of
gyration and end-to-end distance in the
heteropolymeric $(\epsilon_{\rm p})_1^\pm$ interaction scheme.
(a) An example of the energy matrices with matrix elements
$[(\epsilon_{\rm p})_1^+]_{ij}$ (lower triangle) and
$[(\epsilon_{\rm p})_1^-]_{ij}$ (upper triangle) depicted using
a color scale (right) for $n=75$ and $v=3.0$.
(b) Root-mean-square radius of gyration
$\langle R^2_{\rm g}\rangle^{1/2}$ (diamonds) and root-mean-square end-to-end
distance $\langle R^2_{\rm EE}\rangle^{1/2}$ (circles)
of chains with full excluded volume ($\epsilon_{\rm ex}=1.0 k_{\rm B}T$)
for $(\epsilon_{\rm p})_1^+$, $v=|v|>0$ (blue) and
$(\epsilon_{\rm p})_1^-$, $v=|v|>0$, which is equivalent to
$(\epsilon_{\rm p})_1^+$, $v=-|v|<0$ (red) are plotted slightly offset
to the left and to the right, respectively,
of $|v|=0.5, 1.0, 1.5, 2.0, 2.5, 3.0$, and $6.0$.
For $|v|=0.1, 0.5, 1.0$, and $1.5$,
corresponding $\langle R^2_{\rm g}\rangle^{1/2}$ and
$\langle R^2_{\rm EE}\rangle^{1/2}$
for $\epsilon_{\rm ex}=0$ (i.e., with Gaussian-chain baseline)---which
are always smaller than their counterparts for
$\epsilon_{\rm ex}=1.0 k_{\rm B}T$---are also included for comparison.
}
\label{fig17}
\end{figure}
%%%%%%%%%%%%%%%%%%%%%%%%%%%%%%%%%%%%%%%%%%%%%%%%%%%%%%%%%%%%%%%%%%%%%%%%%%%%%%%%

An immediately useful application of the present analytical formulation
is to identify explicit-chain models with
theory-inspired heteropolymeric interactions that exhibit significant 
decoupling of $\langle R_{\bf g}^2\rangle$ and $\langle R_{\bf EE}^2\rangle$.
Fig.~17 provides the $\langle R_{\bf g}^2\rangle^{1/2}$ and
$\langle R_{\bf EE}^2\rangle^{1/2}$ values (Fig.~17b) of
examples of heteropolymers embodying the $(\ep)^\pm_1$ interaction 
scheme (Eq.~\ref{v1-eq-discrete}) illustrated by the heat maps in Fig.~17a.
Consistent with the theoretical prediction in Eq.~\ref{eq:vpm-comp},
Fig.~17b shows that for a given $|v|$, the $\avRgs^{1/2}$ (diamonds) 
of the $v>0$ heteropolymer (blue) is always larger than that of the $v<0$ 
heteropolymer (red). Moreover, the theoretical prediction that a pair of 
such $(\ep)^\pm_1$-heteropolymers with $\pm |v|$ have equal 
$\langle R_{\bf EE}^2\rangle^{1/2}$ values (blue and red circles) 
is also realized approximately by those explicit-chain models in Fig.~17b 
with small to moderate $|v|$ values---i.e., $|v|\leq 3.0$ for SAW 
($\epsilon_{\rm ex}=1.0k_{\rm B}T$) heteropolymers and $|v|\leq 0.5$ 
for $\epsilon_{\rm ex}=0$ heteropolymers without hardcore excluded 
volume---but not for models with higher $|v|$ values probably because 
perturbation theory is less accurate for larger $|v|$. As anticipated by 
theory (Eq.~\ref{eq:vpm-comp}), some of the $(\ep)^\pm_1$-heteropolymers 
have significantly different $\langle R_{\bf g}^2\rangle^{1/2}$ values 
despite having essentially the same $\langle R_{\bf EE}^2\rangle^{1/2}$. 
Examples of such $\avRgs$--$\avREEs$ decoupling include the 
two $|v|=3.0$ SAW $(\ep)^+_1$-heterpolymers with 
$\avREEs^{1/2}\approx 54$~\AA~for which 
$(\avRgs/\avREEs)^{1/2}=23.6$\AA$/54.8$\AA~$=0.43$ 
($\avRgs/\avREEs=0.19$) for $v=+3.0$ (blue symbols) but
$(\avRgs/\avREEs)^{1/2}=20.2/53.7=0.38$ ($\avRgs/\avREEs=0.14$) for $v=-3.0$
(red symbols), and the two $|v|=0.5$ $\epsilon_{\rm ex}=0$,
$(\ep)^\pm_1$-heteropolymers with
$\avREEs^{1/2}\approx 24$~\AA~for which 
$(\avRgs/\avREEs)^{1/2}=12.6/24.4=0.52$ ($\avRgs/\avREEs=0.26$)
for $v=+0.5$ (blue symbols) but
$(\avRgs/\avREEs)^{1/2}=9.0/24.1=0.37$ ($\avRgs/\avREEs=0.14$)
for $v=-0.5$ (red symbols).

%%%%%%%%%%%%%%%%%%%%%%%%%%%%%%%%%%%%%%%%%%%%%%%%%%%%%%%%%%%%%%%%%%%%%%%%%%%%%%%%
\begin{figure}[t]
\vskip -0.3cm
{\includegraphics[height=45mm,angle=0]{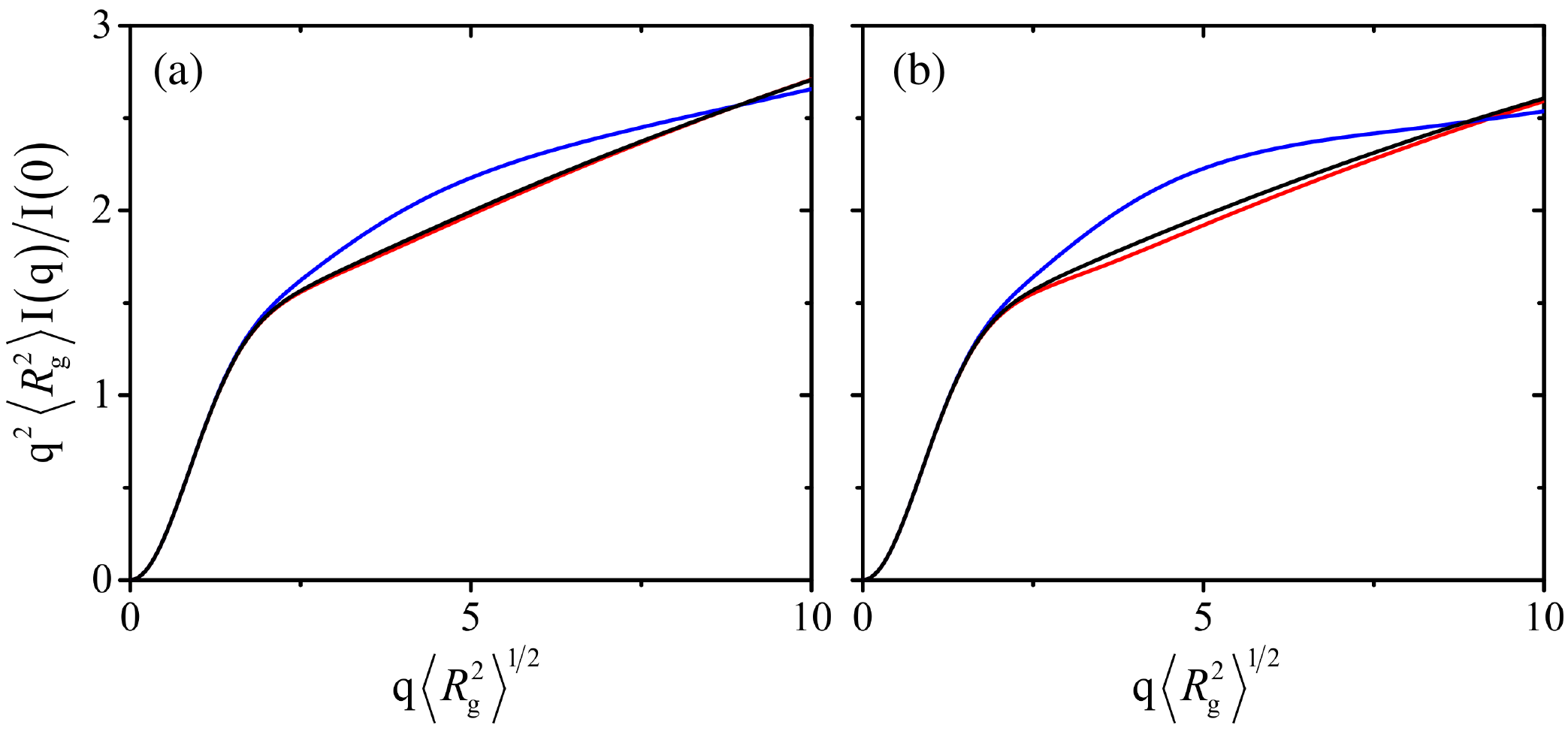}}
\begin{center}
\end{center}
\vspace{-1.3cm}
\caption{Comparing MFFs of homogeneous and energy-matrix-specified
heterogeneous explicit-chain ensembles with full excluded volume
($\epsilon_{\rm ex}=1.0 k_{\rm B}T$) and essentially the same
$\langle R^2_{\rm g}\rangle$.
(a) MFFs (dimensionless Kratky plots) of heteropolymer ensembles governed
by the $(\epsilon_{\rm p})_1^+$, $v=2.5$ interactions (blue) and
the $(\epsilon_{\rm p})_1^-$, $v=0.5$ interactions (red), with
$\langle R^2_{\rm g}\rangle^{1/2}=24.1$ and $24.2$ \AA, respectively,
are compared with that of the homogeneous  $\ep=-0.1$ ensemble with
$\langle R^2_{\rm g}\rangle^{1/2}=24.2$ \AA~(black).
(b) MFFs of heteropolymer ensembles governed
by the $(\epsilon_{\rm p})_1^+$, $v=3.0$ interactions (blue)
and the $(\epsilon_{\rm p})_1^-$, $v=1.5$ interactions (red), both with
$\langle R^2_{\rm g}\rangle^{1/2}=23.6$ \AA, are compared with
that of the homogeneous  $\ep=-0.4$ ensemble with
$\langle R^2_{\rm g}\rangle^{1/2}=23.7$ \AA~(black).
}
\label{fig18}
\end{figure}
%%%%%%%%%%%%%%%%%%%%%%%%%%%%%%%%%%%%%%%%%%%%%%%%%%%%%%%%%%%%%%%%%%%%%%%%%%%%%%%%

Scenarios exist within the $(\ep)^\pm_1$ heteropolymeric interaction scheme 
in Eq.~\ref{v1-eq-discrete} that different heteropolymers sharing
essentially the same $\avRgs^{1/2}$ can exhibit significantly different
SAXS signatures. Two examples are shown in Fig.~18. In both examples,
the MFF of the $(\ep)^\pm_1$-heteropolymer with the smaller $|v|$ (red
curve) is practically indistinguishable from that of a homopolymer (black
curve), which, however, is significantly different from the MFF of the 
$(\ep)^\pm_1$-heteropolymer with the larger $|v|$ (blue curve).  
These results reinforce the above observation (Figs.~6, 7, 9, 10, and 13) 
that SAXS MFFs can sometimes but cannot always distinguish between 
heterogeneous and homogeneous conformational ensembles.

We now turn to another class of heteropolymeric interactions which
is predicted by our perturbation theory to yield pairs of heteropolymers 
that have the same $\langle R_{\bf g}^2\rangle$ but different 
$\langle R_{\bf EE}^2\rangle$s
(in contrast to the above $(\ep)^\pm_1$ scheme for
heteropolymers with the same $\langle R_{\bf EE}^2\rangle$ but
different $\langle R_{\bf g}^2\rangle$s). For this purpose,
we make use of the change in integration variable in 
Eq.~\ref{REE-rewrite-eq} to rewrite
the $O(v_{ij})$ term for $\langle R^2_{\rm g}\rangle$ in 
Eq.~\ref{Rg-hetero-eq} as
\begin{equation}
-\frac {1}{N^2} 
\int_a^{N} d\Delta\tau_{ij}
\frac {1}{\sqrt{\Delta\tau_{ij}}}
\int_0^{N-\Delta\tau_{ij}} d\tau_i \; 
v(\tau_i,\Delta\tau_{ij}) 
{\cal Z}(N,\tau_i,\Delta\tau_{ij}) 
\; ,
\label{Rg-rewrite-eq}
\end{equation}
where
\begin{equation}
{\cal Z}(N,\tau_i,\Delta\tau_{ij}) \equiv 
\tau_i^2 - (N-\Delta\tau_{ij})\tau_i 
+ \frac {\Delta\tau_{ij}}{12}\Bigl(3\Delta\tau_{ij} -4N\Bigr)  
 = 
(\tau_i - \lambda) (\tau_i -\lambda^\prime) \; ,
\label{Q-eq}
\end{equation}
with $\lambda =[N-\Delta\tau_{ij}+\Lambda(\Delta\tau_{ij})]/2>0$ and
$\lambda^\prime =[N-\Delta\tau_{ij}-\Lambda(\Delta\tau_{ij})]/2\le 0$, wherein
$\Lambda(\Delta\tau_{ij})\equiv \sqrt{N(3N-2\Delta\tau_{ij})/3}$.
Within the integration range in Eq.~\ref{Rg-rewrite-eq},
the function ${\cal Z}(N,\tau_i,\Delta\tau_{ij})<0$. For a given
$\Delta\tau_{ij}$ and for $\tau_i$ values within the range 
$[0,N-\Delta\tau_{ij}]$, the function takes its 
maximum (least negative) value of 
${\cal Z}_{\rm max}\equiv
\lambda\lambda^\prime=\Delta\tau_{ij}(3\Delta\tau_{ij}-4N)/12$ 
at $\tau_i = 0$ and $N-\Delta\tau_{ij}$, and its minimum (most negative) value 
of $-\Lambda(\Delta\tau_{ij})^2/4=N(2\Delta\tau_{ij}-3N)/12$ at 
$\tau_i=(N-\Delta\tau_{ij})/2$. These considerations indicate that
the following two model heteropolymer interaction schemes, 
$v(\tau_i,\Delta\tau_{ij})=v^+_2$ and $v^-_2$, defined by
\begin{equation}
v^\pm_2(\tau_i,\Delta\tau_{ij}) = v_0 +
u^\pm(\tau_i,\Delta\tau_{ij})
\frac {{\cal Z}_{\rm max}}{{\cal Z}(N,\tau_i,\Delta\tau_{ij})} \; ,
\end{equation}
where $u^\pm(\tau_i,\Delta\tau_{ij})$ is from Eq.~\ref{v1-eq0}, will give
the same $\langle R^2_{\rm g}\rangle$ in Eq.~(\ref{Rg-hetero-eq}). 
This property of $v_2^\pm$ is readily verified by using either of them for
$v(\tau_,\Delta\tau_{ij})$ in Eq.~\ref{Rg-rewrite-eq}
to yield the same $O(v_{ij})$ term for $\langle R^2_{\rm g}\rangle$. 
As is the case for $v_1^\pm$, $v_2^\pm=v_0\mp v$
at $\tau_i=0$ and $\tau_i=N-\Delta\tau_{ij}$.
The model interactions $v_2^\pm$ are of interest because  
according to Eq.~\ref{REE-vij} they should
lead to two different $\langle R^2_{\rm EE}\rangle$ values, denoted here as
$\langle R^2_{\rm EE}\rangle_{v^+_2}$ and 
$\langle R^2_{\rm EE}\rangle_{v^-_2}$.
Their $O(v_{ij})$ difference is given by
\begin{eqnarray}
\langle R^2_{\rm EE}\rangle_{v^+_2} - \langle R^2_{\rm EE}\rangle_{v^-_2}
&=& 2 \int_a^N dx \; \sqrt{x} (3x-4N) \int_0^{N-x} dy\;
\frac {u^+(y,x)}{{\cal Z}(N,y,x)} 
\nonumber \\
&=& 4v \int_a^N dx \; \sqrt{x} (4N-3x) 
\biggl\{ 
\frac {1}{\Lambda(x)}
\ln\biggl[ \frac {\Lambda(x)+N-x}{\Lambda(x)-N+x} \biggr]
\nonumber \\
&& 
\quad\quad\quad\quad\quad\quad\quad\quad\quad\quad
-\frac {2}{N-x} \ln \biggl [\frac {N(3N-2x)}{x(4N-3x)}\biggr ]
\biggr\}
\nonumber \\
&\approx& -1.67(vN^{3/2}) \quad (a\rightarrow 0) 
\; , 
\label{eq:epsilon2}
\end{eqnarray}
where we have used the variables $x$ and $y$
for $\Delta\tau_{ij}$ and $\tau_i$, respectively.
In the same manner in which we arrived at Eq.~\ref{v1-eq-discrete},
a discrete version of $v_2^\pm$ may be given by pairwise interaction
energies
\begin{equation}
[(\epsilon_{\rm p})_2^\pm]_{ij} \equiv
\frac {\Delta_{ij}[3\Delta_{ij}-4(n-1)]/12}{{\cal Z}(n-1,i-1,\Delta_{ij})}
[(\epsilon_{\rm p})_1^\pm]_{ij}
\; ,
\label{v2-eq-discrete}
\end{equation} 
where the function ${\cal Z}$ is now evaluated for $N=n-1$ at discrete 
values of $i=1,2,\dots,n-\Delta_{ij}$ and
$[(\epsilon_{\rm p})_1^\pm]_{ij}$ is defined by Eq.~\ref{v1-eq-discrete}
above. An example of the $(\epsilon_{\rm p})_2^\pm$
interaction scheme in Eq.~\ref{v2-eq-discrete}
is provided by the heat maps in Fig.~19a.

%%%%%%%%%%%%%%%%%%%%%%%%%%%%%%%%%%%%%%%%%%%%%%%%%%%%%%%%%%%%%%%%%%%%%%%%%%%%%%%%
\begin{figure}[t]
\vskip -0.5cm
{\includegraphics[height=60mm,angle=0]{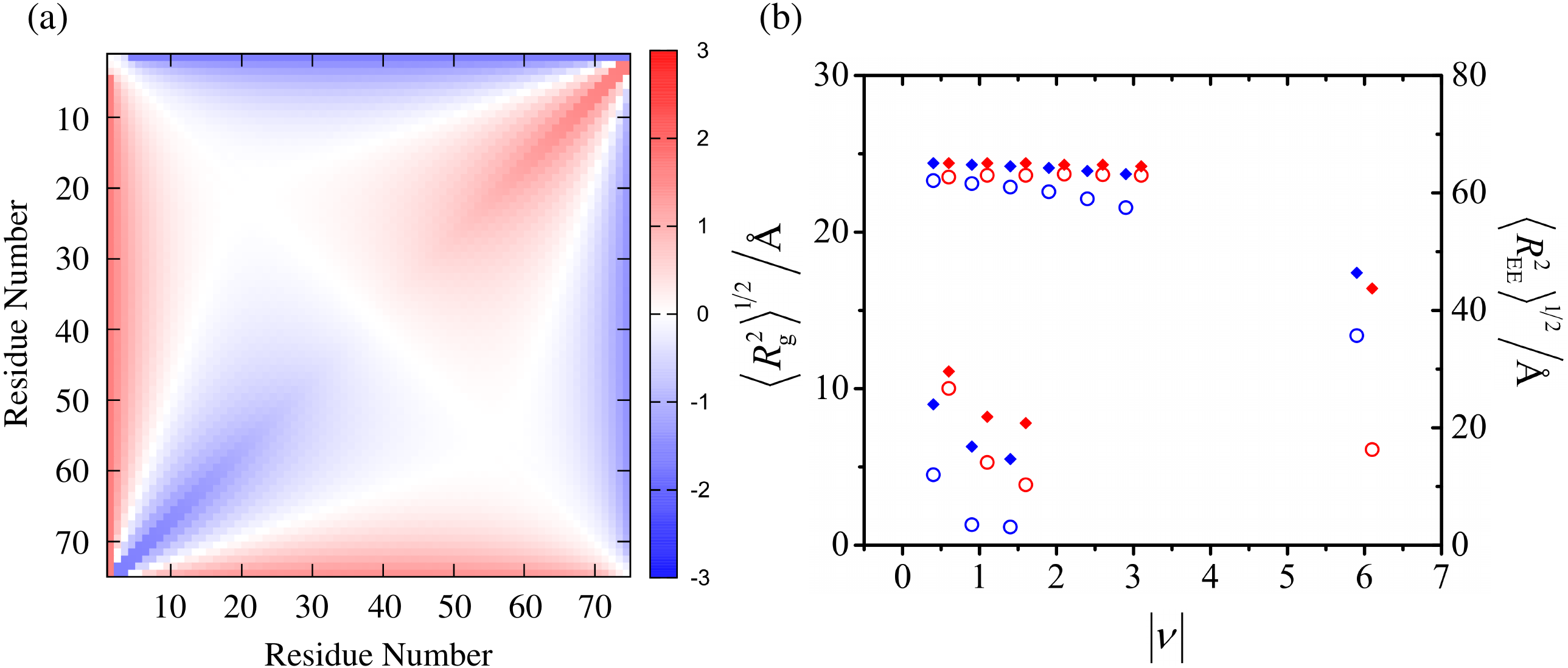}}
\begin{center}
\end{center}
\vspace{-1.3cm}
\caption{Variations of explicit-chain-simulated
$\langle R^2_{\rm g}\rangle^{1/2}$ and $\langle R^2_{\rm EE}\rangle^{1/2}$
in the heteropolymeric $(\epsilon_{\rm p})_2^\pm$ interaction scheme
in Eq.~\ref{v2-eq-discrete}.
(a) Similar to Fig.~\ref{fig17}a for $(\epsilon_{\rm p})_1^\pm$,
here $[(\epsilon_{\rm p})_2^+]_{ij}$ (lower triangle) and
$[(\epsilon_{\rm p})_2^-]_{ij}$ (upper triangle) are shown
for $n=75$ and $v=3.0$.
(b) Using the same notation as that in Fig.~\ref{fig17}b,
$\langle R^2_{\rm g}\rangle^{1/2}$ (diamonds) and
$\langle R^2_{\rm EE}\rangle^{1/2}$ (circles)
for $(\epsilon_{\rm p})_2^+$, $v=|v|>0$ (blue) and
$(\epsilon_{\rm p})_2^-$, $v=|v|>0$ (red) are shown for
full excluded volume ($\epsilon_{\rm ex}=1.0 k_{\rm B}T$)
at $|v|=0.5, 1.0, 1.5, 2.0, 2.5, 3.0$, and $6.0$, and also
for the $\epsilon_{\rm ex}=0$ (Gaussian-chain baseline) case
at $|v|=0.5, 1.0$, and $1.5$.
}
\label{fig19}
\end{figure}
%%%%%%%%%%%%%%%%%%%%%%%%%%%%%%%%%%%%%%%%%%%%%%%%%%%%%%%%%%%%%%%%%%%%%%%%%%%%%%%%

The SAW examples ($\epsilon_{\rm ex}=10k_{\rm B}T$) in Fig.~19b of 
explicit-chain simulated $\avRgs^{1/2}$ and $\avREEs^{1/2}$ values 
of heteropolymers embodying the $(\epsilon_{\rm p})_2^\pm$ interaction 
scheme are largely in line with the theoretical prediction that a pair
of $(\epsilon_{\rm p})_2^\pm$-heteropolymers with the same $|v|$
should have the same $\avRgs^{1/2}$ (diamonds)---which is essentially
the case for $|v|\leq 3.0$ and holds approximately for $|v|=6.0$---but
increasingly different $\avREEs^{1/2}$s (circles) with increasing $|v|$.
(We note, however, that while the sign of the difference in $\avREEs^{1/2}$ 
for the $v>0$ and $v<0$ $(\epsilon_{\rm p})_2^+$-heterpolymers with the
same $|v|$ is consistent with 
Eq.~\ref{eq:epsilon2} for $|v|\leq 3.0$, the sign is opposite for $|v|=6.0$).
Therefore, although the theory does not appear to work well for 
the $\epsilon_{\rm ex}=0$ chains in Fig.~19b (data points at bottom left),
the simulated data on SAW $(\epsilon_{\rm p})_2^+$-heteropolymers 
provide ample examples of $\avRgs$--$\avREEs$ 
decoupling, including the case of $|v|=3.0$ in which
$(\avRgs/\avREEs)^{1/2}=23.7$\AA$/57.5$\AA~$=0.41$
($\avRgs/\avREEs=0.17$) for $v=+3.0$ (blue symbols) but
$(\avRgs/\avREEs)^{1/2}=24.2/63.0=0.38$
($\avRgs/\avREEs=0.15$) for $v=-3.0$ (red symbols) though the
$\avREEs^{1/2}\sim 24$~\AA~of these two heteropolymers are quite similar, 
and the case of $|v|=6.0$ in which
$(\avRgs/\avREEs)^{1/2}=17.4/31.5=0.55$
($\avRgs/\avREEs=0.31$) for $v=+6.0$ (blue symbols) but
$(\avRgs/\avREEs)^{1/2}=16.4/16.3=1.0$ for $v=-6.0$ (red symbols) though
the $\avREEs^{1/2}$ values of $\sim 17$~\AA~for this pair of 
$|v|=6.0$ $(\epsilon_{\rm p})_2^\pm$-heteropolymers are also quite similar.

%%%%%%%%%%%%%%%%%%%%%%%%%%%%%%%%%%%%%%%%%%%%%%%%%%%%%%%%%%%%%%%%%%%%%%%%%%%%%%%%
\begin{figure}[t]
\vskip -0.5cm
{\includegraphics[height=60mm,angle=0]{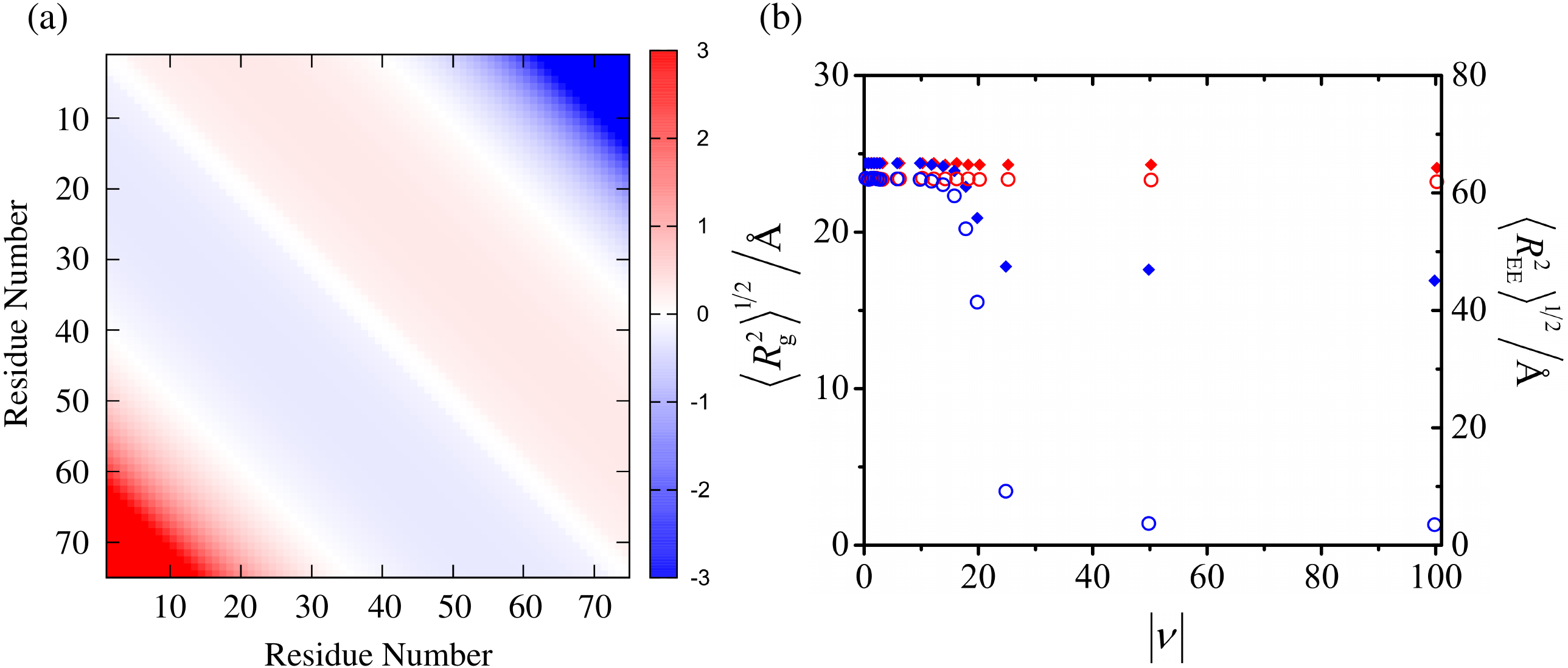}}
\begin{center}
\end{center}
\vspace{-1.3cm}
\caption{Variations of explicit-chain-simulated
$\langle R^2_{\rm g}\rangle^{1/2}$ and $\langle R^2_{\rm EE}\rangle^{1/2}$
in the heteropolymeric $(\epsilon_{\rm p})_3^\pm$ interaction scheme
(Eq.~\ref{v3-eq-discrete}).
(a) Similar to Figs.~\ref{fig17}a and \ref{fig18}a,
here $[(\epsilon_{\rm p})_3^+]_{ij}$ (upper triangle) and
$[(\epsilon_{\rm p})_2^-]_{ij}$ (lower triangle) are shown
for $n=75$ and $v=100.0$. In this case, although
$\min\{[(\epsilon_{\rm p})_3^+]_{ij}\}=-100$ and
$\max\{[(\epsilon_{\rm p})_3^-]_{ij}\}=+100$,
a graded color scale for $[-3,+3]$ (with all
$[(\epsilon_{\rm p})_3^+]_{ij}\leq -3$
and all $[(\epsilon_{\rm p})_3^-]_{ij}\geq +3$
depicted, respectively, by the same deepest blue and red)
is used to visualize variation in pairwise energy because
$|[(\epsilon_{\rm p})_3^\pm]_{ij}|$ is relatively small for most $i,j$.
(b) Using the same notation as that in Figs.~\ref{fig17}b and \ref{fig18}b,
$\langle R^2_{\rm g}\rangle^{1/2}$ (diamonds) and
$\langle R^2_{\rm EE}\rangle^{1/2}$ (circles)
for $(\epsilon_{\rm p})_3^+$, $v=|v|>0$ (red) and
$(\epsilon_{\rm p})_3^-$, $v=|v|>0$, which is equivalent to
$(\epsilon_{\rm p})_3^+$, $v=-|v|<0$ (blue), are shown for
full excluded volume ($\epsilon_{\rm ex}=1.0 k_{\rm B}T$)
at $|v|=0.5, 1.0, 1.5, 2.0, 2.5, 3.0$, $6.0$, $10.0$, $12.0$, $14.0$,
$16.0$, $18.0$, $20.0$, $25.0$, $50.0$, and $100.0$.
Unlike Figs.~\ref{fig17}b and \ref{fig18}b,
no data for the Gaussian-chain, $\epsilon_{\rm ex}=0$ baseline case
is shown.
}
\label{fig20}
\end{figure}
%%%%%%%%%%%%%%%%%%%%%%%%%%%%%%%%%%%%%%%%%%%%%%%%%%%%%%%%%%%%%%%%%%%%%%%%%%%%%%%%

Another class of heterogeneous interactions 
predicted by our perturbation theory to keep $\avRgs$ unchanged
while changing $\avREEs$ is given by 
$v_{ij}=v(\tau_i,\tau_j)=v(\Delta\tau_{ij})$, which is 
a function of $\Delta\tau_{ij}$ alone to be defined below. 
In other words, in this interaction scheme,
all intrachain contacts of the same order have the same energy but 
intrachain contacts with different contact orders have different energies. 
Consider
\begin{equation}
v_3^{\pm}(\Delta\tau_{ij})\equiv v_0\mp {\tilde v}_3
\frac{\sqrt{\Delta\tau_{ij}}}{(N-\Delta\tau_{ij})(2N^2-\Delta\tau_{ij}^2)}
\Bigl(2\Delta\tau_{ij}-N-a\Bigr) \;
\label{v3-eq0}
\end{equation}
where ${\tilde v}_3$ is an adjustable energy scale.
Under $v_3^{\pm}$ and for a given $|{\tilde v}_3|$,
$\langle R^2_{\rm g}\rangle$ is the same irrespective of whether
the minus or plus sign is taken for the $\mp$ sign in Eq.~\ref{v3-eq0}
because the $O(v_{ij})$ term for $\langle R^2_{\rm g}\rangle$ 
(Eq.~\ref{Rg-rewrite-eq}) with $v_3^{\pm}$ is equal to
\begin{equation}
-\frac {1}{N^2} 
\int_a^{N} d\Delta\tau_{ij}
\frac {v_3^{\pm}(\Delta\tau_{ij})}{\sqrt{\Delta\tau_{ij}}}
\int_0^{N-\Delta\tau_{ij}} d\tau_i \; 
{\cal Z}(N,\tau_i,\Delta\tau_{ij}) = 0
\; .
\label{v3-eq1}
\end{equation}
In contrast, the sign choice would affect $\langle R_{\rm EE}^2\rangle$, 
as can be readily verifed
that the $O(v_{ij})$ difference in $\langle R_{\rm EE}^2\rangle$ calculated
using $v_3^+$ verus that calculated using $v_3^-$ is given by
\begin{eqnarray}
\langle R^2_{\rm EE}\rangle_{v^+_3} - \langle R^2_{\rm EE}\rangle_{v^-_3}
&=& 
\frac {{\tilde v}_3}{\sqrt{2}}
\biggl \{ [(2\sqrt{2}+1) + a/N]\ln \biggl (
\frac {\sqrt{2}+1}{\sqrt{2}+a/N} \biggr )
\nonumber \\
&& 
\quad\quad \quad\quad \quad\quad
+ [(2\sqrt{2}-1) - a/N]\ln \biggl (
\frac {\sqrt{2}-1}{\sqrt{2}-a/N} \biggr )
\biggr \} \; ,
\label{v3-eq2}
\end{eqnarray}
which equals $-0.140{\tilde v}_3$ in the $a/N\rightarrow 0$ limit.
We obtain a discretized version of this interaction scheme by
setting $N\rightarrow n = N+1$ in Eq.\ref{v3-eq0} to arrive at
\begin{equation}
[(\epsilon_{\rm p})_3^\pm]_{ij} =
\mp v
\frac {\sqrt{\Delta_{ij}}(2\Delta_{ij}-n-a)}{(n-\Delta_{ij})(2n^2-\Delta_{ij}^2)}
\left[
\frac{n^2+2n-1}{\sqrt{n-1}\; (n-2-a)}
\right]
\;  
\label{v3-eq-discrete}
\end{equation} 
and by choosing $a=3$. Accordingly, the above interaction scheme 
$(\epsilon_{\rm p})_3^\pm$ in Eq.~\ref{v3-eq-discrete} is defined for 
$\Delta_{ij}=a=3,4,\dots,n-1$, where the expression inside the square 
bracket is a normalization factor such that $|v|$ is the magnitude of 
the interaction for the largest contact order $\Delta_{ij}=n-1$
(Fig.~20a).

The SAW examples ($\epsilon_{\rm ex}=10k_{\rm B}T$) in Fig.~20b of
explicit-chain simulated $\avRgs^{1/2}$ and $\avREEs^{1/2}$ values
of heteropolymers embodying the $(\epsilon_{\rm p})_3^\pm$ interaction
scheme show little variation for $|v|\lesssim 12$.
For $|v|\gtrsim 14$, $\avREEs^{1/2}$ of
$(\epsilon_{\rm p})_3^+$-heteropolymers with $v>0$ (blue circles)
 decreases steeply with increasing $|v|$. Although the corresponding 
$\avRgs^{1/2}$ (blue diamonds) also decreases concomitantly in a trend
that differs from the perturbation theory prediction of unchanged
$\avRgs^{1/2}$, the rate of decrease
of $\avRgs^{1/2}$ with increasing $|v|$ is more gradual than that
of $\avREEs^{1/2}$ as one might intuitively anticipate from an approximate 
theory. In this regard,
Fig.~20b provides a few examples of dramatic $\avRgs$--$\avREEs$ 
decoupling, including a pair of
$|v|=20.0$ $(\epsilon_{\rm p})_3^+$-heteropolymers for which
$(\avRgs/\avREEs)^{1/2}=20.9$\AA$/41.4$\AA~$=0.50$
($\avRgs/\avREEs=0.25$) for $v=+20.0$ (blue symbols) but
$(\avRgs/\avREEs)^{1/2}=24.3/62.3=0.39$
($\avRgs/\avREEs=0.15$) for $v=-20.0$ (red symbols), and 
a pair of $|v|=50.0$ $(\epsilon_{\rm p})_3^+$-heteropolymers for which
$(\avRgs/\avREEs)^{1/2}=17.6/3.7=4.8$
($\avRgs/\avREEs=22.6$) for $v=+50.0$ (blue symbols), which is much higher
than that of any homopolymer ensemble shown in the inset of Fig.~3a, but
$(\avRgs/\avREEs)^{1/2}=24.3/62.2=0.39$
($\avRgs/\avREEs=0.15$) for $v=-50.0$ (red symbols).
\\

\noindent
{\large\bf DISCUSSION}\\ 

{\bf The logic of inferring conformational ensembles from MFFs.}
Considered in aggregate, the results presented above show that having 
a homopolymer-like MFF is insufficient to guarantee an underlying 
homogeneous conformational distribution, although having a 
non-homopolymer-like MFF is a good indication of a heterogeneous ensemble.
Proteins are heteropolymers. Therefore, physically, it is most
intuitive to expect disordered conformational ensembles of proteins to be 
heterogeneous even if, somehow, their MFFs turn out to be homopolymer-like. 
Indeed, recent explicit-chain simulations using a coarse-grained potential 
with a simple sidechain representation for real IDPs have shown that such 
behavior is possible---as seen also in the mathematically constructed
examples we showcase above---in that conformational ensembles with 
asphericity and another shape parameter significantly different from those
of homopolymer can nonetheless exhibit scattering intensities similar to 
those of homopolymers\cite{DT2019} (though the simulation-experiment 
agreement is apparently closer for $I(q)/I(0)$--vs--$q$ plots than for 
$q^2I(q)/I(0)$--vs--$q$ (Kratky) plots reported, 
respectively, in Fig.~2 and Fig.~S1 of this reference).
In view of this basic consideration, it would not be prudent to invoke 
Occam's razor and simply infer a homogeneous ensemble as the
most parsimonious interpretation of the SAXS data when confronted with a 
homopolymer-like MFF for a disordered protein ensemble when complementary
experimental techniques are available to gain further insight into
the structural properties of the 
ensemble.\cite{lemke2017,raleighPNAS2019,raleighBiochem2020}
In this regard, a recent statistical survey of ensembles
of model heteropolymers (which are, by construction, heterogeneous
ensembles) with intrachain hydrohobic-polar (HP)-like interactions exhibiting
conformational averages of $R_{\rm g}$ and $R_{\rm EE}$ coupled
approximately in a manner similar to the $\avRgs$--$\avREEs$ 
coupling in homopolymers as well as having MFFs not so dissimilar to those for 
homopolymers\cite{tobinPNAS2019} provides additional illustrations for
our thesis that MFFs alone are insufficient to clearly
distinguish between heterogeneous and homogeneous ensembles.
At the same time, it should be emphasized that HP-like interactions cover
only a subset of intraprotein interactions. In fact, HP-like interactions
alone cannot produce the high degree of folding cooperativity (see below) 
observed experimentally for small, single-domain 
proteins.\cite{chanetal2004,Chan2000}
As discussed above, using our analytical theory-inspired heteropolymeric 
interactions---although that still neglect a lot of other modes of 
intraprotein interactions, we are able to provide ample examples of 
$\avRgs$--$\avREEs$ decoupling in heterogeneous ensembles.
From a biological/evolutionary perspective, one should expect that any
feature of conformational heterogeneity of disordered protein 
ensembles---including $\avRgs$--$\avREEs$ decoupling---can be potentially 
exploited for biological function. As researchers, it would be unwise to
self-impose {\it a priori} boundaries to box in 
our imagination of conformational possibilities.
\\

{\bf Conformational heterogeneity and physical pictures of folding 
cooperativity.}
SAXS was instrumental in revealing an important aspect of two-state-like 
folding cooperativity of small, single-domain proteins by observing 
directly, in a time-resolved manner, that the unfolded state of Protein L 
is consisted of relatively open conformations (large $\avRgs$) even under 
strongly folding conditions.\cite{Plaxco1999}
This finding was remarkable from a theoretical perspective because
common protein chain models at the time predicted a substantial
decrease of $\avRgs$ of the unfolded conformational ensemble 
when its solvent environment is changed from 
strongly unfolding (as in high denaturant) to strongly 
folding (as in low or no denaturant). 
Two-state folding/unfolding cooperativity
has since been found to be intimately related\cite{chanetal2004}
to contact-order-dependent folding rates\cite{Plaxco1998} and linear chevron 
plots, as protein chain models with reduced cooperativity---hallmarked 
by an appreciable decrease in unfolded-state $\avRgs$ with increasingly
strong folding conditions---failed to reproduce these experimental
features.\cite{chanetal2004,ChanNat1998,kaya2003d}
This early success in applying folding cooperativity---with unfolded states 
with large conformational dimensions minimally affected by folding 
conditions---as a basic, unifying rationalization of the defining
experimentally observed 
features in the folding of small, single-domain proteins has led some
researchers to an extreme view of two-state folding cooperativity known 
as the topomer search model of protein folding.\cite{topomer2003}
In this view, the unfolded state of a globular protein is
envisioned as behaving like---and thus modeled by---homopolymeric Gaussian 
chains without excluded volume, and folding is stipulated to be achieved by 
random diffusive searching of a ``native topomer'' among these 
conformations.\cite{topomer2003}
Apparently, the rationale of this perspective is that experimental
observations, such as SAXS data, that the authors interpreted as
implying a homogeneous ensemble of open unfolded conformations---with
excluded volume or not---should take precedence over any theoretical
concern as to how the existence of such an ensemble may follow from
current understanding of the driving forces for protein folding.
In other words, if commonly accepted physical interactions cannot
account for what the authors viewed as experimental facts about the
homopolymer-like unfolded-state ensemble, the fault is with common
theoretical understanding; and, therefore, rather than casting doubt on
the topomer search model as physically unrealizable, it is the commonly 
accepted theoretical understanding of protein folding driving forces 
that needs to be revised and improved.

A contrasting philosophy, which might be common among 
theoreticians, is to place more trust on current theoretical notions about 
protein folding driving forces and accept them as approximately correct. 
Researchers subscribing to this line of thinking tend to emphasize 
interpreting experimental data in terms of what is perceived to be 
physically plausible; thus the accuracies of experiments that appear
inconsistent with pre-conceived notions of physical interactions are often
questioned. As it stands, explicit-chain heteropolymeric protein models 
embodying current notions of physical
interactions---even when common structurally-specific native-centric models are
included---possess less overall folding cooperativity than that of 
many real, small, single-domain proteins.\cite{KayaPRL2003} In particular, 
these heteropolymer models entail unfolded conformations with decreased average 
$R_{\rm g}$ with stronger folding conditions.\cite{DT2008}
While this explicit-chain heteropolymer model-predicted picture was 
apparently at odds with inferences from SAXS experiment\cite{Plaxco1999} 
and in-depth understanding of calorimetric two-state folding 
cooperativity,\cite{chanetal2004,chanetal2011,kayaProteins2000,mk2006}
it was apparently supported by smFRET experiments on Protein L for which 
decreasing $\avRgs$ with decreasing denaturant was inferred using
an interpretation of smFRET data (which measured $\langle R_{\rm EE}\rangle$
but not $\avRgs$ directly) that assumed homopolymeric
$\avRgs$--$\avREEs$ coupling.\cite{haran2006} 

In our estimation, an awareness of this historical background is contextually
useful for appreciating the different investigative logic and contrasting 
conceptual emphases in the 
controversy surrounding the apparent mismatch of smFRET- versus SAXS-determined 
conformational dimensions of protein disordered 
states.\cite{tobinkevinJMB2012,tobinkevinPNAS2015,haran2012,dt2009,Songetal2015}
\\

{\bf A ``near-Levinthal'' scenario of cooperative protein folding.}
A synthesis of the useful insights from the two above-described approaches
needs to consider the following. First, as a proposed physical mechanism,
the topomer search model---which assumes that an unfolded protein state is
a noninteracting homogeneous ensemble---is untenable because this hypothetical
mechanism entails a kinetically impossible Levinthal search when excluded volume
of the chains are (as it should be) taken into account.\cite{WallinChan2005}
Nonetheless, the discourse inspired by the model's emphasis on the 
high degrees of folding cooperativity is valuable because a comprehensive 
theoretical understanding of physical interactions in protein folding should 
be able to account for this experimental property. 
Second, explicit-chain heteropolymer models of protein disordered states
are valuable in underscoring the sequence-dependent heterogeneous nature 
of these conformational ensemble. However, it is important to recognize
the limitations of current understanding of the solvent-mediated protein
interactions.\cite{DavidShaw2,cosb15} A case in point is that
even native-centric models with only pairwise interactions cannot
reproduce the high degree of folding cooperativity 
of Protein L,\cite{TaoPCCP} and that yet-undiscovered 
non-parwise-additive many-body effects, such as local-nonlocal coupling, 
are possibly implicated in protein folding 
cooperativity.\cite{chanetal2011,kaya2003d}

A conceptual picture of cooperative folding that takes all these considerations 
into account is referred to as a ``near-Levinthal'' scenario.\cite{kaya2003c}
This picture of folding expects, because of basic physics,
that the unfolded state is a heterogeneous ensemble and that conformational 
search during folding must experience a certain bias toward the native 
structure because in the absence of any bias, folding would 
be kinetically impossible (Levinthal's paradox). At the same time, it is
stipulated that the bias, though not nonexistent, is relatively weak 
because if the bias is strong---implying that intrachain interactions
are strongly favorable---the unfolded state under folding conditions
would be a relatively compact conformational ensemble
rather than the open conformational
ensemble observed experimentally. Finally, in this ``near-Levinthal'' 
picture of folding, the folded state is envisioned to be thermodynamically 
stabilized by certain strongly favorable interactions, perhaps via many-body
mechanisms such as the proposed local-nonlocal coupling effects, that 
are operative only after the protein has crossed to the folded side of
the transition-state free energy barrier but not in the unfolded 
state.\cite{chanetal2011}
Recent advances in smFRET data interpretation and SAXS experiments lend
credence to this picture: First, it is now recognized that a 
larger $\avRgs^{1/2}$ can be consistent with smFRET data on protein 
unfolded state because of possible $\avRgs$--$\avREEs$ decoupling in 
heterogeneous ensembles.\cite{Songetal2017} Second, more accurate SAXS 
measurements indicate a small contraction of unfolded-state ensembles 
when denaturant concentration is reduced,\cite{tobinSci2017}
indicating that denaturant-dependent favorable intrachain interactions
are present in the unfolded state. Given that these interactions should
be sequence-dependent, this SAXS observation is likely
indicative of a heterogeneous unfolded-state ensemble.  
\\

{\bf Water as solvent for proteins.}
The SAXS-observed phenomenon that the unfolded states of many globular 
proteins remain quite open even under strongly folding conditions (as in water),
which is a hallmark of protein folding cooperativity,
has recently been cast in terms of ``water is a good solvent for 
the unfolded states of many proteins''.\cite{TobinJMBrev2020}
While this narrative may serve to highlight the different solvation
properties of homopolymeric versus evolved heteropolymeric disordered 
protein states and the rhetoric may focus attention on the important 
ramifications of folding/unfolding cooperativity as 
emphasized in the above discussion (and in references thereof) as well as
by the authors of this review,\cite{TobinJMBrev2020}
application of ``solubility'' to a state, i.e., a subpopulation, of a 
molecular species rather than the total population of the molecular species
as a whole may not enhance conceptual clarity.
In the conventional meaning of the term, solubility of a solute in a solvent
refers to the global miscibility of the solvent and the solute, not
a subpopulation of the solute. 
While the unfolded chains of a soluble protein is, by definition, solvated
to various degree, only a very small fraction of the chain population is 
unfolded under the strongly folding conditions of pure water.  Now, if one 
applies the same nomenclatural logic in ref.~\citen{TobinJMBrev2020} 
that water is a good solvent of unfolded states of some proteins 
to a more mundane system of nonpolar solutes and water,
one can arrive at a tautological, and thus uninformative, proposition that
water is a good solvent for the solvated state of the nonpolar solute
(i.e., the minute solvated fraction of the nonpolar substance) even when 
the overwhelming population of the substance is not solvated and the 
substance is practically insoluble. 
In this regard, describing the role of hydrophobicity in protein folding by 
the traditional dictum ``water is a poor solvent for the unfolded state of 
globular proteins'' is less problematic because, semantically consistent with 
the ``poor solvent'' characterization, an overwhelming majority of the 
chain population is folded (i.e., not ``solvated'' as unfolded chains) 
in water. Nonetheless, the ``poor solvent'' narrative can also be misleading.
This is because in this case poor solvation in water applies only to
part of the protein molecule (e.g., nonpolar residues) but not every
part of the heteropolymeric protein chain. In fact, globular proteins are 
soluble as individual folded molecules in water. In other words, water is 
a good---not poor---solvent for the individual protein molecules as a whole. 
Unqualified usage of the ``poor solvent'' description may therefore mask 
important biophysical and functional distinctions among disordered protein 
states of different compactness,\cite{TobinJMBrev2020} differences that are 
crucial for categorizing noncooperative and various types of cooperative 
folding.\cite{chanetal2004,chanetal2011,kayaProteins2000}
Taking all the above into consideration, it is apparent that invocation 
of aqueous solubility of subpopulations of a protein's conformational 
ensemble risk unnecessary confusion and such narratives are not always 
conducive to a clear conceptualization of protein folding.
\\

%%%%%%%%%%%%%%%%%%%%%%%%%%%%%%%%%%%%%%%%%%%%%%%%%%%%%%%%%%%%%%%%%%%%%%%%%%%%%%%%
\begin{figure}[t]
\vskip -0.2cm
{\includegraphics[height=60mm,angle=0]{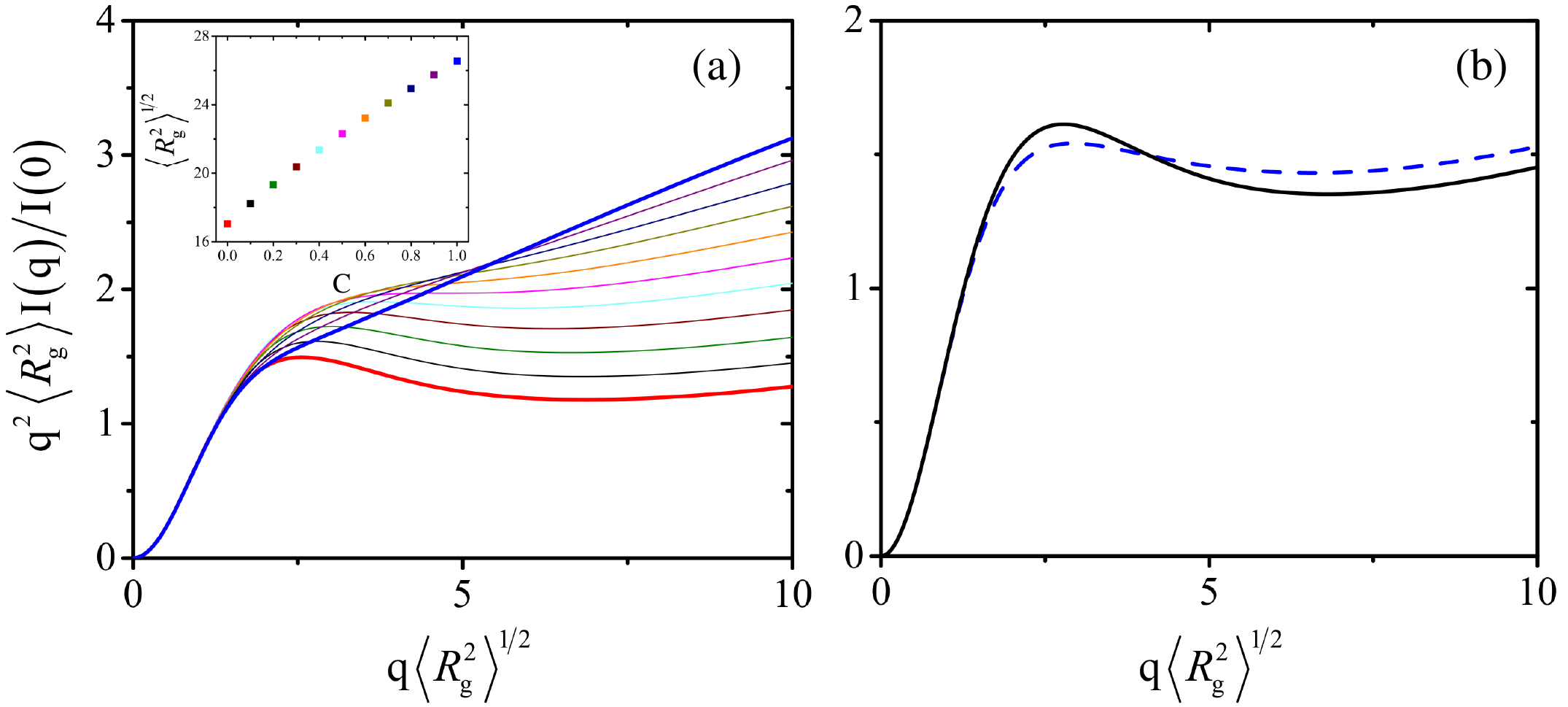}}
\begin{center}
\end{center}
\vspace{-1.3cm}
\caption{MFFs of composite ensembles.
(a) Dimensionless Kratky plots of the $C=0.1,0.2,\dots,0.9$
composite ensembles with $\ep^{(1)}=+2.0$ and $\ep^{(2)}=-2.0$
(Eq.~\ref{eq:compositeI}) with different
overall $\langle R^2_{\rm g}\rangle^{1/2}$ values (inset)
are plotted using the same color code.
Dimensionless Kratky plots for the homogeneous
$\ep^{(1)}=+2.0$ ($C=1.0$, dark blue)
and the homogeneous $\ep^{(2)}=-2.0$ ($C=0$, red) ensembles
are shown by thicker curves.
(b) Dimensionless Kratky plot for $C=0.1$ (solid curve) heteropolymeric
composite ensemble (overall $\langle R^2_{\rm g}\rangle^{1/2}=18.3$ \AA)
is compared with that for the $\ep=-1.8$ homogeneous ensemble
with $\langle R^2_{\rm g}\rangle^{1/2}=18.5$~\AA~(dashed curve).
The $I(q)$s of the $C$-dependent heterogeneous ensembles in (a)
are also compared using other plotting formats in Fig.~S7 of
Supporting Information.
}
\label{fig21}
\end{figure}
%%%%%%%%%%%%%%%%%%%%%%%%%%%%%%%%%%%%%%%%%%%%%%%%%%%%%%%%%%%%%%%%%%%%%%%%%%%%%%%%

{\bf Composite ensembles and experimental uncertainties.}
As emphasized above, in addition to the cases studied so far by us and 
by others, one expects that there are many other scenarios in which 
heterogeneous ensembles are not readily distinguishable from homopolymer
ensembles by MFFs alone. As a further example, consider a heterogeneous 
ensemble which is a superposition of two homogeneous ensembles 
defined by two different $\epsilon_{\rm p}$ values symbolized 
as $\epsilon_{\rm p}^{(1)}$ and $\epsilon_{\rm p}^{(2)}$. 
A situation similar to this hypothetical scenario readily arise when 
there is a mixture of folded and unfolded protein conformations in 
an overall ensemble,\cite{Plaxco1999,raleighBiochem2020} 
or when there is a bimodal-like distribution of 
conformational properties of an IDP, as has been suggested
computationally.\cite{tanja2010} Now, let the $\avRgs$ 
of the two component homogeneous ensembles be 
$\langle R_{\rm g}^2\rangle^{(1)}$ and $\langle R_{\rm g}^2\rangle^{(2)}$, 
respectively, and the corresponding be scattering intensities be 
$I^{(1)}(q)$ and $I^{(2)}(q)$. The $\avRgs$ of
a composite conformational ensemble with weights of $C$ and $1-C$ for the 
$\epsilon_{\rm p}^{(1)}$- and $\epsilon_{\rm p}^{(2)}$-ensembles 
($0\le C\le 1$), respectively, is then given by
\begin{equation}
\langle R_{\rm g}^2\rangle =
C\langle R_{\rm g}^2\rangle^{(1)}+(1-C)\langle R_{\rm g}^2\rangle^{(2)} \; ,
\label{eq:C_comp0}
\end{equation}
and the corresponding scattering intensity is given by
\begin{equation}
I(q)=CI^{(1)}(q)+ (1-C)I^{(2)}(q) \; ,
\label{eq:compositeI}
\end{equation}
and $I_0=CI^{(1)}(0)+ (1-C)I^{(2)}(0)$.
The SAXS signatures of such a system are shown in Fig.~21. 
A comparison is highlighted in Fig.~21b between the MFF of a $C=0.1$ 
heterogeneous composite ensemble (solid curve) and that of a homopolymer 
ensemble with the same $\avRgs^{1/2}$ (dashed curve). The result 
indicates that the two theoretical MFFs are distinct and therefore 
distinguishable in principle, especially in situation where a mixture
of conformational species is expected as in the study of protein 
folding/unfolding.\cite{Plaxco1999,raleighBiochem2020}
At the same time, the similarity between the MFFs also suggests that the 
heterogeneous and homogeneous ensembles may not be easily discriminated 
by their MFFs alone without prior knowledge of a complex conformational 
distribution and in the face of possible experimental 
uncertainties.\cite{DT2008} The theoretical approach we have taken here
is agnostic as to experimental accuracy, using a coarse-grained chain
model without atomistic details and ignoring any explicit consideration of 
solvation effects for the purpose of exploring conceptual principles.
Nonetheless, it is worth mentioning 
that some of the fitted homopolymeric MFFs for disordered protein ensembles 
reported in the literature exhibit significant deviations from experimental
data for $q$ values not much larger 
than those in the shoulder region of the dimensionless Kratky plot 
(e.g., the [KCl] = 0.15 M data in Fig.~3C of ref.~\citen{tobinSci2017} and
Fig.~S5 in ref.~\citen{tobinPNAS2019}). Other effects, such as those
associated with the hydration shell around the protein,\cite{skepo2018} may 
add further uncertainties in matching simulated and experimental MFFs and 
thus render the inference of conformational ensembles from MFFs more 
challenging.
\\

\noindent
{\large\bf CONCLUSIONS}\\ 

To recapitulate, basic physics stipulates that disordered conformational
ensembles of unfolded proteins and IDPs are, in general, heterogeneous 
because proteins are heteropolymers of amino acid residues.
MFFs obtained from SAXS of disordered protein ensembles are valuable
because they yield more structural information about conformational ensembles
from higher-$q$ parts of $I(q)$, whereas this information is unavailable 
from small-$q$ Guinier analyses which afford only an overall
$\avRgs$ (refs.~\citen{tobinSci2017,SciComment1,SciResponse2,SciComment3}).
Nonetheless, structural information from MFFs is still highly averaged.
As exemplified by recent simulation of several IDPs,\cite{DT2019}
the systematic analysis conducted in the present work has shown that 
the MFF of a heterogeneous ensemble is not always distinct from that of a 
homogeneous ensemble. In not a few instances,
MFFs cannot reliably distinguish between heterogeneous 
from homogeneous disordered conformational ensembles with the same $\avRgs$. 
Clearly, then, the conformational ensemble to MFF correspondence is
practically a many-to-one mapping. Not only that, this 
mapping is likely not smooth---meaning that while 
small changes in conformational ensemble lead to only
small changes in MFF, small changes or experimentally indiscernible minute
changes in MFF can map to structurally drastically different conformational 
ensembles. We have now demonstrated, mathematically, the many-to-one and 
non-smooth features of the (ensemble $\rightarrow$ MFF) mapping by 
identifying various heterogeneous ensembles with MFFs very similar to 
those of homopogeneous ensembles. These heterogeneous ensembles with 
homopolymer-like MFFs include some of the conformationally highly restricted
$(R_{\rm g},R_{\rm EE})$ subensembles\cite{Songetal2015,Songetal2017,DICE}
which may be viewed as members of a conformational-space basis 
set (Figs.~7--9), computationally selected reweighted ensembles with 
exceptionally similar MFFs but significanty different distributions of 
$R_{\rm g}^2$, $R^2_{\rm EE}$, and asphericity (Fig.~13) as well as
intuitively plausible ensembles with narrower-than-SAW distributions 
of $R^2_{\rm g}$ and compact-open composite ensembles (Figs. 10 and 21).
These counterexamples to any preconception or tacit assumption of a 
one-to-one ensemble-MFF mapping highlight the intrinsic logical 
uncertainty when one attempts to infer a conformational ensemble 
from a given MFF. 

In view of the limited current knowledge about the effects of 
the sequence-dependent\cite{tanja08,PatriciaPNAS2020}
physical interactions on conformational dimensions and other 
structural properties of IDPs and unfolded states of cooperatively-folding 
globular proteins,\cite{chanetal2011,DavidShaw2,cosb15} it is not possible 
to determine {\it  a priori} which ensemble/structural properties
of disordered proteins are physically encodable by amino acid sequences
and which properties are not.  As our ability to imagine what is physically 
possible for disordered protein conformations is so limited because of
this lack of knowledge, aside from obvious cases, one cannot preclude many of 
the heterogeous ensembles 
with homopolymer-like MFFs as physically unrealizable. Therefore, the
logico-mathematical uncertainty that we have established regarding 
any inference of conformational ensemble that is based solely on MFFs
translates into a practical uncertainty as well.
With this understanding in mind, caution should be used in not 
over-emphasizing the intrachain distance scaling exponent $\nu$--- which,
as it stands, is a fitting parameter that is not measured directly---as a 
proxy for $R_{\rm g}$
in SAXS data analysis of protein disordered ensembles\cite{tobinSci2017}
because it may lead to a misplaced impression that the ensembles in
question are homopolymer-like in every respect (Fig.~8), a view whose validty 
is physically unlikely if not corroborated by other 
measurements.\cite{raleighBiochem2020}
Indeed, our simulations presented above of analytical theory-inspired 
explicit-chain models with heteropolymeric interactions (Figs.~17--20) 
have provided ample examples of $R_{\rm g}$--$R_{\rm EE}$ decoupling in 
disordered conformational ensembles,\cite{Songetal2017,lemke2017} 
even though our coarse-grained
modeling with only short spatial range contact-like interactions has not 
yet explored potentially much more complex and diverse impacts on
conformational preference that may emerge from sidechain sterics,
hydrogen bonding, $\pi$-related interactions, electrostatic interactions
with longer spatial ranges, and solvation effects.
Taken together, our findings underscore the importance of using
multiple complementary experimental 
probes to gain insight into disordered protein 
ensembles, as exemplified by recent efforts in using combined data from SAXS, 
NMR and smFRET to provide more extended structural information about
such ensembles which turn out---according to multiple-probe analyses---to be 
heterogeneous.\cite{benPNAS2016,lemke2017,raleighPNAS2019,GregJACS2020,raleighBiochem2020}
Considering the vast possibilities of amino acid sequence-encoded
conformational ensembles,
these promising advances have only begun to uncover the intricacy 
of disordered biomolecular configurations and how they might
underpin biological functions.
\\

\vfill\eject

\noindent
{\large\bf Supporting Information Description}\\
Relationship between the present heteropolymeric
scattering intensity in Eqs.~\ref{Iq-eq0}, \ref{Fz-eq} and previous 
results for homopolymers, and supporting figures.
\\

\noindent
The authors declare no competing financial interest.
\\

\noindent
{\large\bf Acknowledgements}\\
We thank Yawen Bai, Upayan Baul, Robert Best, Osman Bilsel, Patricia Clark, 
Julie Forman-Kay, Kingshuk Ghosh,
Gregory-Neal Gomes, Claudiu Gradinaru, Elisha Haas, Alex Holehouse,
Ed Lattman, Edward Lemke, Yi-Hsuan Lin, Jeetain Mittal, Rohit Pappu, 
Kevin Plaxco,
Joshua Riback, Ben Schuler, Andrea Soranno, Tobin Sosnick, Dev
Thirumalai, Sara Vaiana, and Wenwei Zheng for helpful discussions,
some over many years.
Part of these fruitful exchanges took place during workshops organized
by the Protein Folding and Dynamics Research Coordination Network 
funded by the U.S. National Science Foundation (NSF grant MCB 1516959
to C. R. Matthews) and Biophysical Society 
Meetings.\cite{Songetalabs1,Songetalabs2}
Financial support of this work was provided by Canadian Institutes of Health 
Research grants MOP-84281, NJT-155930, Natural Sciences and Engineering 
Research Council of Canada Discovery grant RGPIN-2018-04351 to H.S.C.,
and National Natural Science Foundation of China grant 21674055 to J.S.
Part of our computational resources has been afforded generously by 
Compute/Calcul Canada.
\\

%%%%%%%%%%%%%%%%%%%%%%%%%%%%%%%%%%%%%%%%%%%%%%%%%%%%%%%%%%%%%%%%%%%%%%%%%%%%%
%YYY%
%\vfill\eject \input FRETrefs_SAXS
%\vfill\eject
%\input test1_figs
%\vfill\eject
%\input SAXS_SI

%%%%%%%%%%%%%%%%%%%%%%%%%%%%%%%%%%%%%%%%%%%%%%%%%%%%%%%%%%%%%%%%%%%%%%%%%%%%%%%%

$\null$

\noindent
{\Large\bf References}\\

%%%%%%%%%%%%%%%%%%%%%%%%%%%%%%%%%%%%%%%%%%%%%%%%%%%%%%%%%%%%%%%%%%%%%%%%%%%%%%%
%
\vfill\eject
\input Songetal_SAXS_SI

%%%%%%%%%%%%%%%%%%%%%%%%%%%%%%%%%%%%%%%%%%%%%%%%%%%%%%%%%%%%%%%%%%%%%%%%%%%%%%%%

%%%%%%%%%%%%%%%%%%%%%%%%%%%%%%%%%%%%%%%%%%%%%%%%%%%%%%%%%%%%%%%%%%%%%%%%%%%%%%%%
\vfill\eject\end{document}

%% file: Songetal_SAXS_SI.tex
\vfill\eject
$\null$
\renewcommand{\thepage}{S\arabic{page}}% <--- Insert letter S to page number
\setcounter{page}{1}
\renewcommand{\thefigure}{S\arabic{figure}}% <--- Insert letter S to a fig.
\renewcommand{\thetable}{S\arabic{table}}%  <--- Insert letter S to a table.
\renewcommand{\theequation}{{\rm S}\arabic{equation}}
\setcounter{figure}{0}

\vskip 1cm

\begin{center}

{\Huge\bf Supporting Information}\\
\vskip 0.15cm

{\Large{\textit{\textbf{for}}}}\\
\vskip 0.15cm

{\Large\bf ``Small-Angle X-Ray Scattering Signatures of}\\
\vskip 0.15cm

{\Large\bf Conformational Heterogeneity and}\\
\vskip 0.15cm

{\Large\bf Homogeneity of Disordered Protein Ensembles''}\\

\vskip .3in

{\large\bf Jianhui S{\footnotesize{\bf{ONG}}}}$^{1*}$,
{\large\bf Jichen L{\footnotesize{\bf{I}}}}$^{1}$ and
{\large\bf Hue Sun C{\footnotesize{\bf{HAN}}}}$^{2*}$

$\null$
{\large
$^1$ School of Polymer Science and Engineering,
Qingdao University of\\ Science and Technology,
53 Zhengzhou Road, Qingdao 266042, China;\\
$^2$ Department of Biochemistry,
University of Toronto,\\
Toronto, Ontario M5S 1A8, Canada
\\
$\null$\\

$^*$Corresponding authors.\\
Email: {\tt chan@arrhenius.med.utoronto.ca} (H.S.C.)\\ 
%{\phantom{Email:\ }}
{\tt jhsong@qust.edu.cn} (J.S.)
}

\end{center}

\vfill\eject

\noindent
{\bf Relationship Between The Heteropolymer Results In Eqs.~12 and 13
of the Main Text and the Homopolymer Results in Ohta et al.,
{\textit{\textbf Phys. Rev. A}} {\bf 1982}, 
{\textit{\textbf 25}}, 2801--2811.}\\ 

\noindent
({\it The reference numbers used in the paragraph 
below are those in the main text. The cited references are 
listed again at the bottom of this page.})\\

A detailed comparison of Eq.~\ref{Iq-eq0} 
and Eq.~\ref{Fz-eq} of the main text
for the special case of $v_{ij}=v_0$ against 
Eqs.~(3.3)--(3.8) in Ohta et al. \cite{ohta82} indicates 
that the two sets of results are consistent provided that several possible
typographical errors in ref.~\citen{ohta82} are corrected, as explained 
below. In this comparison, we recognize that the present $v_0$ is 
defined (see comparison in the main text 
with refs.~\citen{chanJCP89,chanJCP90}) to correspond to 
$v_0/(2\pi)^{d/2}=v_0(2\pi)^{-2+\epsilon/2}$ in ref.~\citen{ohta82}, where 
$d$ is the number of spatial dimensions, and $\epsilon\equiv 4-d$ is a
renormalization group expansion parameter, not to be confused with
the intrachain interaction energy $\epsilon_{\rm p}$.
Moreover, our $\alpha N=Nq^2/6$ is equivalent to 
the variable $\beta_0=N_0k^2/2$ in ref.~\citen{ohta82} because
our chain length $N$ corresponds to their ``bare'' chain length
$N_0$ before renormalization but our $q^2/6$ is equivalent to their
$k^2/2$ ($k^2$ = square of wave vector magnitude) because, as 
stated in the main text, unlike ref.~\citen{ohta82} and also 
unlike refs.~\citen{chanJCP89}~and~\citen{chanJCP90}, 
here we do not rescale the spatial
coordinates ${\bf r}(\tau)$ to ${\bf c}(\tau)=\sqrt{d}{\bf r}(\tau)$. 
The possible typographical errors in ref.~\citen{ohta82} are: 
(i) the term $-(\vec{\bf k}-\vec{\bf k}^\prime)z^2/2$ in the third
line of Eq.~(3.3) of ref.~\citen{ohta82} should read
$-|\vec{\bf k}-\vec{\bf k}^\prime|^2 z^2/2$, 
(ii) an overall multiplicative factor of $-v_0$ is likely missing
in the expression for $\mu$ on the right hand side of the
first line of Eq.~(3.4) in ref.~\citen{ohta82},
(iii) an overall
multiplicative factor of $x^{\epsilon/2}$ is likely missing in
the integrand on the
right hand side of Eq.~(3.5) for $V_2$ in ref.~\citen{ohta82}, and (iv)
the $\int_0^{x}dy$ integral in Eq.~(3.6) 
for $V_{31}$ in ref.~\citen{ohta82}
should likely be $\int_0^{1-x}dy$. If these four suggested corrections
in Eqs.~(3.3)--(3.6) of ref.~\citen{ohta82} are made,
the term independent of $v_{ij}$ (first term) in our expression for $I(q)$ 
in the main-text Eq.~\ref{Iq-eq0} would be equal to two 
times the expression given by Eq.~(3.2) of ref.~\citen{ohta82}; and 
the $v_{ij}=v_0$ case of the term proportional to $v_{ij}$ 
in our main-text Eq.~\ref{Iq-eq0} can be verified by straightforward though
somewhat tedious algebra to be equal to
$2(U_1+U_2+U_3+U_4+U_5+U_6)$ for $d=3$ ($\epsilon=1$), 
where the $U$s are defined in Eq.~(3.3) 
of ref.~\citen{ohta82}. It should be noted that the overall factor of 2 in our
expression for $I(q)$ arises from counting a pair of scattering positions 
twice instead of once, as in Eq.~(2) of ref.~\citen{ohta81}.
Such an overall factor is inconsequential in the ratio $I(q)/I(0)$.
We have also verified the linear and logarithmic $a\rightarrow 0$ 
divergent terms in Eq.~(3.8) of Ohta et al.\cite{ohta82} by
considering the $d=4$, $v_{ij}=v_0$ version of the $O(v_{ij})$ term 
in our main-text Eq.~\ref{Iq-eq0}---which entails substituting the
single $\Delta\tau_{ij}^{-3/2}$ factor by $\Delta\tau_{ij}^{-2}$---and 
performing the integrals $\int_a^N d\tau_j \int_0^{\tau_j-a}d\tau_i$
as $\int_0^{N-a} d\tau_i \int_a^{N-\tau_i} d\Delta\tau_{ij}$.
The resulting $a\rightarrow 0$ ($N/a\rightarrow\infty$) divergent term 
is exactly equal to two times the expression given in Eq.~(3.8) of 
Ohta et al. \cite{ohta82} when our $v_0$ corresponds to their 
$u_0/4\pi^2$ for $d=4$ ($\epsilon=0$, see above).

\noindent
---------------------------------------------------------------------------------------------------------------------------------------\\
{\footnotesize{
$^{104}$ Ohta, T.; Oono, Y.; Freed, K. F. {\it Macromolecules}
{\bf 1981}, {\it 14}, 1588--1590.
\vskip -1mm
\noindent
$^{106}$ Ohta, T.; Oono, Y.; Freed, K. F. {\it Phys. Rev. A} {\bf 1982}
{\it 25}, 2801--2811.
\vskip -1mm
\noindent
$^{108}$ Chan, H. S.; Dill, K. A. {\it J. Chem. Phys.} {\bf 1989},
{\it 90}, 492--509; {\it ibid}. {\bf 1992}, {\it 96}, 3361.
\vskip -1mm
\noindent
$^{109}$ Chan, H. S.; Dill, K. A. {\it J. Chem. Phys.} {\bf 1990},
{\it 92}, 3118--3135; {\it ibid}. {\bf 1997}, {\it 107}, 10353.
}}

\vfill\eject
\centerline{\large\bf Suppporting Figures}

\input arXiv_SI_figs

%% file: arXiv_SI_figs.tex
%\noindent {\large\bf Supporting Figures} \\ $\null$\\

$\null$\\
%\usepackage{caption}
%\captionsetup[figure]{labelfont={bf},name={SFig.},labelsep=period}

\begin{figure}[h]
{\includegraphics[height=145mm,angle=0]{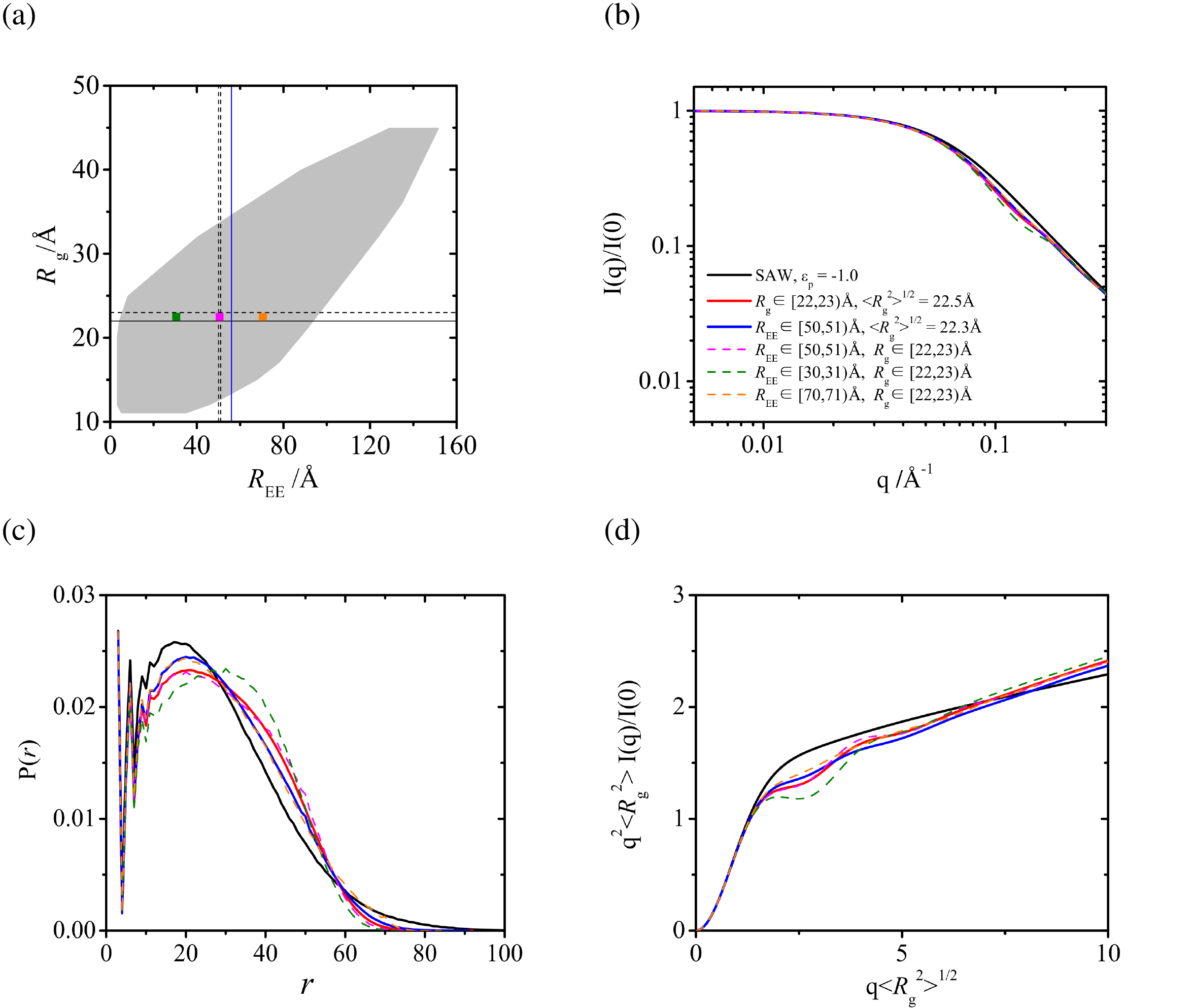}}
\begin{center}
\end{center}
\vspace{-1.0cm}
\caption{%{\bf SFig.1.}$\quad$
Comparing SAXS properties of the $\ep=-1.0$ homopolymer ensemble and
$(R_{\rm g},R_{\rm EE})$ subensembles of $\ep=0$ SAW chains.
(a) The homogeneous (homopolymer) ensemble is represented by the
gray area corresponding to the overall profile for $\ep=-1.0$ in Fig.~2c 
of the main text; the $\langle R_{\rm EE}^2\rangle^{1/2}$ of this ensemble 
is indicated by the vertical blue solid line; the notation
follows that in Fig.~7 of the main text otherwise. Note that the subensembles
marked in (a) and analyzed in (b)--(d) here are sampled from the pure SAW
($\ep=0$) ensemble, {\it not} from the $\ep=-1.0$ ensemble.
}
\label{sfig1}
\end{figure}

\vfill\eject

\begin{figure}[t]
{\includegraphics[height=145mm,angle=0]{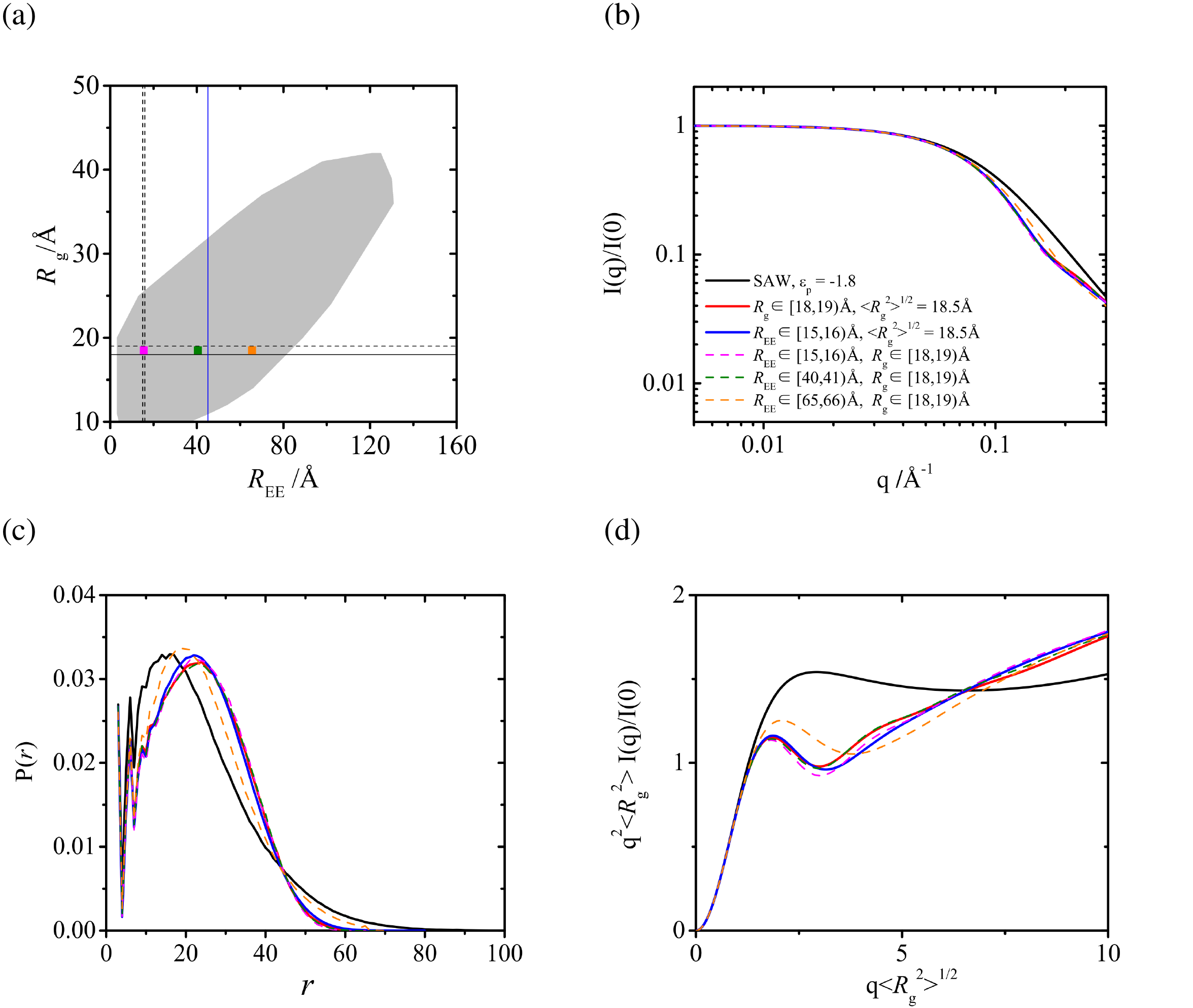}}
\begin{center}
\end{center}
\vspace{-1.0cm}
\caption{%{\bf SFig.2.}$\quad$
Comparing SAXS properties of the $\ep=-1.8$ homopolymer ensemble and
$(R_{\rm g},R_{\rm EE})$ subensembles of $\ep=0$ SAW chains.
Same notation as Fig.~S1 except the full homogeneous (homopolymer) ensemble here
is for $\ep=-1.8$.
}
\label{sfig2}
\end{figure}

\vfill\eject

\begin{figure}[t]
{\includegraphics[height=145mm,angle=0]{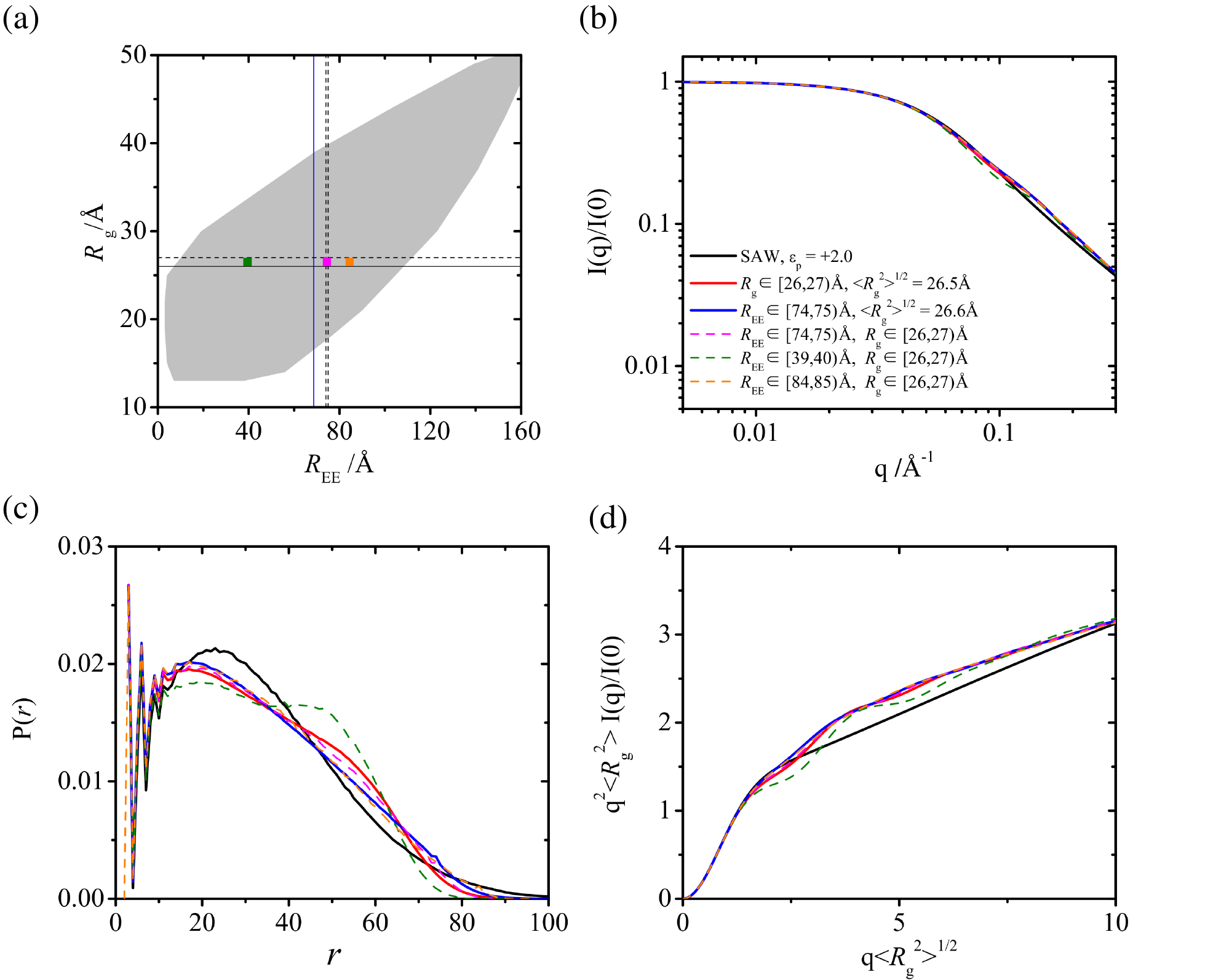}}
\begin{center}
\end{center}
\vspace{-1.0cm}
\caption{%{\bf SFig.3.}$\quad$
Comparing SAXS properties of the $\ep=+2.0$ homopolymer ensemble and
$(R_{\rm g},R_{\rm EE})$ subensembles of $\ep=0$ SAW chains.
Same notation as Fig.~S1 except the full homogeneous (homopolymer) ensemble here
is for $\ep=+2.0$.
}
\label{sfig3}
\end{figure}

\vfill\eject

\begin{figure}[t]
{\includegraphics[height=125mm,angle=0]{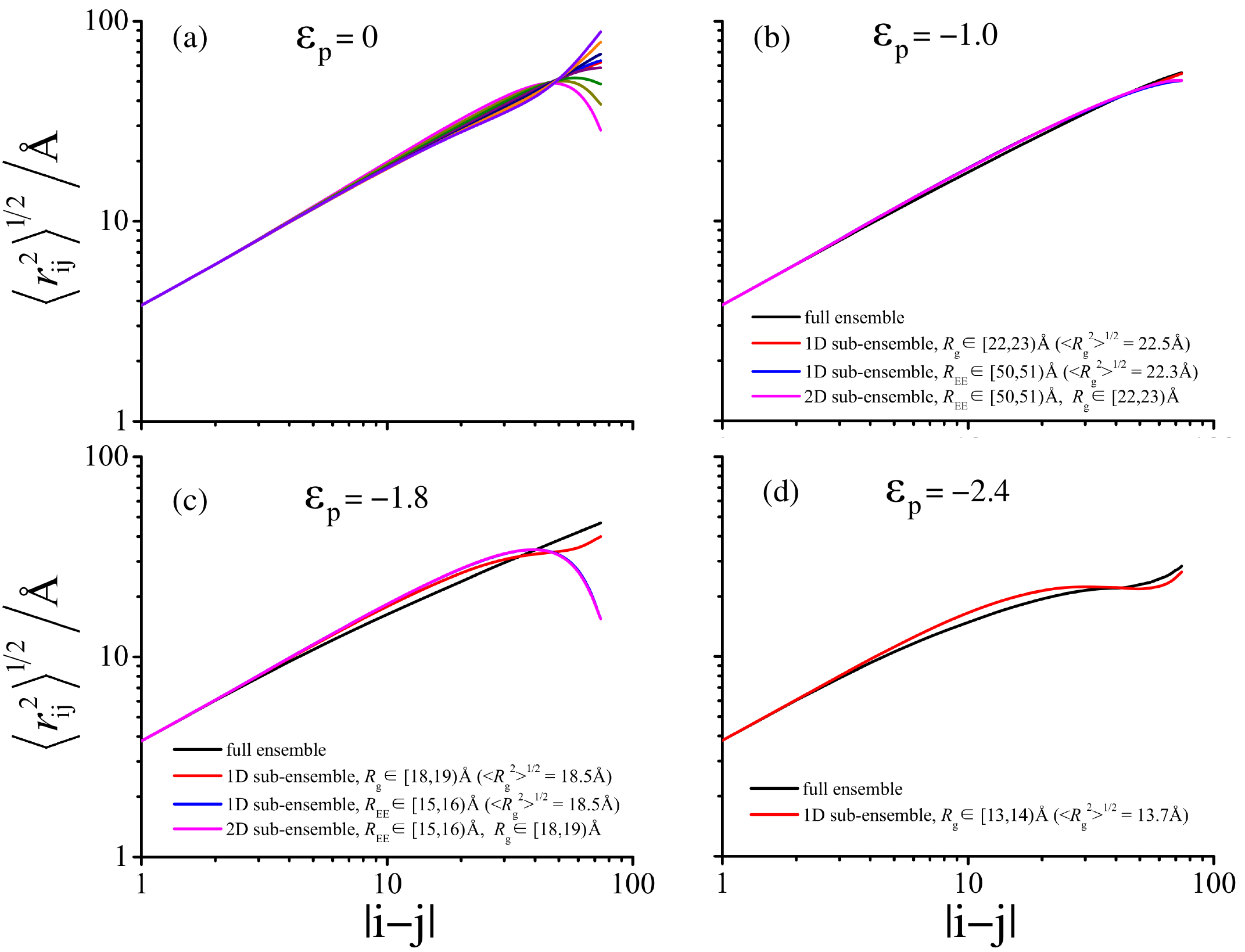}}
\begin{center}
\end{center}
\vspace{-1.0cm}
\caption{%{\bf SFig.4}$\quad$
Comparing the $\langle r^2_{ij}\rangle^{1/2}$ vs $|i-j|$ relationships
of homogeneous (homopolymer) ensembles and heteropolymeric 
$(R_{\rm g},R_{\rm EE})$ subensembles with
essentially identical root-mean-square radii of gyration 
$\langle R^2_{\rm g}\rangle^{1/2}$.
(a) Full $\ep=0$ SAW homopolymer 
ensemble and its subensembles considered in Fig.~8
of the main text. Using the same color code for the lines, the present 
plot provides their $\langle r^2_{ij}\rangle^{1/2}$ as functions of $|i-j|$ 
over a wider range of $|i-j|$ than that of the main-text figure.
(b)--(d) Select full $\ep\neq 0$ homogeneous ensembles (from Fig.~3b of
the main text) are compared with select heteropolymeric subensembles 
sampled from the pure $\ep=0$ SAW homopolymer ensemble 
({\it not} from an $\ep\neq 0$ 
ensemble) as specified by the legends. Here, a ``1D subensemble''
refers to a subensemble with a narrow range of either $R_{\rm g}$ 
or $R_{\rm EE}$ but not both, and is unrestricted otherwise; whereas
a ``2D subensemble'' is a subensemble with 
narrow ranges of $R_{\rm g}$ as well as $R_{\rm EE}$.
}
\label{sfig4}
\end{figure}

\vfill\eject

\begin{figure}[t]
{\includegraphics[height=140mm,angle=0]{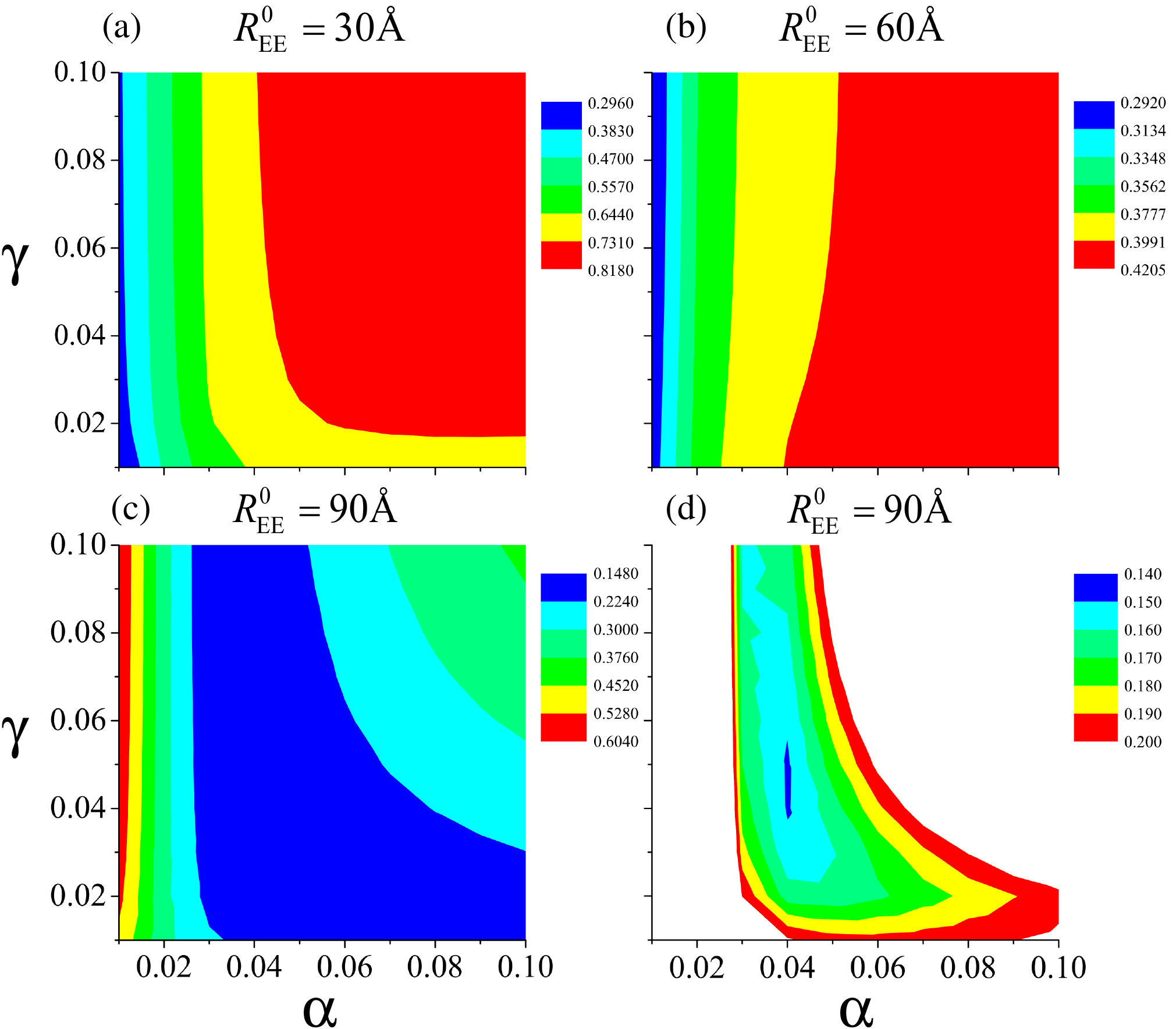}}
\begin{center}
\end{center}
\vspace{-1.0cm}
\caption{%{\bf SFig.5}$\quad$
Extensive cataloging of SAXS properties of reweighted heterogeneous 
ensembles. $\Delta_{\rm Kratky}$ between $(R^0_{\rm EE},\alpha,\gamma)$-defined
reweighted ensembles (see main text for details) and the homogeneous SAW 
($\ep=0$) homopolymer ensemble is computed 
for three select $R^0_{\rm EE}$ values:
(a) 30 \AA, (b) 60 \AA, and (c,d) 90 \AA, each for
a $10\times 10$ grid ($0.01$ increments) of $\alpha,\gamma$ values.
Shown here (a--d) are the resulting contour plots, with (c) and (d) 
for the same $R^0_{\rm EE}=90$ \AA~data plotted with different contour 
increments.
For the systems considered, the $(R^0_{\rm EE}/{\rm \AA},\alpha,\gamma)$ 
at the minimum $\Delta_{\rm Kratky}$ values encountered (approximate
minimum $\Delta_{\rm Kratky}$ in curly brackets) are as follows:
(30, 0.01, 0.01) \{0.30\},
(60, 0.01, 0.1) \{0.30\}, and
(90, 0.04, 0.05) \{0.14\}.
Similar analyses are performed for $\ep=-1.0$, $-1.8$, and $+2.0$
(contour plots not shown), the resulting minimum-$\Delta_{\rm Kratky}$
$(R^0_{\rm EE}/{\rm \AA},\alpha,\gamma)$ parameters and approximate
minimum $\Delta_{\rm Kratky}$ values are (same notation as above):
For $\ep=-1.0$,
(22.7, 0.01, 0.01) \{0.36\},
(52.7, 0.01, 0.01) \{0.31\}, and
(82.7, 0.06, 0.06) \{0.14\};
for $\ep=-1.8$,
(11.8, 0.02, 0.01) \{0.30\},
(41.8, 0.01, 0.01) \{0.31\}, and
(71.8, 0.07, 0.03) \{0.16\};
and for $\ep=+2.0$,
(35.0, 0.01, 0.01) \{0.30\},
(65.0, 0.01, 0.01) \{0.30\}, and
(95.0, 0.04, 0.1) \{0.21\}.
}
\label{sfig5}
\end{figure}

\vfill\eject

\begin{figure}[t]
{\includegraphics[height=38mm,angle=0]{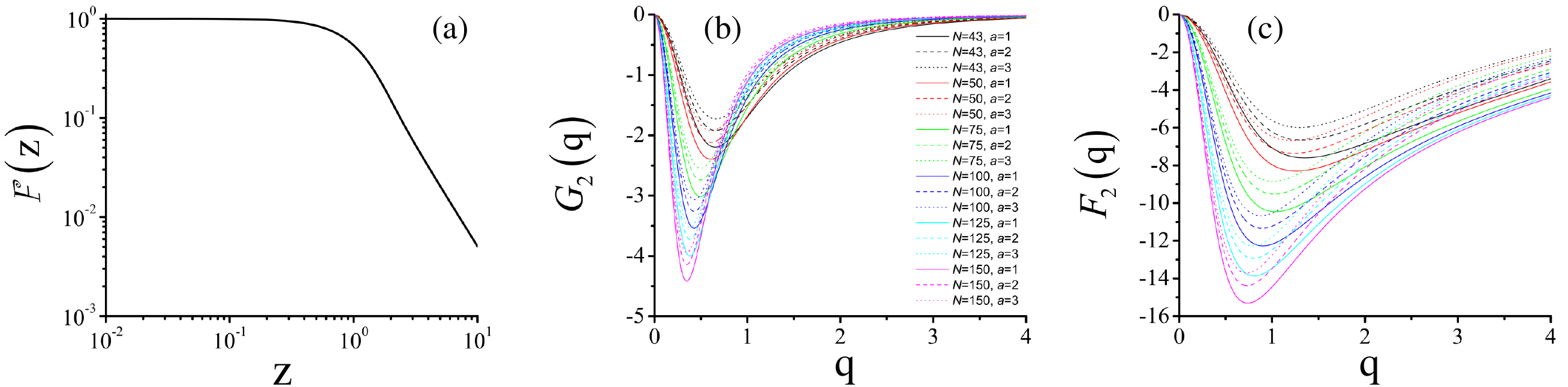}}
\begin{center}
\end{center}
\vspace{-1.0cm}
\caption{%{\bf SFig.6}$\quad$
Useful mathematical functions in the present perturbative 
treatment of scattering intensities.
(a) ${\cal F}(z)$ is the function defined in Eq.~\ref{Fz-eq} of the main text.
(b) $G_2(q)$ is the $G_2(N,q;a)$ function in
Eq.~\ref{eq:G2-def} of the main text, shown here for a variety of $N$ 
and $a$ values as specified by the inset legend.
(c) $F_2(q)$ is the $F_2(N,q;a)$ function in Eq.~\ref{eq:F2} of the main text
for the same set of $N$ and $a$ values in (b) plotted in the same line styles
as those in (b).
}
\label{sfig6}
\end{figure}

\vfill\eject

\begin{figure}[t]
{\includegraphics[height=42mm,angle=0]{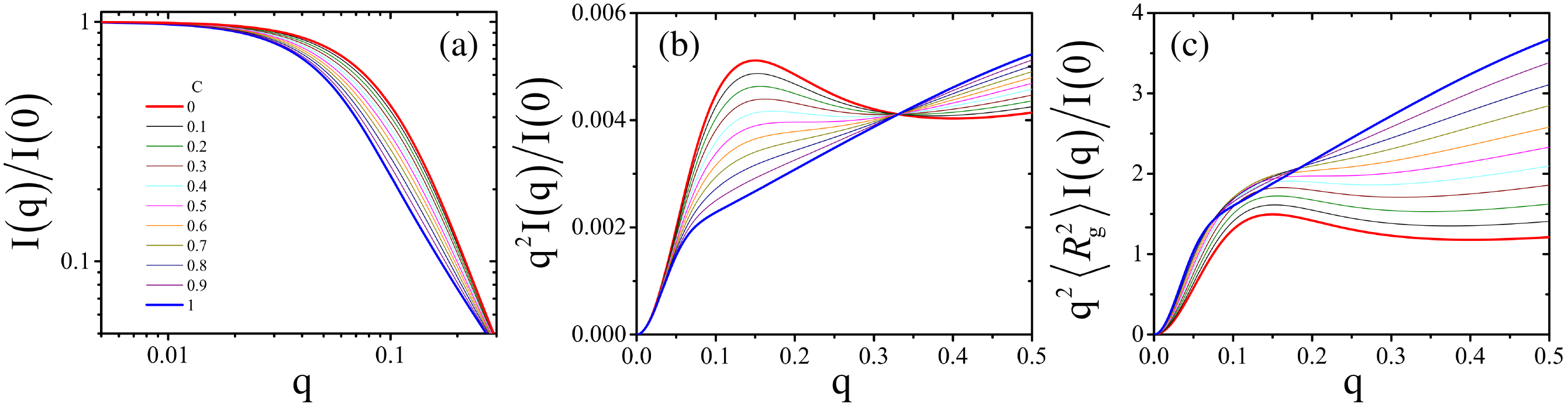}}
\begin{center}
\end{center}
\vspace{-1.0cm}
\caption{%{\bf SFig.7}$\quad$
SAXS properties of the homogeneous ($C=0,1$) and
composite ($C\neq 0,1$) ensembles with $\ep^{(1)}=+2.0$ and $\ep^{(2)}=-2.0$
in Fig.~21 of the main text.
The color code for different $C$ values is identical to that in the
main-text figure.
}
\label{sfig7}
\end{figure}

\vfill\eject